\documentclass[11pt]{article}
\pdfoutput=1
\usepackage{graphicx}
\graphicspath{{./figures/}}
\usepackage{appendix}
\usepackage{latexsym,amsmath,amsfonts,amssymb,booktabs}
\usepackage[font=small]{caption}
\usepackage{slashed,upgreek,amscd,cancel,tensor,color}
\usepackage{adjustbox}
\usepackage{soul}
\usepackage[normalem]{ulem}
\usepackage[numbers,compress,square]{natbib}
\usepackage{epsfig,latexsym}
\usepackage{amsmath}
\usepackage[pdfencoding=auto]{hyperref} 
\usepackage{url}
\numberwithin{equation}{section}
\usepackage{doi}
\usepackage{subcaption}
\definecolor{MyBlue}{rgb}{0.15,0.15,0.70}
\definecolor{linkblue}{rgb}{0,0,0.8}
\definecolor{linkgreen}{rgb}{0,0.5,0}

\hypersetup{
colorlinks=true,
citecolor=linkgreen,
linkcolor=linkblue,
urlcolor=linkblue
}

\input epsf
\newcommand{\vk}{\vec{k}}
\newcommand{\vq}{\vec{q}}

\newcommand{\vx}{\vec{x}}
\def\xfl{{\vec x_{\rm fl}}}
\newcommand{\sfrac}[2]{{\textstyle\frac{#1}{#2}}}

\usepackage{amssymb}
\usepackage{amsmath}
\usepackage{amsfonts}
\usepackage{upgreek}
\usepackage{latexsym}
\usepackage{stfloats}
\usepackage{afterpage}

\numberwithin{equation}{section}

\newcommand{\bea}{\begin{eqnarray}}
\newcommand{\eea}{\end{eqnarray}}
\newcommand{\be}{\begin{equation}}
\newcommand{\ee}{\end{equation}}
\newcommand{\fr}[2]{\frac{ #1}{#2}}

\newcommand{\cH}{\mathcal{H}}

\newcommand{\knl}{k_{\rm NL}}
\newcommand{\kvec}{\vec{k}}
\newcommand{\qvec}{\vec{q}}
\newcommand{\xvec}{\vec{x}}

\newcommand{\half}{\frac{1}{2}}

\newcommand{\eqn}[1]{Eq.~(\ref{#1})}

\newcommand{\unitsk}{\, h \, { \rm Mpc^{-1}}}
\newcommand{\pd}{\partial}
\newcommand{\om}{\Omega_{\rm m}}
\newcommand{\omn}{\Omega_{\rm m,0}}
\newcommand{\deltaone}{\tilde \delta^{(1)}}

\newcommand{\mpl}{M_{\rm Pl}}
\newcommand{\andd}{\ , \quad \text{and}  \quad}
\newcommand{\Om}{\Omega_{\rm m}}
\newcommand{\secref}[1]{Sec.~\ref{#1}}

\newcommand{\appref}[1]{App.~\ref{#1}}

\newcommand{\tr}{\text{tr}}

 \definecolor{MattOrange}{rgb}{1.0,0.4,0.2}

\newcommand{\Comment}[1]{{}}

\begin{document}
\def\thefootnote{\fnsymbol{footnote}}
%\begin{titlepage}
\setcounter{page}{1} \baselineskip=15.5pt \thispagestyle{empty}

\vspace*{-25mm}

\begin{flushright}
{\small NUHEP-TH/22-09}
\end{flushright}

\vspace{0.5cm}

\begin{center}

{\Large \bf The one-loop bispectrum of galaxies in redshift space\\[0.2cm] from the Effective Field Theory of Large-Scale Structure}  \\[0.7cm]
{\large   Guido D'Amico${}^{1,2}$,  Yaniv Donath${}^{3}$, Matthew Lewandowski${}^{4,5}$,\\[0.3cm] Leonardo Senatore${}^{5}$, and  Pierre Zhang${}^{{5},6,7,8}$ \\[0.7cm]}

\end{center}

%matt
\begin{center}

\vspace{.0cm}

\begin{small}

{ \textit{  $^{1}$ Department of Mathematical, Physical and Computer Sciences,\\ University of Parma, 43124 Parma, Italy}}
\vspace{.05in}

{ \textit{  $^{2}$ INFN Gruppo Collegato di Parma, 43124 Parma, Italy}}
\vspace{.05in}

{ \textit{ $^{3}$ Department of Applied Mathematics and Theoretical Physics,\\
University of Cambridge, Cambridge, CB3 OWA, UK}}
\vspace{.05in}

{ \textit{  $^{4}$ Department of Physics and Astronomy,\\ Northwestern University, Evanston, IL 60208}}
\vspace{.05in}

{ \textit{  $^{5}$ Institut fur Theoretische Physik, ETH Zurich,
8093 Zurich, Switzerland}}
\vspace{.05in}

{ \textit{ $^{6}$ Department of Astronomy, School of Physical Sciences, \\
University of Science and Technology of China, Hefei, Anhui 230026, China}}
\vspace{.05in}

{ \textit{ $^{7}$ CAS Key Laboratory for Research in Galaxies and Cosmology, \\
University of Science and Technology of China, Hefei, Anhui 230026, China}}
\vspace{.05in}

{ \textit{ $^{8}$ School of Astronomy and Space Science, \\
University of Science and Technology of China, Hefei, Anhui 230026, China}}
\vspace{.05in}

\end{small}
\end{center}

\vspace{0.5cm}

\begin{abstract}
We derive {the Effective Field Theory of Large-Scale Structure kernels and counterterms} for the one-loop bispectrum of dark matter and of biased tracers in real and redshift space. This requires the expansion of biased tracers up to fourth order in fluctuations. In the process, we encounter several subtleties related to renormalization. One is the fact that, in renormalizing the momentum, a local counterterm contributes non-locally. A second subtlety is related to the renormalization of local products of the velocity fields, which need to be expressed in terms of the renormalized velocity in order to preserve Galilean symmetry. We check that the counterterms we identify are necessary and sufficient to renormalize the one-loop bispectrum at {leading and} subleading order in the derivative expansion. The kernels that we originally present here have already been used for the first analyses of the one-loop bispectrum in BOSS data~\cite{DAmico:2022osl,DAmico:2022gki}.
\end{abstract}

\newpage

%\newpage
\tableofcontents

\vspace{.5cm}
%\newpage

\def\thefootnote{\arabic{footnote}}
\setcounter{footnote}{0}

%%%%%%%%%%%%%%%%%%
%
%
%     Introduction
%
%
%%%%%%%%%%%%%%%%%

%%%%%%%%%%%%%%%%%%
%
%
%

\section{Introduction and Conclusion}  \label{sec:intro}

The Effective Field Theory of Cosmological Large-Scale Structure (EFTofLSS)~\cite{Baumann:2010tm,Carrasco:2012cv} describes the long distance dynamics of matter and galaxies in the universe.
It is quite a complex endeavor. The dark matter~\cite{Baumann:2010tm,Carrasco:2012cv} and baryons~\cite{Lewandowski:2014rca,Braganca:2020nhv} are described through their density and momenta, and satisfy some equations of motion that resemble those of fluids. Galaxies are described as composite operators in terms of the dark-matter long-wavelength fields (see e.g.~\cite{ Senatore:2014eva, Mirbabayi:2014zca, Angulo:2015eqa, Fujita:2016dne, Perko:2016puo, Nadler:2017qto,Donath:2020abv}, and also~\cite{McDonald:2009dh}). The effect on short fluctuations of long wavelength modes that represent displacements needs to be resummed~\cite{Senatore:2014vja,Baldauf:2015xfa,Senatore:2017pbn,Lewandowski:2018ywf,Blas:2016sfa}. Predictions for observables, such as correlation functions of galaxies in redshift space, typically involve all of these ingredients. It took quite a large and long endeavor to develop all of this formalism (see~\cite{DAmico:2022osl} for a recent, more complete, list of references on the various steps of development of the theory). 

Starting from~\cite{DAmico:2019fhj,Ivanov:2019pdj,Colas:2019ret}, the EFTofLSS has been successfully applied to large-scale structure data, specifically to BOSS data \cite{BOSS:2016wmc}, where the analysis of the full shape of the power spectrum has led to the measurement of all the $\Lambda$CDM parameters using just a prior from Big Bang Nucleosynthesis. Since then, many applications to data have followed (see again~\cite{DAmico:2022osl} for a recent, more complete, list of references on the various applications of the EFTofLSS to data).

Recently, in~\cite{DAmico:2022osl,DAmico:2022gki}, the analysis of the BOSS data using the one-loop prediction of the bispectrum of galaxies in redshift space was performed. Doing such an analysis required the development of the kernels for biased tracers in redshift space up to fourth order, and the real and redshift space counterterms up to second order in the fluctuations and at subleading order in the derivatives ({\it i.e.} at order $k^2/\knl^2$ with $k$ being the typical wavenumber of interest, and $\knl$ the wavenumber associated to the non-linear scale). As referred to in the same papers~\cite{DAmico:2022osl,DAmico:2022gki}, those kernels are originally derived here.

While naively it might appear that there is only a computational challenge facing us, {\it i.e.} the need to write all possible operators for biased tracers in redshift space up to fourth order, in reality there are also two conceptual subtleties we will need to face, and that we now explain. The first such subtlety stems from the non-local Green's function associated to the momentum operator, while the second is associated to the renormalization of composite operators involving the velocity. 

%\vspace{0.5cm}

\paragraph{The local counterterm that contributes non-locally:} Let us start from the first subtlety, focussing initially on the case of dark matter. So far, the counterterms in the EFTofLSS have been explored at high order only for dark matter {in real space}. This fact has prevented the emergence of a subtlety that, on hindsight, is rather straightforward. The equations in the Newtonian limit contain the Poisson equation, whose solution is famously not local in space. This is mapped for example in the non-locality of the perturbative kernels. In fact, even though the absence of a tree-level speed of sounds makes the kernels just space dependent (rather than spacetime dependent), the dependence on the spatial wavenumber is not analytic, so that, once written in real space, they are non-local. For example, the solution of the locally-observable tidal tensor of the gravitational field, $\Phi$, due to a density perturbation, $\delta$, is, schematically,
\be
\pd_i\pd_j\Phi(\vec x,t)\sim H^2\frac{\pd_i\pd_j}{\pd^2}\delta ( \xvec , t ) \sim H^2 \int d^3x' \frac{1}{|\vec x-\vec x'|} \frac{\pd}{\pd x'^i}\frac{\pd}{\pd x'^j} \delta(\vec x',t)\ ,
\ee
with $H$ being the Hubble constant. This is non-local unless $i=j$ and we sum over $i$. 

As we {discuss in more detail} later, counterterms are local, {\it i.e.} the response of the stress tensor to the long wavelength fields is local. But the way the counterterms contribute to the fields is through a convolution with the Green's functions of the fields themselves, which, as we mentioned, are not local. {This subtlety does not show up for linear counterterms, though.} At that order, for the dark matter overdensity, only the divergence of the momentum matters, which in turn is affected, {at linear level}, by $\Phi$ only through $\pd^2\Phi\sim \delta$. So, the linear equation and the resulting Green's function are accidentally local. In this way, once one uses the counterterms at linear order ({\it i.e.} not multiplied by other fields), one obtains a local contribution. 

But this local result is an artifact of the density field and of low order in the perturbative series, which limits the available tensorial structures.  Already once one looks at the momentum,~$\pi^i$, one finds that the traceless part of $\pd_i\pi^j$, which is observable, is affected at linear order by the traceless part of $\pd_i\pd_j\Phi$, which, as argued above, is non-local. So the associated Green's function will be non-local. Therefore, unless accidental cancellations happen, one should expect  the local counterterms to contribute non locally. This is the situation we will encounter in this paper, as the momentum is important for redshift space distortions where, additionally, the anisotropy induced by the line of sight provides a richer tensorial structure where accidental cancellations are more rare.

Explicitly, we find that for the momentum, $\pi^i$, we need a counterterm that contributes in a way schematically given by 
\begin{align} 
 \pi^i &\supset \frac{1}{H} \frac{\partial_i \partial_j \partial_k}{\partial^2}     \tau^{jk}_{(2)} \ ,
\end{align} 
where $ \tau^{jk}$ is the stress tensor, {and the subscript ${}_{(n)}$ or superscript ${}^{(n)}$ indicates $n$-th order in perturbations}. Now, among the second-order response terms for the stress tensor, we have terms such as
\begin{align}
\begin{split} \
 \tau^{ij}_{(2)} \supset & \quad \frac{\bar \rho}{\knl^2 H^2} {\partial_i \partial_k \Phi} {\partial_k  \partial_j \Phi} \ ,
\end{split}
\end{align}
with $\bar \rho$ being the background density. This is indeed local. This term affects non-locally the gradient of the momentum as
\begin{align} 
\pd_j \pi^i &\supset \frac{\bar \rho}{\knl^2 H^2}  \frac{\partial_j \partial_i \partial_k \partial_m}{H \partial^2}  \left(  {\partial_k \partial_l \Phi} {\partial_l  \partial_m \Phi}\right)\ .
\end{align} 
In turn, in redshift space, the dark matter overdensity, $\delta_{r}$, at second order is affected as 
\be
\delta_{r} ( \xvec)  \supset \frac{1}{H \bar \rho} \hat z^i \hat z^j \partial_i   \pi^j ( \xvec)  \sim   \hat z^i \hat z^j  \frac{\partial_i\partial_j \partial_k \partial_m }{\knl^2 \partial^2} \left(  \frac{\partial_k \partial_l}{H^2} \Phi ( \xvec)  \frac{\partial_l  \partial_m}{H^2} \Phi ( \xvec)   \right) \ .
\ee
The $1/\pd^2$ does not simplify in the final expression: this is a counterterm that contributes non-locally to the observable $\delta$ in redshift space. There are several such terms in the perturbative expansion, and similar terms appear also  when considering the stochastic counterterms. {As discussed in more detail later, these terms are linked to the generation of vorticity in the cosmological fluid.} {As a validation of the above, we} explicitly {find} that these terms are needed to renormalize the redshift-space matter overdensity bispectrum at one-loop order.\footnote{At this point, one might wonder why counterterms are local to start with. The terms that we have identified have the property that the region that can non-locally affect a mode is at most of order of the wavelength of the mode itself. For counterterms, for example in $\tau^{ij}$, we are integrating out short modes, and so this can affect at most regions within $1/\knl$, which is equivalent to a normal local response.}

When passing to biased tracers, a further subtlety arises. In the EFTofLSS, the biased-tracer density in redshift space, which is constructed as combinations of the biased-tracer density and momentum,  is written as a spatially {\it local} linear combination  of composite operators of the matter field~\cite{Senatore:2014eva}. We do not have the equations of motion and the associated Green's function for them. This was the way that we identified the non-locally-contributing counterterm in dark matter: by simply solving the equations of motion in the presence of a local {stress tensor}. But because of the local relation to dark matter, it is expected that the non-locally-contributing  counterterms should be completely determined by the one of dark matter. This is so even for the momentum of biased tracers, which, for dark matter, was the operator being affected by the non-locally-contributing counterterm.  Indeed, by the equivalence principle, biased tracers should have the same velocity as the underlying dark matter field at leading order in derivatives.  In fact, there is a symmetry argument based on this idea  connecting the two: the non-locally-contributing  counterterm for biased tracers has the same functional form and coefficient as for dark matter.  {We confirm and discuss this argument in detail later, and we} check again that this result is sufficient for renormalizing the one-loop bispectrum of tracers in redshift space.

\paragraph{Renormalization of local products of the velocity field:} Let us now pass to the second main subtlety that we encounter in this paper. In redshift space, there appear several contact operators involving the long-wavelength velocity. Contact operators are operators made of products of long-wavelength fields at the same location. Performing a product of long wavelength fields at the same location is a process sensitive to arbitrary short-distance fluctuations, and so needs to be renormalized~\cite{Senatore:2014vja, Lewandowski:2015ziq}. For operators involving the velocity $v^i$, care must be taken in preserving the non-trivial transformations under the Galilean group (which is nothing but the non-relativistic limit of the group of diffeomorphisms), and this is complicated by the fact that the velocity is itself a contact operator and so needs to be renormalized~\cite{Carrasco:2013sva,Mercolli:2013bsa}. In order to have the correct transformation properties under the Galilean transformation $v^i \rightarrow v^i + \chi^i$, we wish to have, for example,
\begin{align}
\begin{split}
& [v^i]_R \rightarrow [v^i]_R + \chi^i \ , \\
& [v^i v^j ]_R \rightarrow [v^i v^j]_R + [v^i]_R \chi^j + [v^j]_R \chi^i + \chi^i \chi^j \ , \\
\end{split}
\end{align}
where $ [v^i]_R$  and $ [v^i v^j ]_R$ are respectively the renormalized velocity and the renormalized velocity-squared.
While satisfying this constraint is quite straightforward for the velocity, we see that the velocity-squared needs to {have a transformation involving} the renormalized velocity itself. One way to write renormalized quantities satisfying the above in terms of the non-renormalized fields is therefore to write the renormalized velocity-squared {in terms of} the renormalized velocity and additional counterterms,
\begin{align}
\begin{split} 
& [v^i]_R = v^i + \mathcal{O}_v^i \ ,   \\
& [v^i v^j]_R = [v^i]_R [v^j]_R + \mathcal{O}_{v^2}^{ij} \ ,  \\
\end{split}
\end{align}
where all of the $\mathcal{O}$ terms are Galilean scalars.  In our calculation of the one-loop bispectrum, we will have to implement this procedure for products up to four powers of the velocity or {four} powers of the velocity and one power of the overdensity.  
 
 \vspace{0.5cm}

On top of addressing these two conceptual challenges, the rest of the paper is devoted to developing the full calculation of the kernels of the one-loop bispectrum in redshift space both for dark matter and for tracers, including the relevant counterterms, and finally to checking that indeed all the ultraviolet (UV) dependence of the loop diagrams can be cancelled by a suitable choice of the resulting effective field theory (EFT) parameters. {Indeed, we find that both including the non-locally-contributing counterterms and implementing the correct redshift space renormalization procedure are crucial for matching the UV limits of the loops.} We will perform first the study for dark matter, and then repeat it for biased tracers. As mentioned, the resulting kernels have been instrumental in performing the first one-loop analysis of the bispectrum of galaxies in large-scale structure~{\cite{{DAmico:2022osl,DAmico:2022gki}}}, {and the relevant counterterm kernels were already presented in those works}. 

 One final observation that is worthwhile to make is the following. The generic expression of biased tracers is non-local in time~\cite{Senatore:2014eva}. If $\delta_h$ is the tracer overdensity, we have
\be
\delta_h(\vec x,t)= \int^t dt'  \sum_i {\rm Ker}_i(t,t') \ {\cal{O}}_i(\vec x_{\rm fl}(\vec x,t,t'),t') 
\ee
where ${\cal{O}}_i(\xvec,t') $ are all the scalar operators that can be built from the long wavelength fields, $\vec x_{\rm fl}(\vec x,t,t')$ represents the position at time $t'$ of the fluid element that at time $t$ is at location $\vec x$, and $ {\rm Ker}_i(t,t')$ are generic kernels, assumed to have a time scale of order Hubble. Up to fourth order in the operators, we checked that this expression is accidentally degenerate with an analogous expression where one assumes that $ {\rm Ker}_i(t,t')\propto \delta_D(t-t')$, {\it i.e.} as if the biases where local in time.

{Finally, accompanying this paper, we also provide a Mathematica file with all of the expressions for the biased tracer kernels in redshift space up to fourth order, UV limits of the loops, EFT counterterms, and values of EFT parameters that match the UV limits of the loops.  }

%%%%%%%%%%%%%%%%%%
%
%
%
\section{Dark-matter equations and notation}

Here we collect the relevant background equations and notation for the cold dark-matter field.  We assume a background $\Lambda$CDM expansion, with metric $ ds^2 = - dt^2 + a(t)^2 d \xvec^2 $, where $a(t)$ is the scale factor, which will often be used as the time variable.  The dark-matter field is described in terms of the mass density $\rho ( \xvec , a ) $ and the velocity field $v^i( \xvec , a )$.  The background expansion is driven by a non-relativistic, time-dependent, background mass density $\bar \rho ( a )$ which is given by
\be
\bar \rho ( a )  = \bar \rho_0 \left( \frac{a}{a_0} \right)^{-3} \ ,
\ee
where subscripts ${}_0$ refer to current-day values, and the Hubble rate is $H  = \dot a  / a $ (we use the dot to denote time derivatives, {\it i.e.} $\dot g = \partial g / \partial t$ for generic functions $g$).

We describe scalar perturbations in the metric with the metric potentials $\Phi$ and $\Psi$,\footnote{Anisotropic stress is small and can be neglected for our purposes, in which case the Einstein equations imply $\Phi = \Psi$, which we assume throughout this work.} by writing
\be
d s^2 = - ( 1 + 2 \Phi  ) dt^2 + a(t)^2 ( 1 - 2 \Psi ) d \xvec^2  \ . 
\ee
In terms of the momentum density $\pi^i$, defined by
\be \label{pidef}
\pi^i ( \xvec , a ) \equiv \rho ( \xvec , a ) v^i ( \xvec , a ) \ , 
\ee
the equations of motion for dark matter are (see e.g. \cite{Baumann:2010tm, Carrasco:2012cv, Carrasco:2013mua}) 
\begin{align}
\begin{split} \label{eom1}
& \dot \rho + 3 H \rho + a^{-1} \partial_i \pi^i = 0  \ , \\
& \dot \pi^i + 4 H \pi^i + a^{-1} \partial_j \left(  \frac{\pi^i \pi^j}{\rho} \right) + a^{-1} \rho \, \partial_i \Phi = - a^{-1} \partial_j \tau^{ij}  \ , 
\end{split}
\end{align}
along with the Poisson equation
\be \label{poisson}
a^{-2} \partial^2 \Phi = \frac{3}{2} \om H^2 \delta  \ ,
\ee
which is in terms of the overdensity $\delta$, given by
\be
\delta ( \xvec , a ) \equiv ( \rho ( \xvec , a ) - \bar \rho ( a ) ) / \bar \rho ( a )  \  , 
\ee
and the time-dependent matter fraction $\om ( a )$.\footnote{This is defined by $\om (a) \equiv \bar \rho ( a ) / (3 \mpl^2 H(a)^2) $, where $\mpl$ is the Planck mass, related to Newton's constant $G_N$ by $\mpl^2 = 1 / ( 8 \pi G_N)$.  In $\Lambda$CDM, the Hubble rate can be parameterized by $H(a)^2 / H_0^2 =  \omn ( a / a_0)^{-3} + ( 1 - \omn )  $.}   The quantity $\tau^{ij}$ appearing in \eqn{eom1} is the EFTofLSS stress tensor \cite{Carrasco:2012cv}, which we will describe in much more detail later.   Using the Poisson equation \eqn{poisson}, we can write 
\be
\bar \rho \,  \delta \,  \partial_i \Phi = 2 \mpl^2 a^{-2} \partial_j \left( \partial_i \Phi \partial_j \Phi - \half \delta_{ij} ( \partial \Phi)^2 \right)  \ , 
\ee
and consequently the equations of motion \eqn{eom1} become
\begin{align}
\begin{split} \label{piieom1}
& \bar \rho\,  \dot \delta + a^{-1} \partial_i \pi^i = 0 \ ,  \\
& \dot \pi^i + 4 H \pi^i + a^{-1} \bar \rho \, \partial_i \Phi = - a^{-1} \partial_j \left(  2 \mpl^2 a^{-2}\left( \partial_i \Phi \partial_j \Phi - \half \delta_{ij} ( \partial \Phi)^2 \right)   + \frac{\pi^i \pi^j}{\rho} + \tau^{ij}   \right)  \ . 
\end{split}
\end{align}

Next, we can decompose the momentum density into a scalar and a vector part
\be \label{pisdef}
\pi_S \equiv \partial_i \pi^i \andd \pi^i_V \equiv \epsilon^{ijk} \partial_j \pi^k \ , 
\ee
 which gives
 \be \label{piidef}
 \pi^i = \frac{\partial_i}{\partial^2 } \pi_S - \epsilon^{ijk} \frac{\partial_j}{\partial^2} \pi^k_V \ ,
 \ee
 where $\epsilon^{ijk}$ is the three-dimensional totally antisymmetric Levi-Civita symbol (with $\epsilon^{123} = 1$).  With this decomposition, we can write the equations of motion in terms of the scalar and vector parts
\begin{align}
\label{greateom}
& \bar \rho \, \dot \delta + a^{-1} \pi_S = 0 \ ,  \nonumber \\
& \dot \pi_S + 4 H \pi_S +   \frac{3}{2} a \bar \rho \, \om H^2 \delta = - a^{-1} \partial_i \partial_j \left(  2 \mpl^2 a^{-2}\left( \partial_i \Phi \partial_j \Phi - \half \delta_{ij} ( \partial \Phi)^2 \right)   + \frac{\pi^i \pi^j}{\rho} + \tau^{ij}   \right)  \ ,  \nonumber \\
& \dot \pi^i_V + 4 H \pi^i_V = - a^{-1} \epsilon^{ijk} \partial_j \partial_l \left( 2 \mpl^2 a^{-2} \partial_k  \Phi \partial_l \Phi + \frac{\pi^k \pi^l}{\rho} + \tau^{k l} \right)  \ . 
\end{align}
 We see that the scalar part $\pi_S$ is determined entirely by $\delta$, and the vector part $\pi^i_V$ is only sourced non-linearly.  For reference, the full differential equation for $\delta$ is 
\be \label{fulldiffeqdelta}
a^2 \delta'' + \left( 2 + \frac{a \cH'}{\cH} \right) a \delta' - \frac{3}{2} \om  \delta = \frac{\partial_i \partial_j}{\cH^2 \bar \rho}   \left(  2 \mpl^2 a^{-2}\left( \partial_i \Phi \partial_j \Phi - \half \delta_{ij} ( \partial \Phi)^2 \right)   + \frac{\pi^i \pi^j}{\rho} + \tau^{ij}   \right)  \  ,
\ee
where $\cH = a H$, and the prime denotes a derivative with respect to the scale factor, {\it i.e.} $ g' = \partial g / \partial a$ for generic functions $g$.  For dark matter in real space, all renormalization and counterterms enter through the stress tensor $\tau^{ij}$, which is a Galilean scalar, a tensor under spatial rotations, {and a local-in-space and non-local-in-time function of second derivatives of the metric and gradients of the velocity (because of the equivalence principle)}, in the equations of motion above.  We will return to the stress tensor in much more detail below.  {We also note that  while the equation of motion for $\pi^i$ is non-local (because of the appearance of $\partial_i \Phi$ on the left-hand side of \eqn{piieom1}), the equations of motion for $\pi_S$ and $\pi^i_V$ \eqn{greateom} are local.}

In this work, we use the following notation 
\be
\int_{\kvec_1 , \dots , \kvec_n} \equiv \int \frac{d^3 k_1 }{(2 \pi)^3} \cdots \frac{d^3 k_n}{(2 \pi)^3 } \ ,  \quad \int_{\kvec_1 , \dots , \kvec_n}^{\kvec} \equiv \int_{\kvec_1 , \dots , \kvec_n} ( 2 \pi)^3 \delta_D ( \kvec - \sum_{i = 1}^n \kvec_i )  \ , 
\ee
where $\delta_D$ is the Dirac delta function, and our Fourier conventions are
\be
f ( \xvec , t ) = \int_{\kvec} f (\kvec ,  t ) \,  e^{i \kvec \cdot \xvec}  \ . 
\ee
For a three-dimensional vector $\kvec$, we write $k \equiv | \kvec |$ for the magnitude, and $\hat k \equiv \kvec / k$ for the unit vector parallel to $\kvec$.  We use Latin letters like $i,j,k,l$ to denote spatial indices, in general we do not distinguish between upper and lower spatial indices, and repeated indices imply summation.  We also use the prime on correlation functions, $\langle \cdot \rangle' $ to denote the correlation function with the factor of $(2\pi)^3$ and Dirac delta function of {translation invariance} stripped off.

%%%%%%%%%%%
%
%
\subsection{Perturbative solutions and observables in SPT} \label{sptsolssec}

In this work, we use the so-called Einstein-de Sitter (EdS) approximation to solve the above equations which allows us to separate the time dependence from the spatial (momentum) dependence and is known to be accurate to percent level \cite{Takahashi:2008yk, Donath:2020abv, Zhang:2021uyp} {(we give details about the EdS Green's function in \appref{edsgreensfunctionsec}).}  Let us start with the standard perturbation theory (SPT) contribution, which is the solution ignoring EFTofLSS counterterms, {\it i.e.} with $\tau^{ij} = 0$.  First, the solution to the linear equation for $\delta$ is called the growth factor $D(a)$, which solves
\be \label{growthfactor}
a^2 D'' + \left( 2 + \frac{a \cH'}{\cH} \right) a D' - \frac{3}{2} \om D = 0  \ .
\ee
For the perturbations, we write
\be
\delta ( \kvec , a ) = \sum_n \delta^{(n)} ( \kvec , a ) \andd v^i(\kvec , a ) = \sum_n v^i_{(n)} ( \kvec , a )
\ee
where, assuming that the velocity field is irrotational,\footnote{In the absence of counterterms, it can be shown that an initially irrotational velocity remains so. In our universe, the initial vorticity is negligible. EFT counterterms induce a vorticity, though, matching what is observed in simulations \cite{Carrasco:2013mua}, {and we discuss this in more detail in \secref{irlimitcheckssec}.}} 
\begin{align}
\begin{split} \label{dmkernels}
& \delta^{(n)} ( \kvec,  a ) =  D(a)^n \int_{\kvec_1 , \dots , \kvec_n}^{\kvec} F_n ( \kvec_1 , \dots , \kvec_n ) \tilde \delta^{(1)}_{ \kvec_1} \cdots \tilde \delta^{(1)}_{ \kvec_n } \ ,  \\
& v^i_{(n)} (\kvec , a ) = i \frac{k^i}{k^2} \cH(a) f(a)  D(a)^{n} \int_{\kvec_1 , \dots , \kvec_{n}}^{\kvec} G_n ( \kvec_1 , \dots , \kvec_n ) \tilde \delta^{(1)}_{ \kvec_1} \cdots \tilde \delta^{(1)}_{ \kvec_n }  \ ,
\end{split}
\end{align}
$\tilde \delta^{(1)}_{\kvec}$ is the time-independent initial field,\footnote{{We normalize $D(a_{\rm in}) = 1$ for some initial time $a_{\rm in}$ in matter domination where initial conditions are given, so that $P_{11}$ is the linear power spectrum at $a_{\rm in}$.}} $F_n$ and $G_n$ are the standard symmetric kernels for dark matter (see \cite{Goroff:1986ep,Jain:1993jh}, for example), and the growth rate $f$ is defined by
\be
f(a ) \equiv \frac{ a D'(a)}{D(a)} \ . 
\ee
In general, we use the tilde to denote time-independent fields, in particular
\be
\tilde \delta^{(n)} \equiv \frac{\delta^{(n)} ( a ) }{D(a)^n} \ ,
\ee
and we will often drop spatial or momentum arguments when the understanding is clear.  The above expressions for $\delta$ and $v^i$ solve the equations of motion under the approximation 
\be \label{edscondition}
\om ( a ) \approx \left( \frac{a D'(a)}{D(a)} \right)^2 \ , 
\ee
which is approximately true in our universe \cite{Takahashi:2008yk, Donath:2020abv, Zhang:2021uyp}.   For the one-loop bispectrum, we need to consider up to $n =4$.  The SPT expression for $\pi^i$ can be derived from $\delta$ and $v^i$ using \eqn{pidef}.

In this work, we are eventually interested in computing the one-loop power spectrum and the one-loop bispectrum of {galaxies in redshift space, but we start in this section with dark matter in real space}.  In Fourier space, the power spectrum $P$ and bispectrum $ B $ are defined by
\begin{align}
\begin{split} \label{genps}
\langle \delta ( \kvec , a  ) \delta ( \kvec'  , a ) \rangle & = ( 2 \pi)^3 \delta_D ( \kvec + \kvec' )  P( k ,a  ) \ , \\ 
\langle \delta ( \kvec_1 , a ) \delta ( \kvec_2 , a )  \delta ( \kvec_3 , a  )\rangle & = ( 2 \pi)^3 \delta_D ( \kvec_1 + \kvec_2 + \kvec_3 )   B ( k_1 , k_2 , k_3 , a) \ .
\end{split}
\end{align}
The total one-loop power spectrum is 
\be
P_{1\text{-loop tot.}} ( k , a )  =  D(a)^2 P_{11} ( k )  + D(a)^4 ( P_{22} ( k ) + P_{13} ( k ) ) \ , 
\ee
where {$\langle \tilde \delta_{\kvec}^{(1)} \tilde \delta_{\kvec'}^{(1)} \rangle = ( 2 \pi)^3 \delta_D ( \kvec+ \kvec') P_{11} ( k)$} defines the linear power spectrum $P_{11}$ at the initial time $a_{\rm in}$, and the one-loop terms are
\begin{align}
\begin{split} \label{loopexpressions}
&P_{22} ( k )  = 2 \int_{\qvec} F_2 ( \qvec , \kvec - \qvec)^2 P_{11} ( q ) P_{11}( | \kvec - \qvec|) \ , \\
& P_{13}(k)  = 6 P_{11} ( k ) \int_{\qvec} F_3 ( \qvec , - \qvec , \kvec) P_{11}(q) \  . 
\end{split}
\end{align}
The total one-loop bispectrum is 
\be
B_{1\text{-loop tot.}}  = D(a)^4 B_{211}  + D(a)^6 \left(  B_{222}  + B_{321}^{(I)} + B_{321}^{(II)} + B_{411}   \right)   \ , 
\ee
where the tree-level bispectrum is 
\be
B_{211} ( k_1, k_2, k_3 ) = 2   F_2 ( \kvec_1 , \kvec_2 ) P_{11}(k_1 ) P_{11} ( k_2 ) + \text{ 2 perms.} \ , 
\ee
and the one-loop contributions are 
\begin{align}
\begin{split} \label{oneloopbisp}
& B_{222}(k_1, k_2, k_3 ) = 8 \int_{\qvec} P_{11}(q) P_{11}(| \kvec_2 - \qvec| ) P_{11} ( |\kvec_1 + \qvec|) \\
& \hspace{2in} \times F_2 ( - \qvec , \kvec_1 + \qvec ) F_2 ( \kvec_1 + \qvec , \kvec_2 - \qvec) F_2 ( \kvec_2 - \qvec , \qvec)  \ , \\
& B_{321}^{(I)} ( k_1 , k_2 , k_3 ) = 6 P_{11}(k_1) \int_{\qvec} P_{11}(q) P_{11}(| \kvec_2 - \qvec|)    \\
& \hspace{2in} \times F_3 ( - \qvec , - \kvec_2 + \qvec , - \kvec_1 ) F_2 ( \qvec , \kvec_2 - \qvec )  + \text{ 5 perms.} \ , \\
& B_{321}^{(II)} ( k_1 , k_2 , k_3 ) = 6 P_{11}(k_1 ) P_{11}(k_2) F_2 ( \kvec_1 , \kvec_2 ) \int_{\qvec} P_{11}(q) F_3 ( \kvec_1 , \qvec , - \qvec ) + \text{ 5 perms.} \ , \\ 
& B_{411} ( k_1 , k_2 , k_3 )  = 12 P_{11} ( k_1 ) P_{11}(k_2) \int_{\qvec} P_{11}(q) F_4 ( \qvec , - \qvec , - \kvec_1 , - \kvec_2 ) + \text{ 2 perms.}  \ .
\end{split}
\end{align}

%%%%%%%%%%%%%%
%
%
\subsection{Dark-matter counterterm contributions}  \label{dmctsec}

As is well known \cite{Carrasco:2012cv}, the loop contributions \eqn{loopexpressions} and \eqn{oneloopbisp} are UV sensitive because they depend on momenta $q$ much larger than the non-linear scale of structure formation where the theory is out of perturbative control.  The role of the EFT counterterms in $\tau^{ij}$ is to cure this UV sensitivity and allow the theory to match reality.  Concretely, we write (suppressing spatial dependence for convenience)
\be
\tau^{ij} ( a ) = \tau^{ij}_{\Lambda_{\rm UV}} ( a ) + \tau^{ij}_{\rm finite} ( a ) \ .
\ee
 Here, the piece $\tau^{ij}_{\Lambda_{\rm UV}} ( a )$ must give the same time dependence as in \eqn{loopexpressions} and \eqn{oneloopbisp} in order to cancel the dependence of the loop integrals on the UV cutoff $\Lambda_{\rm UV}$.  The other piece, $\tau^{ij}_{\rm finite} ( a )$, does not have a fixed time dependence in general, and its role is to give the correct amount of $\Lambda_{\rm UV}$-independent contribution that matches observations.

For simplicity in this paper, we focus on the contribution $\tau^{ij}_{\Lambda_{\rm UV}} ( a )$ which has a fixed time dependence.  All of our main points will be evident in this case, and we will be able to do explicit calculations with explicit numerical factors.  Additionally, we want to check that the general form of $\tau^{ij}$ that we write is able to capture all of the UV behavior present in the loops \eqn{loopexpressions} and \eqn{oneloopbisp}, and this requires assuming the same time dependence as the loops.  {Inclusion of a general $\tau^{ij}_{\rm finite} ( a )$ is straightforward, as it is based on the same $k$-dependent kernels.}

In general, the EFTofLSS is local in space, but non-local in time \cite{Carrasco:2013mua}.  To obtain the EFT expansion one expands the stress tensor as a local function of second spatial derivatives of the gravitational potential $\Phi$ and gradients of the velocity (because of the equivalence principle), along with stochastic fields, organized, as in any EFT, in an expansion in powers of the fields and spatial derivatives, integrated along the past trajectory of the fluid element \cite{Carrasco:2013mua, Senatore:2014eva}.  More specifically, we have 
\be \label{nonlocaltau}
\tau^{ij} ( \xvec , a ) = \int^a \frac{d a'}{a'} \, \sum_\alpha \kappa_\alpha ( a , a' ) T^{ij}_\alpha ( \xvec_{\rm fl} ( \xvec , a , a'   ) , a' )  \ ,
\ee
where the fluid element is defined by
\be
\xvec_{\rm fl} ( \xvec , a , a' ) = \xvec + \int_{a}^{a'} \frac{d a''}{(a'')^2 H(a'')}  \vec{v} ( \xvec_{\rm fl} ( \xvec ,  a , a'') , a'' )  \ , 
\ee
the $T^{ij}_\alpha ( \xvec , a )$ are all {local-in-time} Galilean scalars (and tensors under rotations on the $i$ and $j$ indices), and the $\kappa_\alpha ( a , a')$ are unknown EFT kernels describing the non-locality in time.  The $T^{ij}_\alpha ( \xvec , a )$ are then organized in a local spatial-derivative expansion of the long-wavelength fields ($\partial_i \partial_j \Phi ( \xvec , a)$, $\partial_i v^j ( \xvec , a )$, etc.) and of stochastic fields $\epsilon^{ij} ( \xvec , a )$.  Since we do perturbation theory in this work, we can write each scalar as a sum over perturbative orders
\be
T_{\alpha}^{ij} ( \xvec , a ) = \sum_n T_{\alpha,(n)}^{ij} ( \xvec , a ) \ , 
\ee
and because the linear solutions are scale independent, we have a simple scaling time dependence for the $n$-th order perturbative pieces
\be
T^{ij}_{\alpha,(n)} ( \xvec , a' )  = \left( \frac{D(a')}{D(a)} \right)^{p_{\alpha,n}} T^{ij}_{\alpha,(n)} ( \xvec , a ) \ ,
\ee
for some power $p_{\alpha,n}$.  This means that \eqn{nonlocaltau} becomes 
\begin{align}
\begin{split} \label{localintimeapprox}
\tau^{ij} ( \xvec , a ) & =  \sum_\alpha   \sum_n  K_\alpha^{p_{\alpha,n}} ( a )  T^{ij}_{\alpha,(n)} ( \xvec , a )   \\
& +\sum_\alpha \sum_{n,m} \frac{1}{m} \left( K_\alpha^{p_{\alpha,n-m}} ( a ) - K_\alpha^{m+p_{\alpha,n-m}} ( a )    \right)  \frac{\partial_k \theta_{(m)} ( \xvec , a )}{\partial^2} \partial_k T^{ij}_{\alpha ,(n-m)} ( \xvec , a )  + \dots   \ , 
\end{split}
\end{align}
where we have defined
\be \label{timefndef}
K_{\alpha}^p ( a ) \equiv \int^a \frac{d a'}{a'} \, \kappa_{\alpha} ( a , a' ) \left( \frac{D(a')}{D(a)} \right)^{p}  \ , 
\ee
used the definition of the velocity field in \eqn{dmkernels} along with  $\theta = - \partial_i v^i / (f a H)$, and the $\dots$ in \eqn{localintimeapprox} are terms coming from Taylor expanding $T^{ij}_\alpha ( \xvec_{\rm fl} ( \xvec , a , a'   ) , a' )$ around $\xvec$ in \eqn{nonlocaltau}, {\it i.e.} higher powers of $\vec{v}$, {all of which should be included up to the desired order}.  The point is that the integral over $d a'$ coming from expanding $\xfl$ can be formally done as in \eqn{timefndef} and leaves distinct functions of $a$, $K_\alpha^p ( a )$.  The same happens with the higher order terms in the Taylor expansion of $\xvec_{\rm fl} ( \xvec , a , a')$, since $\vec{v}_{(n)}$ also has a simple scaling time dependence, see \eqn{dmkernels}.  This means that, in perturbation theory, the expansion of the stress tensor can be {manipulated so that the time integrals disappear,} {in the way indicated by \eqn{localintimeapprox}}.  {In this paper, we only need counterterms from the stress tensor up to second order, so \eqn{localintimeapprox} is sufficient for our purposes.}  This is the same approach taken for the bias expansion in \secref{biasm}.\footnote{{Note that for the growing mode solutions that we consider in this work, we have $K_\alpha^{p_{\alpha ,2}} ( a ) = K_\alpha^{1+p_{\alpha ,1}} ( a ) $, which shows that \eqn{localintimeapprox} is of the correct form for Galilean scalars discussed in \secref{irlimitcheckssec2}.}}

Given this, we write the contribution to $\delta$ from $\tau^{ij}$ relevant to the one-loop bispectrum, which we call $\delta_\tau$, generally as
\be \label{deltatau}
\delta_\tau ( \kvec , a ) =   \delta^{(1)}_{ct} ( \kvec , a ) +   \delta^{(2)}_{ct} ( \kvec , a ) +  \delta_\epsilon^{(1)} ( \kvec, a )  +  \delta_{\epsilon}^{(2)} ( \kvec , a  )  \ , 
\ee
where the subscript $ct$ denotes the response counterterms, and $\epsilon$ denotes the stochastic (and semi-stochastic) counterterms.  Assuming the time dependence needed to cancel UV loop contributions, we have 
\begin{align}
\begin{split} \label{cttimedep}
& \delta^{(1)}_{ct} ( \kvec , a )= D(a)^3  \tilde \delta^{(1)}_{ct} ( \kvec )  \ ,  \quad \delta^{(2)}_{ct} ( \kvec , a )  = D(a)^4 \tilde \delta^{(2)}_{ct} ( \kvec) \ , \\
  & \delta_\epsilon^{(1)} ( \kvec, a )  = D(a)^2 \tilde \delta_\epsilon^{(1)} ( \kvec)  \andd  \delta_{\epsilon}^{(2)} ( \kvec , a  ) =  D(a)^3 \tilde \delta_{\epsilon}^{(2)} ( \kvec ) \ , 
\end{split}
\end{align}
where we use the tilde to denote the appropriate time-independent factor.  Similarly, we have the solution for $\pi^i$ which we write as 
\be \label{piitau}
\pi^i_\tau ( \kvec , a ) =   \pi^i_{ct,(1)} ( \kvec , a ) +   \pi^i_{ct,(2)}( \kvec , a ) +  \pi^i_{\epsilon,(1)}( \kvec, a )  +  \pi^i_{\epsilon,(2)} ( \kvec , a  )  \ , 
\ee
with 
\begin{align}
\begin{split} \label{cttimedeppi}
& \pi^i_{ct,(1)} ( \kvec , a ) = - a H  \bar \rho   f  D(a)^3  \tilde \pi^i_{ct,(1)} ( \kvec )  \ ,  \quad  \pi^i_{ct,(2)} ( \kvec , a ) = - a H  \bar \rho   f  D(a)^4  \tilde \pi^i_{ct,(2)} ( \kvec )   \ , \\
  &\pi^i_{\epsilon,(1)} ( \kvec , a ) = - a H  \bar \rho   f  D(a)^2  \tilde \pi^i_{\epsilon,(1)} ( \kvec )   \andd  \pi^i_{\epsilon,(2)} ( \kvec , a ) = - a H  \bar \rho   f  D(a)^3  \tilde \pi^i_{\epsilon,(2)} ( \kvec ) \ .
\end{split}
\end{align}
To remove clutter, we will sometimes use the notation $A_*$ to mean both $A_{ct}$ and $A_\epsilon$.  
{We note that in SPT, the time dependence of $\pi^i$, $\pi_S$, and $\pi_V^i$ are all the same for each field above linear perturbations.\footnote{{At linear level, we have $\pi_{V,(1)}^i  = \bar \rho \epsilon^{ijk} \partial_j v^k_{(1)}= 0$ because velocity vorticity is zero.  Then, at higher orders, $\pi_V^i$ becomes non-zero because of the growing mode in $\delta$ in the definition of $\pi^i$.  So, above linear perturbations, $\pi_S$, $\pi_V^i$, and $\pi^i$ all have the same time dependence, which one can deduce from the definition of $\pi^i$ in terms of $\delta$ and $v^i$.} }  However, as can be seen in \eqn{greateom}, if $\partial^i \partial^j \tau^{ij}$ and $\epsilon^{ijk} \partial_j \partial_l \tau^{kl}$ have different time dependence, then $\pi_S$ and $\pi_V^i$ will as well.  For simplicity, in this work, we assume that the counterterms have the same time dependence as needed to cancel UV divergences in the SPT loops, in which case $\pi_S$ and $\pi_V^i$ have the same time dependence as in \eqn{cttimedeppi} for $\pi^i$. }

%%%%%%%%%%%%%
%
%
%
\section{Dark-matter renormalization in real space} \label{dmrenormrssec}

%%%%%%%%%%%%%%
%
%
\subsection{Counterterm solutions up to second order} \label{tauijdmsec}

We now present the solutions for $\delta$ and $\pi^i$ that are sourced by the stress tensor $\tau^{ij}$ in \eqn{greateom} up to second order, leaving the derivation to \appref{rsdmapp}.  We start by writing   
\be \label{sometauij}
\tau^{ij}   = \frac{\om \cH^2 \bar \rho}{\knl^2} \left(D^3 \tilde \tau^{ij}_{ct,(1)} + D^4  \tilde \tau^{ij}_{ct,(2)} + D^2  \tilde \tau^{ij}_{\epsilon,(1)} +  D^3 \tilde  \tau^{ij}_{\epsilon,(2)}  \right) \ , 
\ee
where the quantities with subscript $ct$ source the pure response solutions of $\delta$ and $\pi^i$, the quantities with subscript $\epsilon$ source  the stochastic and semi-stochastic solutions of $\delta$ and $\pi^i$, and the number in parentheses indicates the order in fields.  Again, quantities with tildes are time independent, and the time dependence given above is chosen to lead to \eqn{cttimedep} {(the factor of $\om(a)$ can be seen from \appref{edsgreensfunctionsec})}.  As described in more detail in \appref{rsdmapp}, the second-order solutions here come from two sources.  The first is directly from the second-order stress tensors in \eqn{sometauij}, and the second is from plugging the first-order counterterm solutions back into the equations of motion \eqn{greateom}.  

Then, the first-order response solutions are 
\begin{align}
\begin{split}  \label{linsol1}
\tilde \delta^{(1)}_{ct}  & = \frac{1}{9 \knl^2} \partial_i \partial_j  \tilde \tau^{ij}_{ct, (1)} \andd  \tilde  \pi^i_{ct,(1)}  =   \frac{1}{3 \knl^2}   \frac{\partial_i \partial_j \partial_k}{\partial^2} \tilde \tau^{jk}_{ct, (1)}  \ ,
\end{split}
\end{align}
and the second-order response solutions are 
\begin{align}
\begin{split} \label{delta2ct}
\tilde \delta^{(2)}_{ct}   = \frac{2 \,   \partial_i \partial_j }{33 \knl^2} \left[  \tilde \tau^{ij}_{ct, (2)}  +   \frac{\partial_i \tilde \delta^{(1)}}{\partial^2}  \partial_k \tilde \tau^{jk}_{ct, (1)} -\frac{1}{6} \delta_{ij}   \frac{\partial_k \tilde \delta^{(1)}}{\partial^2}  \partial_l \tilde \tau^{kl}_{ct, (1)}    \right]  \ , 
\end{split}
\end{align}
and
\begin{align} \label{piict2sol1}
\tilde \pi^i_{ct,(2)} & = \frac{2}{9 \knl^2} \Bigg[  \partial_j \tilde \tau^{ij}_{ct,(2)} + \frac{1}{11} \frac{\partial_i \partial_j \partial_k}{\partial^2} \left(   \tilde \tau^{jk}_{ct, (2)} + \frac{\partial_j \tilde \delta^{(1)}}{\partial^2}  \partial_l \tilde \tau^{kl}_{ct, (1)}  \right)   - \frac{2}{11} \partial_i \left(  \frac{\partial_j \tilde \delta^{(1)}}{\partial^2}\partial_k \tilde \tau^{jk}_{ct,(1)}   \right)    \\
& \hspace{.6in}  + \frac{1}{2} \partial_l \left(   \frac{\partial_i \tilde \delta^{(1)}}{\partial^2} \partial_m  \tilde \tau^{lm}_{ct, (1)}  +    \frac{\partial_l \tilde \delta^{(1)}}{\partial^2} \partial_j \tilde \tau^{ij}_{ct,(1)}   \right)  \Bigg]  \ . \nonumber
\end{align} 
{To find the above, we used \eqn{greateom}, solved for $\pi_S$ and $\pi_V^i$ separately (which are given in \appref{rsdmapp}), and then combined them to form $\pi^i$ using \eqn{piidef}.}   Above, as we will justify in the next section, we have assumed that $\partial_i \partial_j \partial_k \tilde \tau^{jk}_{ct,(1)} = \partial^2 \partial_j \tilde  \tau^{ij}_{ct,(1)}$.

Similarly, for the first-order stochastic solutions, we have
\begin{align}
\begin{split}  \label{linsol1st}
\tilde \delta^{(1)}_{\epsilon}    = \frac{2 }{7 \knl^2} \partial_i \partial_j  \tilde \tau^{ij}_{\epsilon, (1)}    \andd \tilde \pi^i_{\epsilon,(1)}   =   \frac{1}{\knl^2} \left[  \frac{4}{7} \frac{\partial_i \partial_j \partial_k}{\partial^2} \tilde \tau^{jk}_{\epsilon,(1)}  - \frac{2}{5} \left( \frac{\partial_i \partial_j \partial_k}{\partial^2} \tilde \tau^{jk}_{\epsilon,(1)}  - \partial_j \tilde \tau^{ij}_{\epsilon,(1)} \right) \right]  \ ,
\end{split}
\end{align}
and for the second-order stochastic solutions, we have
\begin{align}
\begin{split} \label{delta2ctep}
 \tilde \delta^{(2)}_{\epsilon}  & =  \frac{ \partial_i \partial_j }{9 \knl^2 } \Bigg[  \tilde \tau^{ij}_{\epsilon, (2)} +  2 \frac{\partial_i \tilde \delta^{(1)}}{\partial^2} \frac{\partial_j \partial_k \partial_l}{\partial^2} \tilde \tau^{kl}_{\epsilon,(1)}  - \frac{3}{7} \delta_{ij}  \left( \frac{\partial_k \tilde \delta^{(1)}}{\partial^2} \frac{\partial_k \partial_l \partial_m}{\partial^2} \tilde \tau^{lm}_{\epsilon, (1)}   \right)  \\
& \hspace{.6in} - \frac{4 }{5 }  \frac{\partial_i \tilde \delta^{(1)}}{\partial^2} \left(     \frac{\partial_j \partial_k \partial_l}{\partial^2} \tilde \tau^{kl}_{\epsilon,(1)} - \partial_l \tilde \tau^{jl}_{\epsilon,(1)}      \right)      \Bigg]   \ , 
\end{split}
\end{align}
and
\begin{align}
 \label{piict2sol1ep}
\knl^2 \tilde \pi^i_{\epsilon,(2)}  & =   \frac{2}{7} \partial_j \tilde \tau^{ij}_{\epsilon, (2)} + \frac{1}{21} \frac{\partial_i \partial_j \partial_k}{\partial^2} \left(  \tilde \tau^{jk}_{\epsilon, (2)}  + 2   \frac{\partial_j \tilde \delta^{(1)}}{\partial^2} \frac{\partial_k \partial_l \partial_m}{\partial^2} \tilde \tau^{lm}_{\epsilon,(1)}  \right)   \\
& \hspace{.2in}  - \frac{1}{7} \partial_i \left(  \frac{\partial_j \tilde \delta^{(1)}}{\partial^2} \frac{\partial_j \partial_k \partial_l}{\partial^2} \tilde \tau^{kl}_{\epsilon, (1)}   \right) + \frac{2}{7} \partial_l \left(   \frac{\partial_i \tilde \delta^{(1)}}{\partial^2} \frac{\partial_l \partial_m \partial_n}{\partial^2} \tilde \tau^{mn}_{\epsilon,(1)}  +    \frac{\partial_l \tilde  \delta^{(1)}}{\partial^2} \frac{\partial_i \partial_m \partial_n}{\partial^2} \tilde \tau^{mn}_{\epsilon, (1)}   \right) \nonumber  \\
& \hspace{.2in} -  \frac{4}{105} \frac{\partial_i \partial_j \partial_k}{\partial^2} \left( \frac{\partial_j \tilde \delta^{(1)}}{\partial^2}  \left( \frac{\partial_k \partial_l \partial_m}{\partial^2} \tilde \tau^{lm}_{\epsilon,(1)} - \partial_l \tilde \tau^{kl}_{\epsilon,(1)}   \right)   \right) \nonumber   \\
& \hspace{.2in}  - \frac{4}{35} \partial_l \left(  \frac{\partial_i \tilde \delta^{(1)}}{\partial^2} \left( \frac{\partial_l \partial_m \partial_n}{\partial^2} \tilde \tau^{mn}_{\epsilon,(1)} - \partial_m \tilde \tau^{lm}_{\epsilon,(1)}  \right)   + \frac{\partial_l \tilde \delta^{(1)}}{\partial^2} \left( \frac{\partial_i \partial_m \partial_n}{\partial^2} \tilde \tau^{mn}_{\epsilon,(1)} - \partial_m \tilde \tau^{im}_{\epsilon,(1)}  \right)    \right)     \nonumber \ . 
\end{align} 
In the above stochastic expressions, as we will discuss further in the next section, we have \emph{not} assumed that $\partial_i \partial_j \partial_k \tilde \tau^{jk}_{\epsilon,(1)} = \partial^2 \partial_j \tilde  \tau^{ij}_{\epsilon,(1)}$.

As a final note, we point out that all of the various numerical coefficients above come from the linear equations of motion and the assumed time dependence $D(a)^n$ for the various contributions; they represent {different combinations of Green's functions integrated over different kernels} in the EdS approximation.  {We give the EdS Green's function in \appref{edsgreensfunctionsec}.}

%%%%%%%%%%%%%
%
%
\subsection{Explicit expression for the stress tensor} \label{explicitstresstensor}

We can now write down the most general stress tensor {local in second spatial derivatives of $\Phi$ and gradients of the velocity, following the discussion  in \secref{dmctsec},} up to second order, that obeys the symmetries of the problem, which are rotation and Galilean invariance.  We focus on the leading order in derivatives, which, because of mass and momentum conservation, is $\mathcal{O}(k^2 P_{11})$ for $P_{13}$, $\mathcal{O}(k^2 P_{11}^2)$ for $B_{411}$ and $B_{321}^{(II)}$, $\mathcal{O}(k^4)$ for $P_{22}$, $\mathcal{O} (k^6)$ for $B_{222}$, and $\mathcal{O}(k^4 P_{11})$ for $B_{321}^{(I)}$.  For the response terms, we have 
\be \label{tauij1}
\tilde \tau^{ij}_{ct,(1)} = c_1 \frac{\partial_i \partial_j \tilde \delta^{(1)} }{\partial^2}  + c_3 \delta_{ij} \tilde \delta^{(1)} \ , 
\ee
and 
\begin{align}
\begin{split} \label{tauij2}
\tilde \tau^{ij}_{ct,(2)} = & c_1 \frac{ \partial_k  \partial_i \partial_j \deltaone }{\partial^2}   \frac{\partial_k \deltaone}{\partial^2}  + c_3 \delta_{ij} \partial_k \deltaone \frac{\partial_k \deltaone}{\partial^2} \\
& +c_2 \frac{\partial_i \partial_j \tilde \delta^{(2)}}{\partial^2} - c_2    \frac{\partial_k \partial_i \partial_j \deltaone}{\partial^2}   \frac{\partial_k \deltaone}{\partial^2}  + c_4 \delta_{ij} \tilde \delta^{(2)} - c_4 \delta_{ij} \partial_k \deltaone \frac{\partial_k \deltaone}{\partial^2} \\
& + c_5 \frac{\partial_i \partial_j \deltaone}{\partial^2} \deltaone + c_6 \frac{\partial_i \partial_k \deltaone}{\partial^2} \frac{\partial_k  \partial_j \deltaone}{\partial^2} + c_7 \delta_{ij} \deltaone \deltaone  \ ,
\end{split}
\end{align}
where all of the $c_i$ are time independent.  {To determine the above list of operators, we follow the procedure laid out in \secref{dmctsec} (which is the same approach as \cite{Senatore:2014eva} for biased tracers, which we detail in \secref{biasm} and \appref{bias4}); we first write all contractions of $\partial_i \partial_j \Phi$ and $\partial_i v^j$ with the same tensor structure as $\tau^{ij}$ up to second order in fields and zeroth order in derivatives, and then we expand the fluid element and do the remaining time integrals which define the $c_i$.  We then check for degeneracies in the resulting operators, and only use the minimal basis, which is given above.}

For the stochastic terms, we have
\begin{align}
\begin{split} \label{stochstress}
\tilde \tau_{\epsilon,(1)}^{ij}  = \epsilon_1^{ij} \andd \tilde \tau_{\epsilon,(2)}^{ij}  = \partial_k \epsilon_1^{ij} \frac{\partial_k \deltaone}{\partial^2}  + \epsilon_3^{ijkl} \frac{\partial_k \partial_l \deltaone}{\partial^2} \ ,
\end{split}
\end{align}
where, in momentum space, we define the correlation of the stochastic fields $\epsilon_n^{ij\dots}$ as an expansion in powers of $\kvec$ of all of the terms allowed by rotation invariance \cite{Lewandowski:2015ziq}, for example
\begin{align}
\begin{split} \label{epsiloncontract}
\langle \epsilon_{a}^{ij} ( \kvec ) \epsilon_{b}^{kl} (\kvec') \rangle'  = & c_{a,b}^{(1)} \delta^{ij} \delta^{kl} + c_{a,b}^{(2)} ( \delta^{ik} \delta^{jl} + \delta^{il} \delta^{jk} ) \\
& + \knl^{-2} \left( c_{a,b}^{(3)} \delta^{ij} k^k k^l +c_{a,b}^{(4)} \delta^{kl} k^i k^j + c_{a,b}^{(5)} ( \delta^{ik} k^j k^l + \delta^{il} k^j k^k  )  \right) + \dots    \ .
\end{split}
\end{align}
{We also do a similar expansion for three-point functions of stochastic fields, for example}
\begin{align}
\begin{split} \label{threepointstoch}
\langle \epsilon^{ij}_a ( \kvec_1 ) \epsilon^{kl}_b ( \kvec_2 ) \epsilon_c^{mn} ( \kvec_3 ) \rangle' = & c_{a,b,c}^{(1)} \delta^{ij} \delta^{kl} \delta^{mn} + \half c_{a,b,c}^{(2)} \delta^{ij} \left( \delta^{ln }\delta^{km} + \delta^{lm} \delta^{kn}  \right) \\
& + \half c_{a,b,c}^{(3)} \delta^{mn} \left( \delta^{ik} \delta^{jl} + \delta^{il} \delta^{jk} \right)  + \half c_{a,b,c}^{(4)}  \delta^{kl} \left( \delta^{in} \delta^{jm} + \delta^{im} \delta^{jn} \right)  \\
& + \frac{1}{8} c_{a,b,c}^{(5)} \Big(\delta^{im} ( \delta^{jl} \delta^{kn} + \delta^{jk} \delta^{ln})  + \delta^{il} ( \delta^{jn} \delta^{km} + \delta^{jm} \delta^{kn} ) \\
& \quad \quad \quad \quad + \delta^{ik } ( \delta^{jn} \delta^{lm} + \delta^{jm} \delta^{ln}  ) + \delta^{in} ( \delta^{jl} \delta^{km} + \delta^{jk} \delta^{lm} )  \Big)  \ ,
\end{split}
\end{align}
{where we only need the terms up to $k^0$ for the stochastic three-point functions for dark-matter renormalization in this paper.}\footnote{ {The coefficients $c_{a,b,c}^{(i)}$ in the contraction \eqn{threepointstoch} are defined with respect to the specific ordering of the fields on the left-hand side of \eqn{threepointstoch}, {\it i.e.} they are not necessarily symmetric in $\{a,b,c\}$.  However, we can derive relations among the coefficients with different orderings of $\{a,b,c\}$ by permuting the $\{\kvec_i\}$.  For example, since at the order that we work, the right-hand side is independent of the $\{\kvec_i\}$, by permuting the $\{\kvec_i\}$ on the left-hand side, we obtain relations like $c_{a,c,b}^{(1)} = c_{a,b,c}^{(1)}$, $c_{a,c,b}^{(3)} = c_{a,b,c}^{(4)}$, and $c_{b,a,c}^{(4)} = c_{a,b,c}^{(2)}$, so that all permutations of $\{a,b,c\}$ can be related back to the canonical ordering $(a,b,c)$. This ensures that all of the correlations obey the relevant symmetries.} }  {We note that the free coefficients appearing in two-point and three-point functions of stochastic fields can in general be independent, although they can be related under the assumption that the stochastic fields are purely Poissonian, for example.}  {Since we expand the stochastic fields in all possible tensor structures, it is clear that \eqn{stochstress} is the most general expression satisfying the equivalence principle.}

Terms containing $\partial_i \tilde \delta^{(1)} / \partial^2$ are sometimes referred to as leading infrared (IR) terms.  This is because they can lead to contributions $\mathcal{O}( k / q)$ as the momentum $\qvec$ of some field goes to zero.  These terms are completely fixed by Galilean invariance.  The correct terms are generated by expanding in the fluid line element, as in \secref{dmctsec}, \secref{biasm}, and \appref{fluidelementapp}, leading to the so-called flow terms \cite{Senatore:2014eva}.  {We explicitly derive the expressions in \eqn{stochstress} and point out a clarification of \cite{Senatore:2014eva,Angulo:2015eqa} regarding the stochastic fields in \appref{stochflowtermsec}.}

{As expected, the counterterms above, when plugged into the relevant expressions for $\delta$, allow us to absorb all UV divergences in the one-loop power spectrum and bispectrum, and we give explicit values for the free coefficients that absorb the UV divergences of the loops in \appref{uvmatchingdm}.  {In the rest of the paper, we will often refer to the choice of counterterms that cancels UV parts of SPT loops as `UV matching,' and in particular, we always choose signs so that the UV part cancels in the \emph{sum} of the loop and the counterterms.}   Our results become much more interesting when performing renormalization in redshift space, where {non-locally-contributing} counterterms in $\pi^i$ are necessary, and we discuss this in much more detail in \secref{irlimitcheckssec}, \secref{rssresponsesec}, and \secref{stochdmsec}.}

As mentioned in \secref{tauijdmsec}, we assumed that $\partial_i \partial_j \partial_k \tilde \tau^{jk}_{ct,(1)} = \partial^2 \partial_j \tilde  \tau^{ij}_{ct,(1)}$, but $\partial_i \partial_j \partial_k \tilde \tau^{jk}_{\epsilon,(1)} \neq \partial^2 \partial_j \tilde  \tau^{ij}_{\epsilon,(1)}$, and now we can see why.  For the first-order response stress tensor, $\tilde \tau_{ct,(1)}^{ij}$, the only two terms that we can write are $\delta_{ij} \tilde \delta^{(1)}$ and $\partial_i \partial_j \tilde \delta^{(1)} / \partial^2$, which both separately satisfy $\partial_i \partial_j \partial_k \tilde \tau^{jk}_{ct,(1)} = \partial^2 \partial_j \tilde  \tau^{ij}_{ct,(1)}$.  For the stochastic terms, however, this is not the case.  This is essentially because we allow all tensor structures when contracting stochastic fields.  For example, if we have
\begin{align}
\begin{split} \label{stochcontractions}
\langle \epsilon_1^{ij} (\kvec_1) \epsilon_2^{kl} ( \kvec_2)  \rangle' \supset a_1 \delta^{ij} \delta^{kl} + a_2 ( \delta^{ik}\delta^{jl} + \delta^{il} \delta^{jk} )  \ , 
\end{split}
\end{align}
then the combination in question gives
\begin{align}
\begin{split}
  k_1^i k_1^j k_1^k \langle \epsilon_1^{jk} ( \kvec_1) \epsilon_2^{ab} ( \kvec_2) \rangle' - k_1^2 k^j_1 \langle \epsilon_1^{ij} ( \kvec_1) \epsilon_2^{ab} ( \kvec_2) \rangle'  =  a_2 (2  k_1^i k_1^a k_1^b -  k_1^2 \delta^{ia} k_1^b - k_1^2 \delta^{ib} k_1^a)  \ ,
\end{split}
\end{align} 
 which is not zero.   While we find that this makes no difference in real space, we find that the terms proportional to $\partial_i \partial_j \partial_k \tilde \tau^{jk}_{\epsilon,(1)} / \partial^2  - \partial_j \tilde \tau^{ij}_{\epsilon,(1)} $ in stochastic expressions \eqn{linsol1st}, \eqn{delta2ctep}, and \eqn{piict2sol1ep} lead to contributions in the redshift space quantities $B_{321}^{r,(I),\epsilon}$ and $B_{222}^{r,\epsilon}$ {(see \appref{dmrssctkernelssec} for definitions and \secref{stochdmsec} for a discussion)} that have a unique functional form and are not captured by any other terms that we have discussed, and indeed they are necessary to match the UV structures of $B_{321}^{r,(I)}$ and $B_{222}^{r}$.

%%%%%%%%%%%%%%
%
%
%
\subsection{Appearance of non-locally-contributing counterterms} \label{irlimitcheckssec}

As a main result of this work, we would like to draw particular attention to the way that the second-order stress tensors $\tilde \tau^{ij}_{*,(2)}$ enter the counterterm solutions.  Although they enter with a unique derivative structure in $\tilde \delta^{(2)}_*$, {\it i.e.} $\partial_i \partial_j \tilde \tau^{ij}_{*,(2)}$, the second-order stress tensors appear in both of the structures
\be \label{twocombos}
\partial_j \tilde \tau^{ij}_{*,(2)} \andd \frac{\partial_i \partial_j \partial_k}{\partial^2} \tilde \tau^{jk}_{*,(2)} \ , 
\ee
in the second-order solutions $\tilde \pi^i_{*,(2)}$ in \eqn{piict2sol1} and  \eqn{piict2sol1ep}.  The second term above is a {spatially} non-local contribution to $\pi^i$, although it originates from local contributions to $\pi_S$ and $\pi_V^i$, as can be seen in \eqn{greateom}, {and we will often refer to it as the new `non-locally-contributing counterterm.'}  {The non-locality comes simply from the Green's function for the $\pi^i$ equation, as evident in \eqn{piidef}, and we discuss this in more detail in \secref{renormalization} for biased tracers.}

First, we point out that the appearance of these two structures relies on $\pi_S$ and $\pi^i_V$ having two different Green's functions.  To see this, notice that the equations of motion \eqn{greateom} imply that 
\be
\pi_S \sim \alpha_S \partial_i \partial_j {S_\tau^{ij}} \andd \pi^i_V \sim \alpha_V \epsilon^{ijk} \partial_j \partial_l {S_\tau^{kl}} \ , 
\ee
for some constants $\alpha_S$ and $\alpha_V$ coming from the EdS Green's functions (which are simply numerical constants in the EdS approximation, see \appref{edsgreensfunctionsec}), where we have defined
\be \label{sourcetau}
S_\tau^{ij} \equiv \left[ 2 \mpl^2 a^{-2}\left( \partial_i \Phi \partial_j \Phi - \half \delta_{ij} ( \partial \Phi)^2 \right)   + \frac{\pi^i \pi^j}{\rho} \right]_\tau + \tau^{ij}           \ ,
\ee
and used the notation $[\cdot ]_\tau$ to mean that the term inside of the brackets is sourced by at least one insertion of the stress tensor $\tau^{ij}$ in perturbation theory.  
Using the definition of $\pi^i$ in \eqn{piidef}, these expressions lead to
\be \label{newtermspiexp}
\pi^i \sim \alpha_V \partial_j {S_\tau^{ij}} + ( \alpha_S - \alpha_V ) \frac{\partial_i \partial_j \partial_k}{\partial^2} {S_\tau^{jk}}  \ ,
\ee
so we see that the last term would vanish if $\alpha_S = \alpha_V$, {\it i.e.} if the Green's functions for $\pi_S$ and $\pi_V^i$ are the same.  Second, we note that the new term is also absent if $\partial_i \partial_j \partial_k {S^{jk}_{\tau,(2)} } = \partial^2 \partial_j {S^{ij}_{\tau,(2)}}$, which is true for any terms in ${S^{ij}_{\tau,(2)}}$ proportional to $\delta_{ij}$, for example.  Thus, terms in \eqn{tauij2} that lead to the new non-locally-contributing counterterm structure are the ones proportional to $c_5$ and $c_6$ and the flow terms proportional to $c_1$ and $c_2$.  The term $\partial_i \partial_j \partial_k ( \partial_l \tilde \tau^{kl}_{ct,(1)} \partial_j \tilde \delta^{(1)} / \partial^2 ) / \partial^2$  in \eqn{piict2sol1} also allows $c_3$ to contribute to the non-local structure.

The two structures in \eqn{twocombos} have distinct dependence on the momenta of the fields in $\tilde \tau^{ij}_{*,(2)}$, and so can give distinct momentum dependence for EFT counterterms.  {Notice, however, that for correlators that only involve $\delta$, like the power spectrum and bispectrum of $\delta$ in real space in \secref{sptsolssec}, this difference does not show up (as can be seen from \eqn{fulldiffeqdelta} where $\tau^{ij}$ only contributes locally).  The situation is different, though, in redshift space where the structures in \eqn{twocombos} contribute in different ways so that one must include both possibilities in the counterterms to correctly describe the UV physics.  }Indeed, both terms are necessary to match the UV structures of the loop integrals in redshift space, and we describe this in more detail in \secref{rssresponsesec} and \secref{stochdmsec}.   {The bottom line conclusion from renormalization of dark matter in real space is that as long as one correctly solves the equations of motion, all renormalization comes through a stress tensor that is a local-in-space function of $\partial_i \partial_j \Phi$ and $\partial_i v^j$, as expected.  The bottom line conclusion that we will find in \secref{totalrsssec} for dark matter in redshift space is that, as long as one correctly solves the equations of motion (especially for $\pi^i$), renormalization happens through a local-in-space stress tensor and local-in-space redshift space counterterms.}

We would like to point out, as a side note, that renormalization of the dark-matter one-loop bispectrum \emph{requires} the generation of vorticity in the velocity field.  Although absent in SPT, velocity vorticity $\omega^i$ is sourced from the symmetric stress tensor through \cite{Carrasco:2013mua} 
\be
\omega^i \sim \epsilon^{ijk} \partial_j \left(\epsilon^{kmn}v^m \omega^n -  \rho^{-1} \partial_l {\tau^{lk}} \right) \ .
\ee
For simplicity, we focus on the response terms.  In this case, since $\omega_{(1)}^i = 0$, $\omega^i$ starts being sourced at second order from the stress-tensor term $\rho^{-1} \partial_l \tau^{lk}$, which in turn means that the term $\epsilon^{kmn}v^m \omega^n$ starts at third order.  We can thus ignore the latter term for this discussion.  One can then check that the parameters in the stress tensor \eqn{tauij1} and \eqn{tauij2} that source vorticity at second order are $c_1$, $c_2$, $c_5$, and $c_6$, and we have explicitly verified that it is impossible to renormalize the one-loop bispectrum (specifically $B_{411}$) if all of these parameters are zero.

Finally, let us briefly comment on some potentially confusing aspects of the renormalization and UV matching of $B_{321}^{(I)}$ with the stochastic terms, explicitly shown in \eqn{dmstochrenormeqs} and \eqn{dmstochrenormeqs2}.  Although we defined $ B_{321}^{ (I) , \epsilon} $ by summing over all permutations in \eqn{dmstochrenormeqs2},  we could have done the UV matching in terms of our Galilean invariant stress tensor by considering all terms with an external $P_{11} ( k_1)$, {\it i.e.} just symmetrizing over $\kvec_2$ and $\kvec_3$ and considering
\be
\bar B_{321}^{(I),\epsilon} ( k_1 , k_2 , k_3 ) \equiv \langle \tilde  \delta^{(1)} ( \kvec_1  ) \tilde  \delta^{(1)}_\epsilon ( \kvec_2  ) \tilde \delta^{(2)}_\epsilon ( \kvec_3  ) \rangle '  +  \langle \tilde  \delta^{(1)} ( \kvec_1  ) \tilde  \delta^{(1)}_\epsilon ( \kvec_3  ) \tilde \delta^{(2)}_\epsilon ( \kvec_2  ) \rangle '  \ . 
\ee  
{A curious point about absorbing the UV divergences of $B_{321}^{(1)}$ with $\bar B_{321}^{(1),\epsilon}$ is that it is still possible to do even if the leading IR part of the counterterm solution \eqn{delta2ctep} ({\it i.e.} those ensuring the correct Galilean properties) were wrong.  However, this is only true because the IR part of \eqn{delta2ctep} actually does not contribute to $\bar B_{321}^{(I),\epsilon}$  after symmetrization over $\kvec_2 $ and $\kvec_3$.\footnote{This is for a reason very similar to why the LSS consistency relations are trivial for the tree-level bispectrum, because the leading term when $\qvec \rightarrow 0$ has to be zero because of permutation symmetry in $(\qvec , \kvec_2 , \kvec_3)$.} }  However, as we will see in \secref{stochdmsec}, this is no longer the case in redshift space, and the precise form of the IR terms in \eqn{delta2ctep} is crucial to obtaining the correct renormalization and UV matching.   {This clarifies some statements made in \cite{Baldauf:2014qfa} about this renormalization.}

%%%%%%%%%%%%%%%%%%
%
%
%
%

%%%%%%%%%%%%%%
%
%
%
%
\section{Dark-matter renormalization in redshift space} \label{totalrsssec}

\subsection{General redshift-space equations}

The distribution of matter is roughly homogeneous and isotropic in the comoving coordinate $\xvec$, but since we use redshift to assign distances, the coordinate that we actually measure for each galaxy is (see for example \cite{Matsubara:2007wj}) 
\be
\xvec_r = \xvec + \frac{\hat z \cdot \vec{v}}{a H }\hat z \ ,
\ee
where $\hat z$ is the line of sight direction.  Mass conservation implies that $\rho( \xvec_r) d^3 x_r = \rho ( \xvec) d^3 x$, which gives
\be
\delta_r ( \kvec, \hat z ) = \delta ( \kvec ) + \int d^3 x \, e^{- i \kvec \cdot \xvec} \left(  \exp \left[ - i \frac{(\hat z \cdot \kvec)}{aH } ( \hat z \cdot \vec{v} ( \xvec ) ) \right] - 1 \right) (1 + \delta( \xvec ) ) \ ,
\ee
in Fourier space.  Since we will compute up to the one-loop bispectrum, we expand to fourth order, which gives
\begin{align}
\begin{split} \label{rssrealsp}
\delta_r  &  = \delta  - \frac{\hat{z}^i \hat{z}^j}{aH \bar \rho} \partial_i \pi^j + \frac{\hat{z}^i \hat{z}^j \hat{z}^k \hat{z}^l}{2 (a H)^2 \bar \rho } \partial_i \partial_j ( \pi^k v^l ) \\
& \hspace{.5in} - \frac{\prod_{a=1}^6 \hat{z}^{i_a} }{3! (a H)^3 \bar \rho} \partial_{i_1} \partial_{i_2} \partial_{i_3} ( \pi^{i_4} v^{i_5} v^{i_6} ) + \frac{\prod_{a=1}^8 \hat{z}^{i_a} }{4! (a H)^4 \bar \rho} \partial_{i_1} \partial_{i_2} \partial_{i_3}\partial_{i_4}  ( \pi^{i_5} v^{i_6} v^{i_7} v^{i_8} )  + \dots \ ,
\end{split}
\end{align}
in position space.  {We now see the appearance of $\pi^i$ contracted in a non-isotropy-preserving way, since isotropy is broken in redshift space by the preferred direction $\hat z$.}

 {The $n$-th order expression for $\delta_r$ in SPT, {\it i.e.} with $\tau^{ij}=0$ and redshift space EFT counterterms (related to contact operators in \eqn{rssrealsp} that we will discuss in the subsequent sections) set to zero, can be written }
\begin{align}
\begin{split} \label{rsskernels}
& \delta^{(1)}_r ( \kvec , \hat z , a ) = D(a) F_1^r ( \kvec ;  \hat z ) \tilde \delta_{\kvec}^{(1)}  \ , \\
 & \delta^{(n)}_r ( \kvec,  \hat z , a ) =  D(a)^n \int_{\kvec_1 , \dots , \kvec_n}^{\kvec} F^r_n ( \kvec_1 , \dots , \kvec_n ; \hat z) \tilde \delta^{(1)}_{ \kvec_1} \cdots \tilde \delta^{(1)}_{ \kvec_n } \ , 
\end{split}
\end{align}
for $n \geq 2$, where the redshift space kernels $F^r_n$ up to $n=3$ can be found in {\cite{Matsubara:2007wj}, for example}.   
The observables that we are interested in are defined analogously to those in \secref{sptsolssec}.  In the plane-parallel approximation that is common in redshift space {and that we use in this paper}, the power spectrum and bispectrum are defined by 
\begin{align}
\begin{split}
 \langle \delta_r ( \kvec_1 ,  \hat z , a  ) \delta_r ( \kvec_2 ,  \hat z , a  ) \rangle & = ( 2 \pi)^3 \delta_D ( \kvec_1 + \kvec_2 ) P^r ( k_1 , \hat k_1 \cdot  \hat z , a ) \ , \\
 \langle \delta_r ( \kvec_1 ,  \hat z , a  ) \delta_r ( \kvec_2  ,  \hat z , a ) \delta_r ( \kvec_3  ,  \hat z , a ) \rangle & = ( 2 \pi)^3 \delta_D ( \kvec_1 + \kvec_2 + \kvec_3 )B^r ( k_1 , k_2 , k_3, \hat k_1 \cdot \hat z , \hat k_2 \cdot \hat z , a)  \ .
\end{split}
\end{align}
Notice that since translation invariance is preserved, the correlation functions still have the Dirac delta functions for total wavenumber conservation.  However, since isotropy is broken, the spectra can depend on angles with respect to $\hat z$.  

The total one-loop power spectrum is
\be
P^r_{1\text{-loop tot.}} ( k, \hat k \cdot \hat z , a )  =  D(a)^2 P^r_{11} ( k ,  \hat k \cdot \hat z )  + D(a)^4 ( P^r_{22} ( k , \hat k \cdot \hat z) + P^r_{13} ( k , \hat k \cdot \hat z ) ) \ , 
\ee
where 
\be
P_{11}^r ( k , \hat k \cdot \hat z ) = ( 1 + f ( \hat k \cdot \hat z)^2 )^2 P_{11} ( k ) \ ,
\ee
is the famous Kaiser result, and
\begin{align}
\begin{split} \label{loopexpressionsrss}
&P_{22}^r ( k, \hat k \cdot \hat z  )  = 2 \int_{\qvec} F_2^r ( \qvec , \kvec - \qvec ; \hat z )^2 P_{11} ( q ) P_{11}( | \kvec - \qvec|) \ , \\
& P_{13}^r  (k, \hat k \cdot \hat z)  = 6 P_{11} ( k )F_1^r ( \kvec , \hat z )  \int_{\qvec} F_3^r ( \qvec , - \qvec , \kvec ; \hat z ) P_{11}(q) \ .
\end{split}
\end{align}
The total one-loop bispectrum is 
\be
B^r_{1\text{-loop tot.}}  = D(a)^4 B^r_{211}  + D(a)^6 \left(  B^r_{222}  + B_{321}^{r,(I)} + B_{321}^{r,(II)} + B^r_{411}   \right)   \ , 
\ee
where here and below we suppress the argument $(k_1, k_2, k_3, \hat k_1 \cdot \hat z , \hat k_2 \cdot \hat z )$ of the bispectra terms to remove clutter.  The tree-level bispectrum is 
\be
B^r_{211}  = 2  F_1^r( \kvec_1 ; \hat z) F_1^r ( \kvec_2 ; \hat z)  F^r_2 ( \kvec_1 , \kvec_2; \hat z ) P_{11}(k_1 ) P_{11} ( k_2 ) + \text{ 2 perms.} \ , 
\ee
and the one-loop contributions are 
\begin{align}
\begin{split} \label{bispexpressionsrss}
& B^r_{222}  = 8 \int_{\qvec} P_{11}(q) P_{11}(| \kvec_2 - \qvec| ) P_{11} ( |\kvec_1 + \qvec|) \\
& \hspace{1.5in} \times F_2^r ( - \qvec , \kvec_1 + \qvec ; \hat z) F^r_2 ( \kvec_1 + \qvec , \kvec_2 - \qvec ;  \hat z ) F^r_2 ( \kvec_2 - \qvec , \qvec ;  \hat z)  \ , \\
& B_{321}^{r,(I)}  = 6 P_{11}(k_1) F_1^r ( \kvec_1  ; \hat z)  \int_{\qvec} P_{11}(q) P_{11}(| \kvec_2 - \qvec|)  \\
& \hspace{1.5in} \times F_3^r ( - \qvec , - \kvec_2 + \qvec , - \kvec_1;  \hat z ) F^r_2 ( \qvec , \kvec_2 - \qvec ; \hat z )  + \text{ 5 perms.}  \ , \\
& B_{321}^{r,(II)}  = 6 P_{11}(k_1 ) P_{11}(k_2) F_1^r ( \kvec_1 ;  \hat z) F_2^r ( \kvec_1 , \kvec_2 ;  \hat z ) \int_{\qvec} P_{11}(q) F_3^r ( \kvec_1 , \qvec , - \qvec ; \hat z ) + \text{ 5 perms.} \ ,  \\ 
& B_{411}^r  = 12 P_{11} ( k_1 ) P_{11}(k_2) F_1^r ( \kvec_1 ; \hat z) F_1^r ( \kvec_2 ;  \hat z)  \int_{\qvec} P_{11}(q) F_4^r ( \qvec , - \qvec , - \kvec_1 , - \kvec_2 ;  \hat z ) + \text{ 2 perms.}  \ .
\end{split}
\end{align}
{As a final point, we note that we have explicitly displayed and factored out the major source of time dependence, which is through the factors of $D(a)^n$ in \eqn{rsskernels}, in the above equations.  The kernels $F_n^r$ in \eqn{rsskernels} are in fact time dependent as well, coming from factors of $f(a)$ that enter \eqn{rssrealsp} through the factors of $\vec v$.  While we fully take into account this time dependence, we do not explicitly write the time argument in the $F_n^r$ kernels, to remove clutter; all kernels and observables with the redshift space marking `$r$' are understood to contain this time dependence through $f(a)$.} {For details on how to evaluate the above integrals, see \appref{bisploopapp}.}

%%%%%%%%%%%%
%
\subsection{Renormalization of dark matter in redshift space} \label{renormdmrsssec}
Ultimately, we want a renormalized expression for the redshift space overdensity $\delta_r$ in \eqn{rssrealsp}.  The first two terms, containing only $\delta$ and $\pi^j$, have already been renormalized in \secref{dmrenormrssec}, and this is entirely determined by the local stress-tensor counterterms in $\tau^{ij}$.  The non-linear terms in \eqn{rssrealsp}, however, are contact operators ({\it i.e.} UV sensitive) and must be separately renormalized \cite{Senatore:2014vja}, which essentially amounts to adding new counterterms directly to \eqn{rssrealsp}.  Here we present a systematic renormalization that preserves Galilean transformation properties, extending \cite{Lewandowski:2015ziq}, and address some subtleties that appear since we are going to quadratic order in the counterterms.

As can be seen in \eqn{rssrealsp}, we ultimately want to renormalize products like $\pi^i v^j v^k \cdots$.   In order to build up to that, let us start with the renormalization of velocity products, up to $[v^i v^j v^k v^l]_R$, where $[\cdot]_R$ denotes a renormalized quantity.  In order to have the correct transformation properties under the Galilean transformation $v^i \rightarrow v^i + \chi^i$ {(here and elsewhere $\chi^i$ is a constant vector)}, we wish for the renormalized quantities to transform in the same way as the bare operators, so we have
\begin{align}
\begin{split}
& [v^i]_R \rightarrow [v^i]_R + \chi^i \ , \\
& [v^i v^j ]_R \rightarrow [v^i v^j]_R + [v^i]_R \chi^j + [v^j]_R \chi^i + \chi^i \chi^j \ , \\
& [v^i v^j v^k]_R \rightarrow [v^i v^j v^k]_R + ( [v^i v^j]_R \chi^k + \text{2 perms.} ) + ( [v^i]_R \chi^j \chi^k + \text{2 perms.} ) + \chi^i \chi^j \chi^k \ , \\ 
& [v^i v^j v^k v^l]_R \rightarrow [v^i v^j v^k v^l]_R + ( [v^i v^j v^k]_R \chi^l + \text{3 perms.} ) + ( [v^i v^j]_R \chi^k \chi^l + \text{5 perms.} )  \\
& \hspace{1in}  + ( [v^i]_R \chi^j \chi^k \chi^l + \text{3 perms.}) + \chi^i \chi^j \chi^k \chi^l  \ .
\end{split}
\end{align}
One way to write renormalized quantities satisfying the above in terms of the non-renormalized fields is
\begin{align}
\begin{split} \label{vprods}
& [v^i]_R = v^i + \mathcal{O}_v^i \ ,   \\
& [v^i v^j]_R = [v^i]_R [v^j]_R + \mathcal{O}_{v^2}^{ij} \ ,  \\
& [v^i v^j v^k]_R = ([v^i v^j]_R [v^k]_R + \text{2 perms.})-  2 [v^i]_R [v^j]_R [v^k]_R + \mathcal{O}_{v^3}^{ijk} \ ,  \\
&  [v^i v^j v^k v^l]_R  = ([v^i v^j v^k]_R [v^l]_R + \text{3 perms.} ) - ( [v^i v^j]_R [v^k v^l]_R + \text{2 perms} )  + \mathcal{O}_{v^4}^{ijkl} \ , 
\end{split}
\end{align}
where all of the $\mathcal{O}$ terms are Galilean scalars.  The last expression is not unique in the sense that other operators could have been used that are not independent from the ones shown, like $[v^i]_R [v^j]_R [v^k]_R [v^l]_R $ and $[v^i v^j]_R [v^k]_R [v^l]_R$, for example.  Definitions using different bases can differ in their scalar parts $\mathcal{O}$. {Note that $v^i$ is renormalized here because it is the composite operator $\pi^i / \rho$ \cite{Carrasco:2013mua}. }

We can similarly renormalize products involving $\delta$.  Demanding again that renormalized quantities transform in the same way as bare ones under Galilean transformations means that we want
\begin{align} \label{deltavtrans}
& [\delta v^i]_R \rightarrow [\delta v^i]_R + [\delta]_R \chi^i \ , \\
& [\delta v^i v^j ]_R \rightarrow [ \delta v^i v^j]_R + [\delta v^i]_R \chi^j + [\delta v^j]_R \chi^i + [\delta]_R \chi^i \chi^j \ ,\nonumber \\
& [\delta v^i v^j v^k]_R \rightarrow [ \delta v^i v^j v^k]_R + ( [\delta v^i v^j]_R \chi^k + \text{2 perms.} ) + ( [\delta v^i]_R \chi^j \chi^k + \text{2 perms.} ) + [\delta]_R \chi^i \chi^j \chi^k \ ,  \nonumber \\ 
& [\delta v^i v^j v^k v^l]_R \rightarrow [\delta v^i v^j v^k v^l]_R + ( [\delta v^i v^j v^k]_R \chi^l + \text{3 perms.} ) + ( [\delta v^i v^j]_R \chi^k \chi^l + \text{5 perms.} ) \nonumber  \\
& \hspace{1in}  + ( [\delta v^i]_R \chi^j \chi^k \chi^l + \text{3 perms.}) + [\delta]_R \chi^i \chi^j \chi^k \chi^l  \ . \nonumber
\end{align}
One way to write renormalized quantities satisfying the above in terms of the non-renormalized fields is 
\begin{align}\label{deltavprods}
& [\delta]_R = \delta + \mathcal{O}_\delta \ ,   \nonumber \\
& [\delta v^i]_R = [\delta]_R [v^i]_R + \mathcal{O}^i_{v \delta } \ ,  \nonumber \\
& [\delta v^i v^j]_R =  [\delta v^i]_R [v^j]_R + [\delta v^j]_R [v^i]_R - [\delta]_R [v^i]_R [v^j]_R + \mathcal{O}^{ij}_{v^2 \delta }  \ ,  \nonumber \\
& [\delta v^i v^j v^k]_R = ( [\delta v^i v^j]_R [v^k]_R + \text{2 perms.}) - \frac{2}{3} ( [\delta v^i]_R [v^j]_R [v^k]_R + \text{2 perms.} ) \\
& \hspace{1in} + \frac{1}{3} ( [\delta]_R [v^i v^j]_R [v^k]_R + \text{2 perms.} ) - \frac{1}{3} ( [\delta v^i ]_R [v^j v^k]_R + \text{2 perms.}) + \mathcal{O}^{ijk}_{v^3 \delta }  \ ,  \nonumber \\
& [\delta v^i v^j v^k v^l]_R = ( [\delta v^i v^j v^k]_R [v^l]_R + \text{3 perms.} ) -\half ( [\delta v^i]_R [v^j v^k v^l]_R  + \text{3 perms.} )    \nonumber  \\
& \hspace{1in} - ([\delta v^i v^j]_R [v^k]_R [v^l]_R + \text{5 perms.} )   + \half ( [\delta v^i]_R [v^j v^k]_R [ v^l]_R + \text{11 perms.} )  \nonumber  \\
& \hspace{1in} + \frac{1}{2} ( [\delta]_R [ v^i v^j v^k]_R [ v^l]_R + \text{3 perms.} )  -   ( [\delta]_R [ v^i v^j ]_R [ v^k  v^l]_R + \text{2 perms.} ) +  \mathcal{O}^{ijkl}_{v^4 \delta }  \ . \nonumber
\end{align}
The above expressions were determined by imposing the correct transformation law \eqn{deltavtrans} and having the correct limit to \eqn{vprods} when $\delta \rightarrow 1$.  The expression for $[\delta v^i v^j v^k]_R$ is not uniquely determined by these constraints, but the difference is immaterial; any other definitions for $[\delta v^i v^j v^k]_R$ differ by Galilean invariant terms, in terms of free EFT coefficients, which vanish when $\delta \rightarrow 1$.

Now, we can combine the above expressions to get the terms in \eqn{rssrealsp} relevant for redshift space distortions.  Specifically, we write
\be
[\pi^{i_1} v^{i_2} \cdots  v^{i_n}]_R \equiv [\rho v^{i_1}v^{i_2} \cdots v^{i_n} ]_R = \bar \rho (  [v^{i_1} v^{i_2} \cdots v^{i_n}]_R + [\delta v^{i_1} v^{i_2} \cdots  v^{i_n} ]_R )
\ee
to define the renormalized quantities involving $\pi^i$.  After doing that, we find it more convenient to expand the above expressions \eqn{vprods} and \eqn{deltavprods} in terms of the non-renormalized fields and write 
\begin{align}
\begin{split} \label{rhovrenorm}
& [\delta]_R = \delta + \mathcal{O}_\delta \ ,   \\
& [\pi^i]_R  =  \pi^i + v^i  \mathcal{O}_\rho +  \mathcal{O}_{ \pi }^i  \ ,  \\
& [\pi^i v^j ]_R  =  \pi^i v^j  +  v^i v^j \mathcal{O}_\rho + v^i \mathcal{O}_{\pi }^j + v^j \mathcal{O}_{\pi }^i + \mathcal{O}^{ij}_{\pi v}   \ , \\
&[\pi^i v^j v^k]_R  =  \pi^i v^j v^k +   v^i v^j v^k \mathcal{O}_\rho  + ( v^i v^j \mathcal{O}_{\pi }^k + \text{2 perms.}) + ( v^i \mathcal{O}_{ \pi v }^{jk} + \text{2 perms.} ) + \mathcal{O}^{ijk}_{\pi v^2} \ ,  \\
& [\pi^i v^j v^k v^l]_R =  \pi^i v^j v^k v^l + v^i v^j v^k v^l \mathcal{O}_\rho +  ( v^i v^j v^k \mathcal{O}_{\pi }^l + \text{3 perms.} ) \\
& \hspace{1.in} +(v^i v^j \mathcal{O}_{\pi v }^{kl} + \text{5 perms.} ) + ( v^i \mathcal{O}_{\pi v^2}^{jkl} + \text{3 perms.} ) + \mathcal{O}_{\pi v^3}^{ijkl}  \ , 
\end{split}
\end{align}
One can show that all of the $\mathcal{O}_{\pi v^n}$ tensors above (which are all Galilean scalars) can be written in terms of $\delta$, $\mathcal{O}_\delta$, the $\mathcal{O}_{v^n}$, and the $\mathcal{O}_{v^n \delta}$.  Explicitly, we have 
\be \label{opirel}
\mathcal{O}_\rho = \bar \rho \, \mathcal{O}_\delta \ , \quad \mathcal{O}_{\pi }^i = \bar \rho \left( (1 + \delta + \mathcal{O}_\delta) \mathcal{O}_v^i + \mathcal{O}_{v \delta}^i \right) \ ,
\ee  
\be
\mathcal{O}^{ij}_{\pi v} =  \bar \rho \left[ ( 1 + \delta + \mathcal{O}_\delta) \mathcal{O}_v^i \mathcal{O}_v^j   + \mathcal{O}_v^i \mathcal{O}_{v \delta}^j +  \mathcal{O}_v^j \mathcal{O}_{v \delta}^i   + \mathcal{O}_{v^2}^{ij} + \mathcal{O}_{v^2 \delta}^{ij}  \right]  \ , 
\ee
\begin{align}
\begin{split}
& \mathcal{O}^{ijk}_{\pi v^2} =  \bar \rho \Big[ ( 1 + \delta + \mathcal{O}_\delta) \mathcal{O}_v^i \mathcal{O}_v^j \mathcal{O}_v^k + \mathcal{O}_{v^3}^{ijk} + \mathcal{O}_{v^3 \delta}^{ijk}   \\
& \hspace{.8in} +  \mathcal{O}_v^i \left( \mathcal{O}_{v^2}^{jk}  + \mathcal{O}_{v^2 \delta}^{jk}  \right) + \mathcal{O}_v^j \left( \mathcal{O}_{v^2}^{ki} + \mathcal{O}_{v^2 \delta}^{ki}  \right)  +  \mathcal{O}_v^k \left( \mathcal{O}_{v^2}^{ij} + \mathcal{O}_{v^2 \delta}^{ij}  \right) \\
& \hspace{.8in} + \mathcal{O}_{v}^i \mathcal{O}_v^j \mathcal{O}_{v \delta}^k + \mathcal{O}_{v}^j \mathcal{O}_v^k \mathcal{O}_{v \delta}^i  + \mathcal{O}_{v}^k \mathcal{O}_v^i \mathcal{O}_{v \delta}^j   - \frac{1}{3} \mathcal{O}_{v^2}^{ij} \mathcal{O}_{v \delta}^k  - \frac{1}{3} \mathcal{O}_{v^2}^{jk} \mathcal{O}_{v \delta}^i  - \frac{1}{3} \mathcal{O}_{v^2}^{ki} \mathcal{O}_{v \delta}^j  \Big]  \ , 
\end{split}
\end{align}
and  
\begin{align}
 \mathcal{O}^{ijkl}_{\pi v^3} & = \bar \rho \Bigg[  ( 1 + \delta + \mathcal{O}_\delta ) \left( \mathcal{O}_{v}^i \mathcal{O}_{v}^j \mathcal{O}_{v}^k \mathcal{O}_{v}^l -  \mathcal{O}_{v^2}^{ij} \mathcal{O}_{v^2}^{kl} - \mathcal{O}_{v^2}^{ik} \mathcal{O}_{v^2}^{jl}  - \mathcal{O}_{v^2}^{il} \mathcal{O}_{v^2}^{jk}  \right)  \nonumber \\
&\hspace{.3in} +  \mathcal{O}_{v^4}^{ijkl}+  \mathcal{O}_{v^4 \delta }^{ijkl}  - \half \mathcal{O}_{v^3}^{ijk} \mathcal{O}_{v \delta}^l - \half  \mathcal{O}_{v^3}^{ijl} \mathcal{O}_{v \delta}^k -  \half \mathcal{O}_{v^3}^{ikl} \mathcal{O}_{v \delta}^j -  \half \mathcal{O}_{v^3}^{jkl} \mathcal{O}_{v \delta}^i  \\
&\hspace{.3in} + \Bigg( \frac{\mathcal{O}_v^i }{6} \left(  \mathcal{O}_{v^3}^{jkl}+\mathcal{O}_{v^3\delta}^{jkl} - \frac{1}{3} \mathcal{O}_{v^2}^{jk}\mathcal{O}_{v\delta}^l   - \frac{1}{3} \mathcal{O}_{v^2}^{jl}\mathcal{O}_{v\delta}^k - \frac{1}{3} \mathcal{O}_{v^2}^{kl}\mathcal{O}_{v\delta}^j  \right) + \frac{\mathcal{O}_v^i \mathcal{O}_v^j }{4}  \left( \mathcal{O}_{v^2}^{kl} + \mathcal{O}_{v^2 \delta}^{kl}   \right)  \nonumber\\
& \hspace{.3in} + \frac{1}{6} \mathcal{O}_{v}^i  \mathcal{O}_{v}^j  \mathcal{O}_{v}^k  \mathcal{O}_{v \delta }^l    + \text{23 perms. of } \{i,j,k,l\}  \Bigg) \Bigg] \nonumber \ . 
\end{align}

{We note that the counterterms above associated with contact operators are in general local-in-space \cite{Carrasco:2013mua}.  This can be seen by considering two operators $\sigma_1 ( \xvec )$ and $\sigma_2 ( \xvec)$.  In the EFT, the ambiguity in the product $\sigma ( \xvec ) \equiv \sigma_1 ( \xvec ) \sigma_2 ( \xvec)$ comes from the fact that $\tilde \sigma ( \xvec ) \equiv \sigma_1 ( \xvec + \delta \xvec ) \sigma_2 ( \xvec)$ is just as good of a definition as $\sigma ( \xvec)$ as long as $|\delta \xvec|$ is below the EFT length cutoff.  The two definitions differ by higher derivatives of the original $\sigma_1$ and $\sigma_2$ fields.  Thus, the origin of the non-locally-contributing counterterms in $[\pi^i]_R$ is not from the fact that it is a contact operator, but rather from the equations of motion \eqn{greateom} and the definition \eqn{piidef}.} 

An important point about \eqn{rhovrenorm} is keeping track of how lower-order counterterms must enter higher-order renormalized products in order to preserve Galilean invariance.   
 To renormalize the contact operators in \eqn{rssrealsp}, we simply replace all of them with the corresponding renormalized operators in \eqn{rhovrenorm}.   This leads us to the renormalized redshift space overdensity $[\delta_r]_R$ which for renormalization up to the one-loop bispectrum we can write as
\be \label{deltataurss}
[\delta_r]_R ( \kvec ,  \hat z , a )  = \delta_r ( \kvec ,  \hat z , a ) +  \delta^{(1)}_{r,ct} ( \kvec ,  \hat z , a ) +   \delta^{(2)}_{r,ct} ( \kvec , \hat z , a ) +  \delta_{r,\epsilon}^{(1)} ( \kvec,\hat z ,  a )  +  \delta_{r,\epsilon}^{(2)} ( \kvec , \hat z , a  )  \ , 
\ee
where the subscript $ct$ denotes the response counterterms, and $\epsilon$ denotes the stochastic (and semi-stochastic) counterterms.  Assuming the time dependence needed to cancel UV loop contributions, we have 
\begin{align}
\begin{split} \label{cttimedeprss}
& \delta^{(1)}_{r,ct} ( \kvec , \hat z , a )= D(a)^3  \tilde \delta^{(1)}_{r,ct} ( \kvec , \hat z  )  \ ,  \quad \delta^{(2)}_{r,ct} ( \kvec ,\hat z ,  a )  = D(a)^4 \tilde \delta^{(2)}_{r, ct} ( \kvec, \hat z ) \ , \\
  & \delta_{r,\epsilon}^{(1)} ( \kvec,  \hat z , a )  = D(a)^2 \tilde \delta_{r,\epsilon}^{(1)} ( \kvec, \hat z )  \andd  \delta_{r,\epsilon}^{(2)} ( \kvec , \hat z , a  ) =  D(a)^3 \tilde \delta_{r,\epsilon}^{(2)} ( \kvec , \hat z  ) \ , 
\end{split}
\end{align}
where we use the tilde to denote the appropriate time-independent factor.

Given the results of \secref{tauijdmsec} for dark-matter renormalization, the linear terms in \eqn{rssrealsp}, $\delta$ and $\pi^i$, are automatically renormalized by the stress tensor $\tau^{ij}$.  As can be seen in \eqn{rhovrenorm}, though, we still need the explicit expression for  $\mathcal{O}_\rho$ and $\mathcal{O}_\pi^i$ in terms of $\tau^{ij}$ to use in higher product renormalizations.  Explicitly, we have
\be \label{Orhoeq}
\mathcal{O}_\rho = \bar \rho \, \delta_\tau \ , 
\ee
where $\delta_\tau$ is the solution sourced by $\tau^{ij}$, \eqn{deltatau}.  Then, given this, we can solve for 
\be \label{Opieq}
\mathcal{O}^i_\pi = \pi^i_\tau - v^i \mathcal{O}_\rho \ , 
\ee
where $\pi^i_\tau$ is the solution sourced by $\tau^{ij}$, \eqn{piitau}.    The counterterms $\mathcal{O}_{\pi v^n}$ entering the higher products are free functions of Galilean scalars and introduce new counterterms in addition to those coming from $\tau^{ij}$.  We give explicit expressions for the redshift space counterterms relevant to this work in \secref{rssresponsesec} and \secref{stochdmsec}.

%%%%%%%%%%%%%%%%%%%%%%%%%
%
%
%

\subsection{IR-limit checks} \label{irlimitcheckssec2}

Let us briefly pause to make some comments about the counterterm solutions that we found in \secref{tauijdmsec}.  A useful consistency check is to confirm that the expressions have the correct IR behavior for their Galilean transformation types.  In perturbation theory, a Galilean scalar $\Sigma$ (like $\delta$ or $\tau^{ij}$) satisfies
\be \label{galscalarirlim}
\Sigma^{(2)} \big|_{\rm IR} = \partial_i \Sigma^{(1)}   \frac{ \partial_i \delta^{(1)} }{\partial^2} \ ,
\ee
where we use the notation $|_{\rm IR}$ to mean the leading term when the momentum of one of the fields goes to zero, which is a straightforward generalization of  \cite{Jain:1995kx, Creminelli:2013mca, Lewandowski:2019txi}, for example.  We can also find the IR behavior of the momentum counterterms directly from \eqn{Opieq}, since we have $\pi^i_\tau = \mathcal{O}^i_\pi + v^i \mathcal{O}_\rho $. 
Then since $\mathcal{O}_\rho$ and $\mathcal{O}_\pi^i$ are Galilean scalars, we must have
\be
\pi_{\tau , (2) }^i \big|_{\rm IR} = \partial_j \mathcal{O}^i_{\pi,(1)} \frac{\partial_j \delta^{(1)}}{\partial^2} + \bar \rho \, v^i_{(1)} \delta_\tau^{(1)} \ ,
\ee
where we have used \eqn{Orhoeq}. 

First, we point out that the stress tensor that we wrote down in \eqn{tauij1}, \eqn{tauij2}, and \eqn{stochstress} is indeed a Galilean scalar, since
\be
\tilde \tau^{ij}_{*,(2)} \big|_{\rm IR} =  \partial_k \tilde \tau^{ij}_{*,(1)}  \frac{ \partial_k \tilde \delta^{(1)}}{\partial^2} \ ,
\ee
for both the response and stochastic contributions, respectively $*  = ct, \epsilon$.  This is of course by construction, since we introduced the flow terms as in \cite{Senatore:2014eva} {(with the clarification in \appref{stochflowtermsec} for the stochastic terms)} specifically to make $\tau^{ij}$ a Galilean scalar.   

Now we move on to the counterterm solutions for $\delta$ and $\pi^i$ sourced by $\tau^{ij}$.  By inspection, one can indeed see that
\be \label{secondorderir}
\tilde \delta^{(2)}_* \big|_{\rm IR} = \partial_i \tilde \delta^{(1)}_{*} \frac{\partial_i \deltaone}{\partial^2 }  \andd \tilde \pi^i_{*,(2)} \big|_{\rm IR} =   \partial_j \tilde  \pi^i_{*,(1)} \frac{\partial_j \deltaone}{\partial^2} +  \frac{\partial_i \deltaone}{\partial^2}  \tilde \delta^{(1)}_*   \ ,
\ee
which are the correct equations in terms of all of the tilde fields.  These are much more nontrivial checks.  In particular, going through the algebra to check these, one can see how both the second-order stress tensors and the first-order stress tensors plugged back into the equations of motion combine to give the correct answer.  

One can also see this in another way.  Following \cite{Lewandowski:2019txi}, we can start with the linear equation of motion and introduce all of the non-linear leading IR terms directly into the equation of motion.  For example, focusing on the counterterm solutions, we have the linear equation
\be
\ddot \delta^{(1)}_{*} +  2 H \dot \delta^{(1)}_* - \frac{3}{2} \om H^2 \delta^{(1)}_* = \frac{1}{a^2 \bar \rho} \partial_i \partial_j \tau^{ij}_{*,(1)} \ .
\ee
Then, the equation of motion for the leading IR piece of the second-order field is
\be \label{deltact2eqfull}
\left( \ddot \delta^{(2)}_{*} + 2 H \dot \delta^{(2)}_{*} - \frac{3}{2} \om H^2 \delta^{(2)}_{*}  \right) \Big|_{\rm IR} =  \frac{1}{a^2 \bar \rho} \partial_i \partial_j \tau^{ij}_{*,(2)} \big|_{\rm IR}    - a^{-1} ( \dot v^i \partial_i \delta^{(1)}_{*} + 2 v^i \partial_i \dot \delta^{(1)}_{*} + H v^i \partial_i \delta^{(1)}_{*}     )  \ .
\ee
This way, one can see the contribution coming from the IR part of $\tau^{ij}_{*,(2)}$, and the part coming from plugging the linear solution $\delta^{(1)}_*$ back into the equation of motion.  Indeed, one can show that the solution to \eqn{deltact2eqfull} for $\delta^{(2)}_*$ is given by \eqn{secondorderir}.

%%%%%%%%%%%%%%%
%
%

\subsection{Redshift space counterterms: response terms} \label{rssresponsesec}
We now write down the explicit response counterterms (up to second order and number of derivatives discussed above \eqn{tauij1}, which is the same in real space and redshift space) needed to renormalize the product operators in \eqn{rssrealsp}.  We start with $[\pi^i v^j]_R$. Since $\mathcal{O}_\rho$ and $\mathcal{O}_\pi^i$ are already known from \eqn{Orhoeq} and \eqn{Opieq}, we only need $\mathcal{O}^{ij}_{\pi v}$, which we can expand as   
\be
\mathcal{O}^{ij}_{\pi v, (0)} = (a H f)^2 \bar \rho \frac{D^2}{\knl^2} \delta_{ij} c_{\rm DM,0}^{\pi v}  \ , 
\ee
\begin{align}
\begin{split}
\mathcal{O}^{ij }_{\pi v, (1)} =   (a H f )^2 \bar \rho  \frac{D^3}{\knl^2}  \left( c^{\pi v}_{\rm DM,1} \frac{\partial_i \partial_j \tilde \delta^{(1)}}{\partial^2 }  + c_{\rm DM,3}^{\pi v} \delta_{ij} \tilde \delta^{(1)}    \right)  \ , 
\end{split}
\end{align}
and
\begin{align}
\begin{split}
& \mathcal{O}^{ij }_{\pi v, (2)} =   (a H f )^2 \bar \rho  \frac{D^4}{\knl^2}  \Bigg(  c_{\rm DM,1}^{\pi v} \frac{\partial_k \partial_i \partial_j \tilde \delta^{(1)}}{\partial^2} \frac{\partial_k \tilde \delta^{(1)}}{\partial^2}  + c_{\rm DM,3}^{\pi v} \delta_{ij} \partial_k \tilde \delta^{(1)} \frac{\partial_k \tilde \delta^{(1)} }{\partial^2 }   +  c_{\rm DM,2}^{\pi v} \frac{\partial_i \partial_j  \tilde \delta^{(2)} }{\partial^2 } \\
& \hspace{1.5in} - c_{\rm DM,2}^{\pi v} \frac{\partial_k \partial_i \partial_j \tilde \delta^{(1)} }{\partial^2 } \frac{\partial_k \tilde \delta^{(1)} }{\partial^2 } +  c_{\rm DM,4}^{\pi v} \delta_{ij} \tilde \delta^{(2)} - c_{\rm DM,4}^{\pi v} \delta_{ij}  \partial_k \tilde \delta^{(1)}  \frac{\partial_k \tilde \delta^{(1)} }{\partial^2 }   \\
& \hspace{1.5in} + c_{\rm DM,5}^{\pi v} \frac{\partial_i \partial_j \tilde  \delta^{(1)}}{\partial^2} \tilde \delta^{(1)} + c_{\rm DM,6}^{\pi v} \frac{\partial_i \partial_k \tilde \delta^{(1)}}{\partial^2 } \frac{\partial_k \partial_j \tilde \delta^{(1)}}{\partial^2 }  + c_{\rm DM,7}^{\pi v} \delta_{ij} \tilde \delta^{(1)} \tilde \delta^{(1)}  \Bigg)  \ .
\end{split}
\end{align}
{We have arrived at this list of counterterms using the same procedure as in \secref{explicitstresstensor}.  The only difference is that we now allow a constant term, which did not contribute through the stress tensor because there are derivatives acting on it.} 
Next we move to $[ \pi^i v^j v^k ]_R$.  Since we are only going up to second order, and because it is not possible to make a Galilean scalar of the form $\mathcal{O}^{ijk}_{\pi v^2}$ at the order of fields and derivatives that we need, no new counterterms are needed, although the term $v^i \mathcal{O}^{jk}_{\pi v}$ still contributes.  The situation is similar for $[ \pi^i v^j v^k v^l ]_R$.  For the same reasons, only the term $v^i v^j \mathcal{O}_{\pi v}^{kl}$ contributes, and so again, no new counterterms are needed.     

 {We can now build the response counterterm kernels $F_1^{r,ct}$ and $F_2^{r,ct}$, which are defined explicitly in \appref{dmrssctkernelssec}.   
We have found that the parameters $c_1$, $c_2$, $c_6$, and $c^{\pi v}_{\rm DM,0}$ are degenerate in these expressions, so we have set them to zero in the following.  For the first-order kernel, we have
\be \label{f1rctexpression}
F_1^{r,ct} (\kvec ; \hat z) =  - \frac{k^2}{18 \knl^2} \left( 2 c_3   + 3 f ( 2 c_3 + 3 f c^{\pi v}_{\rm DM,3} )( \hat k \cdot \hat z)^2 +  9 f^2 c^{\pi v}_{\rm DM,1} ( \hat k \cdot \hat z)^4 \right)   \ ,
\ee
while for the second-order kernel we have 
\begin{align}
\begin{split} \label{f2rct}
F_2^{r,ct} ( \kvec_1 , \kvec_2 ; \hat z ) = \sum_{i=1}^{11} \alpha^{\rm DM}_i   \, e^{F_2}_i ( \kvec_1 , \kvec_2 ; \hat z)  \ , 
\end{split}
\end{align}
with
\be
\alpha^{\rm DM}_i = \{ c_3 , c_4 , c_5 , c_7 , c^{\pi v}_{\rm DM,1}, c^{\pi v}_{\rm DM,2} , c^{\pi v}_{\rm DM,3}, c^{\pi v}_{\rm DM,4}, c^{\pi v}_{\rm DM,5} , c^{\pi v}_{\rm DM,6}   , c^{\pi v}_{\rm DM,7} \} \ ,
\ee
where the basis functions $e_i^{F_2}$ are defined, and the explicit UV matching to SPT loops is given, in \appref{dmrssctkernelssec}.

This brings us to one of the main results of this paper.  The UV limit of $B_{411}^r$ contains a term 
\begin{align}
\begin{split} \label{b411ruvlim}
B_{411}^r \big|_{\rm UV} & \supset \frac{2 (c_5-c_3)}{99}  \frac{f (k_1^2 - k_2^2)^2 (k_1^2 +k_2^2) ( k_1 \mu_1 + k_2 \mu_2)^2 }{ \knl^2 k_1^2 k_2^2 k_3^2 }P_{11}(k_1) P_{11}(k_2)  \\
& \quad \quad + \text{ 2 perms.} \ ,
\end{split}
\end{align}
where we choose to write the bispectrum using the variables $(k_1 , k_2 , k_3 , \mu_1 , \mu_2)$, the coefficients $c_3$ and $c_5$ are given in \eqn{dmuvmatchingresponse}, and here and elsewhere, $\mu_i \equiv \hat z \cdot \hat k_i$.}  This should be compared to $P_{11}(k_1) P_{11}(k_2) / (k_1^2 k_2^2) + \text{ 2 perms.}$ in \eqn{b411dmuvlimit} for dark matter in real space.  The fact that \eqn{b411ruvlim} contains an explicit factor of $1 / k_3^2$ is a new development, novel to redshift space.\footnote{The reader may notice that if one were to use $\mu_3$ as one of the angle variables in \eqn{b411ruvlim}, the factor $(k_1 \mu_1 + k_2 \mu_2)^2 = k_3^2 \mu_3^2$ in the numerator would cancel the factor of $k_3^2$ in the denominator.  In this case, the novel feature of the expression would be that there is a factor of $\mu_3^2$ upstairs without an accompanying $k_3^2$.}  A term like this can only appear as a counterterm in the bispectrum, {through $B_{411}^{r,ct}$ (expressions for the counterterm solutions and which loops they renormalize are given in  \appref{dmrssctkernelssec}),} because of the appearance of the differential operator $\partial_i \partial_j \partial_k  / \partial^2 $ acting on two fields in $\tilde \pi^i_{ct,(2)}$  in \eqn{piict2sol1}, {and thus proves the necessity of the new non-locally-contributing counterterm.  Indeed, the expression for $F_2^{r,ct}$ in \eqn{f2rct} and the values of $c_3$ and $c_5$ given in \eqn{dmuvmatchingresponse} explicitly cancels the UV contribution in \eqn{b411ruvlim}}.  

To see {how this kind of term can be generated from the counterterms}, let us consider some example counterterms and try to reproduce the form of \eqn{b411ruvlim}.   Given that there is a factor $P_{11} ( k_1 ) P_{11} ( k_2 )$ upstairs, this term should come from a contraction like 
\be
\langle \delta^{(1)} ( \xvec_1)  \delta^{(1)} ( \xvec_2)  \delta_{r,ct}^{(2)} ( \xvec_3) \rangle \ ,
\ee
and, since there is a single factor of $f$, it must come from 
\be
\delta^{(2)}_{r,ct} ( \xvec_3)  \sim f \hat z^i \hat z^j \partial_i \tilde  \pi^j_{ct,(2)} ( \xvec_3)  \ ,
\ee
so let us consider the various contributions to $\tilde \pi^i_{ct,(2)} $ from \eqn{piict2sol1}.  First of all, only $\partial_j \tilde \tau^{ij}_{ct,(2)}$ and the term with  $\partial_i \partial_j \partial_k  / \partial^2$ have a chance of giving a factor of $1/k_3^2$ because they contain a $1 / \partial^2$ acting on two fields whose sum of momenta is $-\kvec_3$.  However, it cannot come from $\partial_j \tilde \tau^{ij}_{ct,(2)}$ because in the only term that had a chance, $\tau^{ij}_{ct,(2)} \sim c_2 \partial_i \partial_j \tilde \delta^{(2)} / \partial^2$, the $1 / \partial^2$ is canceled after being hit by $\partial_j$.  So we must have  
\be \label{anewterm}
\delta^{(2)}_{r,ct} ( \xvec_3)  \sim f \hat z^i \hat z^j \partial_i \frac{\partial_j \partial_k \partial_l }{\partial^2} \left(   \tilde \tau^{kl}_{ct, (2)} ( \xvec_3) + \frac{\partial_k \tilde \delta^{(1)} ( \xvec_3)}{\partial^2}  \partial_m \tilde \tau^{lm}_{ct, (1)}  (\xvec_3)  \right)  \ .
\ee
Even still, some terms in the expression for $ \tilde \tau^{kl}_{ct, (2)}$ in \eqn{tauij2}, like $\partial_k \partial_l \tilde \delta^{(2)} / \partial^2$ and those proportional to $\delta_{kl}$,  will not give what we want when plugged into \eqn{anewterm}, since again, the $1 / \partial^2$ gets canceled.  However, many will.  Consider the term proportional to $c_6$ in \eqn{tauij2}.  This gives 
\be \label{anewterm2}
\delta^{(2)}_{r,ct} ( \xvec_3)  \sim f \hat z^i \hat z^j \partial_i \frac{\partial_j \partial_k \partial_l }{\partial^2} \left(  \frac{\partial_k \partial_m \deltaone ( \xvec_3) }{\partial^2} \frac{\partial_m  \partial_l \deltaone ( \xvec_3) }{\partial^2}  \right)  \ ,
\ee
which in Fourier space, contains the form \eqn{b411ruvlim} that we desired.  This and similar terms, {like $c_3$ and $c_5$ that appear in \eqn{b411ruvlim}}, are the origin of the $1 / k_3^2$ in \eqn{b411ruvlim}, and we have pinpointed that it is due to the differential operator $\partial_i \partial_j \partial_k / \partial^2$ {appearing in the solution for $\pi^i$ and being contracted in an isotropy-breaking way ({\it i.e.} with $\hat z$).   }  {In terms of the basis $e_{i}^{F_2}$ that we use, the new non-locally-contributing counterterm enters in $e_1^{F_2}$ and $e_{3}^{F_2}$ in \eqn{dmrssbasisfns} and originates from the bias basis function $e_7^{K_2}$ in \eqn{basisfns1}.}

 As a final point, although we have freedom in defining the various operators in the second-order stress tensor \eqn{tauij2}, notice that the new term $\partial_i \partial_j \partial_k \tilde \tau^{jk}_{ct,(2)} / \partial^2 $  appears with a fixed coefficient relative to $\partial_j  \tilde \tau^{ij}_{ct,(2)}  $ in \eqn{piict2sol1}.  This is important to preserve Galilean invariance, and in fact it is not possible to match the UV structure of the loops if the relative coefficient is made different, {because Galilean invariance would be broken}.  Overall, we find that for the renormalizations of $P_{13}^r$ and $B_{411}^r$ at one loop, the expressions given above give a total of {15 free coefficients, 11 of which are independent, and {all 11 of those} are needed to match the UV parts of the loops.  }

%%%%%%%%%%%%%%%
%
%
\subsection{Redshift space counterterms: stochastic terms}
 \label{stochdmsec}
 The stochastic terms are the final step in fully renormalizing {the one-loop power spectrum and bispectrum of} dark matter in redshift space, and we follow the same logic as the previous sections.  As in \secref{rssresponsesec}, we start with $[\pi^i v^j]_R$.  Similar to the case in \secref{explicitstresstensor}, the only explicit counterterms that we have to add, for our purposes, are
 \begin{align}
\begin{split} \label{stochrsdmterms}
\mathcal{O}_{\pi v,(1)}^{ij} &= ( a H f)^2 \bar \rho \,  \frac{D^2}{\knl^2}   \epsilon_4^{ij}    \ ,  \\
\mathcal{O}_{\pi v,(2)}^{ij}  & = ( a H f)^2 \bar \rho \, \frac{D^3}{\knl^2} \left(  \partial_k \epsilon_4^{ij} \frac{\partial_k \tilde \delta^{(1)}}{\partial^2}   +   \epsilon_5^{ijkl} \frac{\partial_k \partial_l \tilde \delta^{(1)}}{\partial^2}  \right)    \ .
\end{split}
\end{align}
The contractions of the stochastic fields should be expanded as in \eqn{epsiloncontract} and \eqn{threepointstoch}.  For dark-matter renormalization, expansion up to $k^0$ suffices {because there are two derivatives acting on $\mathcal{O}_{\pi v}^{ij}$ in \eqn{rssrealsp} already}.  {Because we include all possible tensor structures in contractions of the stochastic fields, \eqn{stochrsdmterms} is clearly the most general expression we can have up to second order that obeys the equivalence principle.}  Again, we find that the UV matching is no longer possible if the coefficients in \eqn{piict2sol1ep} are modified, showing the importance of correctly including the terms from the dark-matter stress tensor.  Additionally, as a followup to the discussion at the end of \secref{irlimitcheckssec} for real space, it is now the case in redshift space that UV matching is \emph{not} possible if the numerical coefficients of $\delta_{\epsilon}^{(2)}$ in \eqn{delta2ctep} are modified ({\it i.e.} Galilean invariance is broken).   {As mentioned in \secref{irlimitcheckssec},} the reason that changing the coefficients in $\delta_{\epsilon}^{(2)}$ in real space did not ruin UV matching was due to the {fact that the IR terms simply did not contribute because of invariance of $B_{321}^{(I),\epsilon} ( k_1 , k_2 , k_3 )$ under permutations of $(\kvec_1 , \kvec_2 , \kvec_3)$}.  

  Additionally, new functional forms appear in the UV limits of $B_{321}^{r,(I)}$ and $B_{222}^r$ which can only be captured by the new terms  $\partial_i \partial_j \partial_k \tilde \tau^{jk}_{\epsilon,(2)} / \partial^2 $ (for reasons similar to those in the discussion near \eqn{b411ruvlim})
and $\partial_i \partial_j \partial_k \tilde \tau^{jk}_{\epsilon,(1)} / \partial^2  - \partial_j \tilde \tau^{ij}_{\epsilon,(1)} $ (which is zero for the response terms)  coming from the momentum-density renormalization.  In particular, the UV limits of $B_{321}^{r,(I)}$ and $B_{222}^r$ contain terms, for example, of the form 
\begin{align} 
\begin{split}\label{b321nonlocal}
B_{321}^{r,(I)} \big|_{\rm UV} & \supset \frac{\beta_1 f^3  k_1 \mu_1 k_2^2 \mu_2^2 ( k_1 \mu_1 + k_2 \mu_2)^2}{k_1^2 k_2^2 k_3^2} \Big(  k_1\mu_1  k_2^2   ( k_1^2 - k_2^2 + k_3^2 ) \\
& \hspace{1.8in} + k_2 \mu_2 ( k_2^2 - k_3^2)( k_1^2 - k_2^2 - k_3^2)      \Big) P_{11} ( k_1 )  + \text{ 2 perms.} \ , 
\end{split}
\end{align}
and
\begin{align}
\begin{split} \label{b222nonlocal}
B_{222}^r \big|_{\rm UV} \supset \frac{\beta_2 f^5 k_1^2 \mu_1^2 k_2^2 \mu_2^2 (k_1 \mu_1 + k_2 \mu_2)^2 \left( (k_1 \mu_1 + k_2 \mu_2)^2 + k_3^2 ( \mu_1^2 + \mu_2^2)   \right)}{k_3^2} \ ,
\end{split}
\end{align}
{for some cutoff-dependent (but momentum independent) parameters $\beta_1$ and $\beta_2$}. 
Again, these differ from the real-space expressions in \eqn{b321IUV} and \eqn{b222UV} because they are proportional to an overall $1/k_3^2$.   For $B_{321}^{r,(I)}$, the reason is exactly the same as the one given in \secref{rssresponsesec} for $B_{411}^r$, specifically the differential operator $\partial_i \partial_j \partial_k / \partial^2$ in \eqn{piict2sol1ep} applied to two fields whose momenta add up to $-\kvec_3$.  For $B_{222}^r$, the explanation is slightly different, since the counterterm $B_{222}^{r, \epsilon}$ {(expressions for the counterterm solutions and which loops they renormalize are given in  \appref{dmrssctkernelssec})} is made from the contraction of first-order stochastic fields.  In this case, the $1/ k_3^2$ comes from the fact that $\partial_i \partial_j \partial_k \tilde \tau^{jk}_{\epsilon,(1)} / \partial^2  - \partial_j \tilde \tau^{ij}_{\epsilon,(1)} $  is non-zero for the stochastic terms, as explained below \eqn{stochcontractions}.  {In both cases, the difference comes because isotropy is broken in redshift space and indices can be contracted with the line-of-sight direction $\hat z$.}    

Because dark-matter renormalization of the stochastic terms starts at $\mathcal{O}(k^4)$, there are more independent counterterms than we will find for biased tracers later.  For the sake of space, we do not write their explicit expressions in this paper.  However, the expressions, along with the values of the EFT coefficients that match the UV limits of stochastic loops, can be found in the accompanying Mathematica file.  We write the general formulas, which match the notation of the Mathematica file, here for reference.  For the power spectrum and bispectrum counterterms, we write
\begin{align}
\begin{split}
P_{22}^{r,\epsilon} ( k , \hat k \cdot \hat z ) & = \frac{1}{\bar n_{\rm DM}} \sum_{i=1}^{5} c_{\text{DM},r,i}^{\rm St} e_{\text{DM},r,i}^{(22)} ( k , \hat k \cdot \hat z) \ , \\
\bar B^{r,(I),\epsilon}_{321} ( \kvec_1 , \kvec_2 , \kvec_3 ; \hat z ) & = \frac{1 + f ( \hat k_1 \cdot \hat z)^2}{\bar n_{\rm DM}}  P_{11}(k_1) \sum_{i=1}^{19} c^{\rm St}_{\text{DM},r,i} e^{\rm St}_{\text{DM},r,i} ( \kvec_1 , \kvec_2 , \kvec_3 ; \hat z )  \ , \\
B_{222}^{r,\epsilon} ( \kvec_1 , \kvec_2 , \kvec_3 ; \hat z ) & = \frac{1}{\bar n_{\rm DM}^2} \sum_{i=1}^{10} c_{\text{DM},r,i}^{(222)} e^{(222)}_{\text{DM},r,i} ( \kvec_1 , \kvec_2 , \kvec_3 ; \hat z )  \ , 
\end{split}
\end{align}
where $\bar B_{321}^{r,(I),\epsilon}$ is defined analogously to \eqn{b321epdef2} {and $\bar n_{\rm DM} \sim \knl^{3}$ is the number density of regions of linear size $\knl^{-1}$ with $\delta \sim 1$}.  Overall, for $P_{22}^{r,\epsilon}$ and $B_{321}^{r,(I),\epsilon}$, we find 19 free parameters, all of which are needed to match the UV limits of the loops, and for $B_{222}^{r,\epsilon}$, we find 10 free parameters, all of which are needed to match the UV limit of the loop.

%%%%%%%%%%%%%%%
%
%
%
\section{Biased tracers in redshift space} \label{biasrsssec}

\subsection{General formulas} \label{biasrssgensec}

We start with the general equations needed for the one-loop bispectrum of biased tracers in redshift space (ignoring EFT counterterms for now, which we will return to in \secref{renormalization}), which are up to fourth order in perturbations.    The overdensity of biased tracers in redshift space, $\delta_{r,h}$ is given by
\be \label{rsssbias1}
\delta_{r,h} ( \kvec, \hat z ) = \delta_h ( \kvec ) + \int d^3 x \, e^{- i \kvec \cdot \xvec} \left(  \exp \left[ - i \frac{(\hat z \cdot \kvec)}{aH } ( \hat z \cdot \vec{v} ( \xvec ) ) \right] - 1 \right) (1 + \delta_{h} ( \xvec )   ) \ ,
\ee
where $\delta_h$ is the tracer overdensity in real space defined by $\delta_h ( \xvec )  - \langle \delta_h (\xvec ) \rangle = ( \rho_h ( \xvec ) -  \bar \rho_h  )/\bar \rho_h$, where $\rho_h ( \xvec ) $ is the tracer density, and $\bar \rho_h \equiv \langle \rho_h ( \xvec )  \rangle$ is the background tracer density.  In \secref{biasm} we give our explicit bias expansion for $\delta_h$, which will not satisfy $\langle \delta_h ( \xvec ) \rangle = 0$.  This means that in \eqn{rsssbias1} we should replace $\delta_h \rightarrow \delta_h - \langle \delta_h \rangle$, which we choose to do explicitly in the renormalized $[\delta_h]_R$ in \secref{renormalization}.   {Also, there is no} velocity bias at leading order in derivatives, {\it i.e.} $\vec{v}_h = \vec{v}$ where $\vec{v}$ is the dark matter velocity.  In configuration space this becomes
\begin{align}
\begin{split} \label{rsssbias}
\delta_{r,h}  &  =   \delta_h     - \frac{\hat{z}^i \hat{z}^j}{aH \bar \rho_h} \partial_i  \pi_h^j  + \frac{\hat{z}^i \hat{z}^j \hat{z}^k \hat{z}^l}{2 (a H)^2 \bar \rho_h } \partial_i \partial_j (  \pi_h^k   v^l) \\
& \quad \quad - \frac{\prod_{a=1}^6 \hat{z}^{i_a} }{3! (a H)^3 \bar \rho_h} \partial_{i_1} \partial_{i_2} \partial_{i_3} ( \pi_h^{i_4} v^{i_5} v^{i_6} ) + \frac{\prod_{a=1}^8 \hat{z}^{i_a} }{4! (a H)^4 \bar \rho_h} \partial_{i_1} \partial_{i_2} \partial_{i_3}\partial_{i_4}  ( \pi_h^{i_5} v^{i_6} v^{i_7} v^{i_8} )  + \dots  \ ,
\end{split}
\end{align}
where the tracer momentum density $\pi_h^i$  is defined by
\be \label{pihalodef}
\pi^i_h \equiv \bar \rho_h ( 1 + \delta_h ) v^i \ . 
\ee

As always, we expand in perturbations
\be
\delta_{r,h} ( \kvec , \hat z , a  ) =  \delta^{(1)}_{r,h} ( \kvec , \hat z , a )  + \delta^{(2)}_{r,h} ( \kvec , \hat z , a )  + \delta^{(3)}_{r,h} ( \kvec , \hat z , a )  + \delta^{(4)}_{r,h} ( \kvec , \hat z , a ) + \dots 
\ee
and define the redshift space kernels $K_n^{r,h}$ by 
\begin{align}
\begin{split} \label{biasedkernelsdef}
\delta_{r,h}^{(1)} ( \kvec, \hat z  , a ) & = D(a)  K_1^{r,h} ( \kvec ; \hat z ) \tilde  \delta^{(1)}_{ \kvec }  \ ,  \\
\delta_{r,h}^{(n)} ( \kvec  , \hat z , a  ) & = D(a)^n \int_{\kvec_1 , \dots , \kvec_n}^{\kvec } K_n^{r,h} ( \kvec_1 , \dots , \kvec_n  ;  \hat z ) \tilde \delta^{(1)}_{ \kvec_1 } \cdots \tilde  \delta^{(1)}_{\kvec_n} \ , 
\end{split}
\end{align}
for $n \geq 2$.  The power spectrum $P^{r,h}$ and bispectrum $B^{r,h}$ are defined by 
\begin{align} 
 \langle \delta_{r,h} ( \kvec_1 ,  \hat z , a  ) \delta_{r,h} ( \kvec_2 ,  \hat z , a  ) \rangle & = ( 2 \pi)^3 \delta_D ( \kvec_1 + \kvec_2 ) P^{r,h} ( k_1 , \hat k_1 \cdot  \hat z , a ) \ , \\
 \langle \delta_{r,h} ( \kvec_1 ,  \hat z , a  ) \delta_{r,h} ( \kvec_2  ,  \hat z , a ) \delta_{r,h} ( \kvec_3  ,  \hat z , a ) \rangle & = ( 2 \pi)^3 \delta_D ( \kvec_1 + \kvec_2 + \kvec_3 )B^{r,h} ( k_1 , k_2 , k_3, \hat k_1 \cdot \hat z , \hat k_2 \cdot \hat z , a)  \ . \nonumber
\end{align}
We write the total one-loop power spectrum as
\be
P^{r,h}_{1\text{-loop tot.}} ( k, \hat k \cdot \hat z , a )  =  D(a)^2 P^{r,h}_{11} ( k ,  \hat k \cdot \hat z )  + D(a)^4 ( P^{r,h}_{22} ( k , \hat k \cdot \hat z) + P^{r,h}_{13} ( k , \hat k \cdot \hat z ) ) \ , 
\ee
where 
\be
P_{11}^{r,h} ( k , \hat k \cdot \hat z ) = K_1^{r,h} ( \kvec ; \hat z ) K_1^{r,h} ( - \kvec ; \hat z ) P_{11} ( k ) \ , 
\ee
and
\begin{align}
\begin{split} \label{loopexpressionsrssbias}
&P_{22}^{r,h} ( k, \hat k \cdot \hat z  )  = 2 \int_{\qvec} K_2^{r,h} ( \qvec , \kvec - \qvec ; \hat z )^2 P_{11} ( q ) P_{11}( | \kvec - \qvec|) \ , \\
& P_{13}^{r,h}  (k, \hat k \cdot \hat z)  = 6 P_{11} ( k )K_1^{r,h} ( \kvec , \hat z )  \int_{\qvec} K_3^{r,h} ( \qvec , - \qvec , \kvec ;  \hat z ) P_{11}(q) \ .
\end{split}
\end{align}
The total one-loop bispectrum is 
\be
B^{r,h}_{1\text{-loop tot.}}  = D(a)^4 B^{r,h}_{211}  + D(a)^6 \left(  B^{r,h}_{222}  + B_{321}^{r,h,(I)} + B_{321}^{r,h,(II)} + B^{r,h}_{411}   \right)   \ , 
\ee
where here and below we suppress the argument $(k_1, k_2, k_3, \hat k_1 \cdot \hat z , \hat k_2 \cdot \hat z )$ of the bispectra terms to remove clutter.  The tree-level bispectrum is 
\be
B^{r,h}_{211}  = 2  K_1^{r,h}( \kvec_1 ; \hat z) K_1^{r,h} ( \kvec_2 ; \hat z)  K^{r,h}_2 ( \kvec_1 , \kvec_2; \hat z ) P_{11}(k_1 ) P_{11} ( k_2 ) + \text{ 2 perms.} \ ,
\ee
and the one-loop contributions are 
\begin{align} \label{bispexpressionsrssbias}
& B^{r,h}_{222}  = 8 \int_{\qvec} P_{11}(q) P_{11}(| \kvec_2 - \qvec| ) P_{11} ( |\kvec_1 + \qvec|) \nonumber \\
& \hspace{1.5in} \times K_2^{r,h} ( - \qvec , \kvec_1 + \qvec ; \hat z) K^{r,h}_2 ( \kvec_1 + \qvec , \kvec_2 - \qvec ; \hat z ) K^{r,h}_2 ( \kvec_2 - \qvec , \qvec ; \hat z) \nonumber  \ ,  \\
& B_{321}^{r,h,(I)}  = 6 P_{11}(k_1) K_1^{r,h} ( \kvec_1 ; \hat z)  \int_{\qvec} P_{11}(q) P_{11}(| \kvec_2 - \qvec|)  \\
& \hspace{1.5in} \times K_3^{r,h} ( - \qvec , - \kvec_2 + \qvec , - \kvec_1 ; \hat z ) K^{r,h}_2 ( \qvec , \kvec_2 - \qvec  ; \hat z )  + \text{ 5 perms.} \nonumber \ ,  \\
& B_{321}^{r,h,(II)}  = 6 P_{11}(k_1 ) P_{11}(k_2) K_1^{r,h} ( \kvec_1 ; \hat z) K_2^{r,h} ( \kvec_1 , \kvec_2 ; \hat z ) \int_{\qvec} P_{11}(q) K_3^{r,h} ( \kvec_2 , \qvec , - \qvec ;  \hat z ) + \text{ 5 perms.}  \nonumber \ , \\ 
& B_{411}^{r,h}  = 12 P_{11} ( k_1 ) P_{11}(k_2) K_1^{r,h} ( \kvec_1 ;  \hat z) K_1^{r,h} ( \kvec_2 ; \hat z)  \int_{\qvec} P_{11}(q) K_4^{r,h} ( \qvec , - \qvec , - \kvec_1 , - \kvec_2 ;  \hat z ) + \text{ 2 perms.}  \ . \nonumber
\end{align}
{For details on how to evaluate the above integrals, see \appref{bisploopapp}.}

%%%%%%%%%%%%%%%%
%
%
\subsection{Bias expansion to fourth order}\label{biasm}

With the general expressions for tracers in redshift space in hand, we turn next to the bias expansion, which is the missing piece from \eqn{rsssbias}.  For notational convenience, we define all of the perturbations, kernels, power spectra, and bispectra for tracers in real space using the notation in \secref{biasrssgensec}, but with subscripts or superscripts `$r,h$' replaced by `$h$.'   From \eqn{rsssbias}, we see that the tracer quantities in real space can be obtained from the respective quantities in redshift space by setting $f = 0$.  

		As has been previously laid out in \cite{Senatore:2014eva}, by the equivalence principle the tracer overdensity $\delta_h$ can only depend on second derivatives of the gravitational potential and first derivatives of the velocity field, as well as higher derivative and stochastic terms. Additionally, since the tracer overdensity depends on these fields in a non-local-in-time way, we integrate over time {along the fluid element $\xfl$}. In summary we can schematically write the tracer overdensity as 
		\bea\label{eq:genralb}
		\delta_h(\vx,t) = \int^t dt' H(t')f_h\left(\pd_i \pd_j\Phi(\xfl,t'),\pd_i v^j(\xfl,t'),\fr{\pd^i_{x_{\rm fl}}}{k_M},\epsilon(\xfl,t'),t'\right)\Bigg|_{\xfl = \xfl ( \xvec , t , t') } \ ,
		\eea
		where $f_h$ is some complicated {function describing tracer clustering}, $\xfl$ is given by
		\bea\label{xfl}
		\xfl(\vec x, t,t ') = \vec x + \int_{t}^{t'} \frac{dt''}{a(t'') } \vec{v}(\vec x_{\rm fl}(\vec x, t,t''), t'') \ ,
		\eea 	
{and $k_M$ is the scale controlling the clustering of the tracer}.\footnote{This is to distinguish it from $\knl$, which is the scale controlling dark-matter clustering.  However, for simplicity in this paper, we let $k_M = \knl$.  The difference is easily restored if desired.}

		As has been subsequently shown in \cite{Angulo:2015eqa,Fujita:2016dne}, using the approach of \cite{McDonald:2009dh} ({\it i.e.} defining linear combinations of the dark-matter fields) still results in functional degeneracies, and so we will build our bias expansion straight from the contractions of the underlying fields and systematically remove degeneracies in a second step. For the remainder of this subsection we will only focus on the lowest-derivative, non-stochastic, real-space, bias terms and we include higher-derivative real-space and redshift-space EFT counterterms and stochastic terms in \secref{renormalization} and \secref{uvmatchingsecbias}.

{To make sure we include all possible operators,} we consider all possible scalar contractions of $\pd_i \pd_j \Phi $ and $\pd_i v^j$. With $\mathcal{O}$ {representing any scalar of Galilean transformations and rotations,} such as $\delta$, $\partial_i v^i$, $\partial_i \partial_j \Phi \partial_i \partial_j \Phi$, etc., we have the general expansion
		\bea\label{bias_raw}
		\delta_{h}(\vx,t) = \sum_{\mathcal{O}}\int^t dt' H(t') c_{\mathcal{O}}(t,t')\mathcal{O}(\xfl(\xvec,  t , t') ,t') \ , 
		\eea	
and we give the full list of the operators $\mathcal{O}$ needed to go to fourth order in fields in \eqn{fullexp}. 	 The $c_{\mathcal{O}}(t,t')$ are incalculable (within the EFT) time kernels describing the non-locality in time.  We then Taylor expand the operators evaluated at $\xfl$ around $\vx$. Going up to order $N$ (for this work we are interested in $N=4$), and assuming growing mode solutions, an operator $\mathcal{O}_m$ starting at $m$-th order has Taylor expansion \cite{DAmico:2022osl} 
		\bea\label{opt}
		\mathcal{O}_{m}( \xvec_{\rm fl} ( \xvec, t , t') , t')\Big|_N  = \sum_{\alpha=m}^{N}\left( \frac{D(t' )}{D(t)} \right)^{\alpha}\left(\sum_{n=\alpha}^{N}\mathbb{C}^{(n)}_{\mathcal{O}_m , \alpha-(m-1)}  ( \xvec,t )\right) \ , 
		\eea
 where the notation $|_N$ means expansion up to and including $N$-th order in fields, the $\mathbb{C}^{(n)}_{\mathcal{O}_m,  i } $ are all $n$-th order in fields, {and the index $i$ labels different descendants of $\mathcal{O}_m$ at order $n$}.  The Taylor expansion in the fluid element up to fourth order for general operators $\mathcal{O}$ is derived in \appref{fluidelementapp}.   Furthermore, given that in the above equation, the full $t'$ dependence is isolated in powers of the growth factor, we can symbolically do the time integrals in  \eqn{bias_raw}, allowing us to define 
		\bea\label{eq:param}
		c_{\mathcal{O}_m , \alpha-(m-1)}(t) \equiv \int^t dt' H(t') c_{\mathcal{O}_m}(t,t')\left( \frac{D(t' )}{D(t)} \right)^{\alpha} \ ,
		\eea	
		and therefore the sum \eqn{bias_raw} becomes 
		\bea\label{fin}
		\delta_{h}(\vx,t) \Big|_N = \sum_{\mathcal{O}_m}  \sum_{\alpha=m}^{N}c_{\mathcal{O}_m , \alpha-(m-1)}(t)\left(\sum_{n=\alpha}^{N}\mathbb{C}^{(n)}_{\mathcal{O}_m , \alpha-(m-1)}  ( \xvec,t )\right) \ , 
		\eea	
up to order $N$.

		Doing this for all operators allowed by the equivalence principle (\eqn{fullexp}), we obtain \eqn{schematic}. However this expansion is overcomplete, as the $\mathbb{C}^{(n)}_{\mathcal{O},i}$ are not linearly independent. A full list of degeneracies is given in \eqn{eq:degen}.  We then obtain the final bias expansion up to fourth order 
\begin{align}
\begin{split} \label{finalbias}
		\delta_{h}(\vx, t ) = & b_1 \left(\mathbb{C}_{\delta,1}^{(1)}(\vx , t )+\mathbb{C}_{\delta,1}^{(2)}(\vx , t )+\mathbb{C}_{\delta,1}^{(3)}(\vx , t )+\mathbb{C}_{\delta,1}^{(4)}(\vx , t )\right)  \\
		& +  b_2 \left(\mathbb{C}_{\delta,2}^{(2)}(\vx , t )+\mathbb{C}_{\delta,2}^{(3)}(\vx , t )+\mathbb{C}_{\delta,2}^{(4)}(\vx , t )\right)  +  b_3  \left(\mathbb{C}_{\delta,3}^{(3)}(\vx , t )+\mathbb{C}_{\delta,3}^{(4)}(\vx , t )\right) \\
		& + b_4 \;\mathbb{C}_{\delta,4}^{(4)}(\vx , t )  +  b_5  \left(\mathbb{C}_{\delta^2,1}^{(2)}(\vx , t )+\mathbb{C}_{\delta^2,1}^{(3)}(\vx , t )+\mathbb{C}_{\delta^2,1}^{(4)}(\vx , t )\right)\\ 
		& + b_6 \left(\mathbb{C}_{\delta^2,2}^{(3)}(\vx , t )+\mathbb{C}_{\delta^2,2}^{(4)}(\vx , t )\right) +  b_7   \,  \mathbb{C}_{\delta^2,3}^{(4)}(\vx , t ) + b_8 \left(\mathbb{C}_{r^2,2}^{(3)}(\vx , t )+\mathbb{C}_{r^2,2}^{(4)}(\vx , t )\right)\\
		& + b_9 \, \mathbb{C}_{r^2,3}^{(4)}(\vx , t ) + b_{10}  \left(\mathbb{C}_{\delta^3,1}^{(3)}(\vx , t )+\mathbb{C}_{\delta^3,1}^{(4)}(\vx , t )\right)+ b_{11} \, \mathbb{C}_{r^3,2}^{(4)}(\vx , t ) \\ 
		&+ b_{12} \,  \mathbb{C}_{\delta^3,2}^{(4)}(\vx , t )+ b_{13} \,  \mathbb{C}_{r^2\delta,2}^{(4)}(\vx , t )+ b_{14} \, \mathbb{C}_{\delta^4,1}^{(4)}(\vx , t )+ b_{15}\,  \mathbb{C}_{\delta r^3,1}^{(4)}(\vx , t )  \ ,
\end{split}
\end{align}
which through the procedure defined above, is the minimal set of linearly independent functions for the bias expansion up to fourth order.   This parameterization is the same as the one used in  the data analysis of \cite{DAmico:2022osl}.  The operators up to third order are the same as in the basis of \cite{Angulo:2015eqa,Fujita:2016dne}, except that we use $\mathbb{C}_{r^2,2}^{(3)} = \mathbb{C}_{s^2,2}^{(3)}+\sfrac{1}{3}\mathbb{C}_{\delta^2,2}^{(3)}$, as it was easier to generalize. The new fourth order $\mathbb{C}^{(4)}_{\mathcal{O},i}$ are explicitly given in \appref{explicit}, and expressions for $\mathbb{C}^{(n)}_{\mathcal{O},i}$ for $n\leq 3$  can be found in {\cite{Angulo:2015eqa,Fujita:2016dne}}.  {In this way, our expression \eqn{finalbias} extends the so-called basis of descendants to fourth order.}  The dark-matter kernels are obtained by setting $b_1 = b_2 = b_3 = b_4 = 1$, with all of the other bias parameters equal to zero.\footnote{{The formalism to construct a complete basis of biases was originally constructed in \cite{Senatore:2014eva}.  Subsequently, other procedures, which are expected to be equivalent, were developed.  Following these procedures, at fourth order in real space, bases were given in {\cite{Desjacques:2016bnm,Eggemeier:2018qae}}, and in \appref{degenandlocalapp} we provide the explicit transformation between our basis and the one used in \cite{Desjacques:2016bnm}.}}

{The procedure described above ensures that we include all possible operators, including those related to non-locality in time.  However, it is interesting to compare the basis of functions in \eqn{finalbias} to what one would get assuming a strictly local-in-time expansion, which in \eqn{bias_raw} corresponds to setting {$c_{\mathcal{O}}(t,t')=c_{\mathcal{O}}(t)\delta_D(t-t') / H(t)$} and not including any convective derivatives in the set of operators $\mathcal{O}$ included.  {As we show explicitly in \appref{degenandlocalapp},} up to fourth order, the two expansions are mathematically equivalent.}\footnote{{We disagree with some references, including \cite{Mirbabayi:2014zca,Desjacques:2016bnm}, which have claimed that the local-in-time expansion in terms of $\partial_i \partial_j \Phi$ and $\partial_i v^j$ is not sufficient as a fourth-order basis.  For example, they mention a term that they write as $\text{tr} ( \Pi^{[1]} \Pi^{[3]} ) $ which they claim cannot be written in the local-in-time basis.  {However, in \appref{degenandlocalapp}, we explicitly show that this term can be written in terms of the basis of descendants that we use in this work.  In \appref{degenandlocalapp}, we also show that, at fourth order, the basis of descendants is equivalent to the local-in-time basis, which means that $\text{tr} ( \Pi^{[1]} \Pi^{[3]} ) $ can be written in the local-in-time basis (which by definition is also local in space).}}}

In terms of the bias parameters in \eqn{finalbias}, we have the following dependencies of perturbative contributions
\begin{align}
\begin{split}
& P_{11}^{r,h} [ b_1] \ , \quad  P_{13}^{r,h}   [ b_1 , b_3 , b_8  ]  \ , \quad P_{22}^{r,h} [  b_1 , b_2 , b_5  ]   \ , \quad B_{321}^{r,h,(I)} [ b_1 , b_2 , b_3, b_5 , b_6, b_8, b_{10}  ]  \ , \\
& B_{211}^{r,h} [  b_1 , b_2 , b_5 ] \ , \quad B_{321}^{r,h,(II)}  [  b_1 , b_2 , b_3, b_5 , b_8  ]  \ , \quad B_{411}^{r,h} [ b_1 , \dots , b_{11} ] \ , \quad   B^{r,h}_{222}  [ b_1, b_2 , b_5  ] \ ,
\end{split}
\end{align}
where 
\be 
P_{11}^{r,h} ( k , \hat k \cdot \hat z) = (b_1 + f (\hat k \cdot \hat z )^2  )^2  P_{11} ( k )  \ , 
\ee
is the famous Kaiser result from linear theory, for example.  In general, we have the following dependancies of the kernels on the bias parameters
\begin{align}
\begin{split} \label{kernelbiasdep}
& K_{1}^{r,h} [ b_1 ] \ , \quad  K_{2}^{r,h} [ b_1 , b_2 , b_5 ] \ , \quad  K_{3}^{r,h} [ b_1 , b_2 , b_3, b_5 , b_6, b_8, b_{10}  ]  \andd  K_4^{r,h} [ b_1, \dots , b_{15}]  \  . 
\end{split}
\end{align}
Notice that the diagrams $P_{13}^{r,h} $, $B_{321}^{r,h,(II)} $, and $ B_{411}^{r,h} $ depend on less bias parameters than the kernels in \eqn{kernelbiasdep} would suggest.  This happens because, in the particular {momentum} configuration of the kernels that enter the {loops} in \eqn{loopexpressionsrssbias} and \eqn{bispexpressionsrssbias}, {some bias parameters can be removed with bias redefinitions.} Let us explain this in detail.

To the order that we work in this paper, we have
\be \label{deltahaverage}
\langle \delta_{h} ( \xvec ) \rangle \approx   \langle \delta_{h}^{(2)} ( \xvec ) \rangle =   \int_{\qvec} K_2^{h} ( \qvec , - \qvec ) P_{11} ( q ) =   \frac{ - b_1 + b_2 + b_5 }{2 \pi^2} \int d q \, q^2 P_{11} ( q ) \ ,
\ee  
which as mentioned below in \secref{renormalization}, we will explicitly subtract when renormalizing $\delta_h$.  We also note that since number {and momentum} are not conserved for tracers, the loops $P_{13}^{r,h}$, $B_{411}^{r,h}$, and $B_{321}^{r,h,(II)}$ start at $k^0$ (as opposed to $k^2$ for dark matter) as $k \rightarrow 0$.  As described in \cite{McDonald:2006mx, McDonald:2009dh}, this is best understood as the renormalization of lower-order bias parameters.  We explicitly find that the redefinitions
\be
b_1 \rightarrow b_1  +\frac{13 b_1 + 34 b_2 - 47 b_3 + 42 b_5 - 110 b_6 - 82 b_8 - 63 b_{10}}{42 \pi ^2} \int dq \, q^2 P_{11} ( q )  \ , 
\ee
\begin{align}
\begin{split}
& b_2 \rightarrow  b_2  - \frac{1}{1260 \pi ^2} \big(-469 b_1-96 b_2-1099 b_3+1664 b_4-1260 b_6+3554 b_7-2520 b_8 \\ 
& \hspace{.9in}  +5150 b_9+6489 b_{11}+1890 b_{12}+5691 b_{13}+2205 b_{15} \big) \int dq \, q^2 P_{11} ( q )  \ , 
\end{split}
\end{align}
and
\begin{align}
\begin{split}
& b_5 \rightarrow  b_5 + \frac{1}{8820 \pi ^2} \big( -1001 b_1+3876 b_2+1729 b_3-4604 b_4+5460 b_5+14280 b_6  \\
& \hspace{.9in} -34214 b_7  -7140 b_8-7250 b_9+13230 b_{10}+17451 b_{11}-56070 b_{12}  \\
& \hspace{.9in} -1911 b_{13}-26460   b_{14}+6615 b_{15} \big)  \int dq \, q^2 P_{11} ( q )  \ , 
\end{split}
\end{align}
absorb the $k^0$ UV limits of these loops.  {After these redefinitions, $b_{12}$, $b_{14}$, and $b_{15}$ are eliminated from our observables.  {Additionally, once the $k^0$ pieces are removed from the terms proportional to $b_7$ and $b_{13}$ in particular, these two operators become degenerate, and so $b_{13}$ can also be set to zero.}}

%%%%%%%%%%%%
%
%
\subsection{Renormalization of biased tracers in redshift space }\label{renormalization}

In some ways, renormalization of biased tracers is more straightforward than dark matter because there are no equations of motion to solve, one simply writes all of the terms relevant for the final expressions.  On top of this, we express the contact operators of redshift-space distortions in terms of the long-wavelength fields.  We start directly with the renormalized operators that enter redshift space (the same as \eqn{rhovrenorm} but with $\bar \rho \rightarrow \bar \rho_h$ and $\delta \rightarrow \delta_h$)  
\begin{align}
\label{rhovrenormbias}
& [\delta_h]_R =\delta_h + \mathcal{O}_{\delta_h}  \ ,  \nonumber \\
& [\pi^i_h]_R  =  \rho_h v^i + v^i  \mathcal{O}_{\rho_h} +  \mathcal{O}_{ \pi_h }^i  \ ,  \\
& [\pi^i_h v^j ]_R  =  \rho_h v^i v^j  +  v^i v^j \mathcal{O}_{\rho_h} + v^i \mathcal{O}_{\pi_h }^j + v^j \mathcal{O}_{\pi_h }^i + \mathcal{O}^{ij}_{\pi_h v}   \ ,  \nonumber \\
&[\pi^i_h v^j v^k]_R  =  \rho_h v^i v^j v^k +   v^i v^j v^k \mathcal{O}_{\rho_h}  + ( v^i v^j \mathcal{O}_{\pi_h }^k + \text{2 perms.})   + ( v^i \mathcal{O}_{ \pi_h v }^{jk} + \text{2 perms.} ) + \mathcal{O}^{ijk}_{\pi_h v^2}  \ , \nonumber\\
& [\pi^i_h v^j v^k v^l]_R   =  \rho_h v^i v^j v^k v^l + v^i v^j v^k v^l \mathcal{O}_{\rho_h} +  ( v^i v^j v^k \mathcal{O}_{\pi_h }^l + \text{3 perms.} ) \nonumber\\
& \hspace{1.8in} +(v^i v^j \mathcal{O}_{\pi_h v }^{kl} + \text{5 perms.} ) + ( v^i \mathcal{O}_{\pi_h v^2}^{jkl} + \text{3 perms.} ) + \mathcal{O}_{\pi_h v^3}^{ijkl}  \ ,\nonumber
\end{align}
where again, all of the $\mathcal{O}$ terms are Galilean scalars, and we specifically have $\mathcal{O}_{\rho_h} \equiv \bar \rho_h \mathcal{O}_{\delta_h}${ , which is the higher-derivative halo bias or counterterms.}  {Although $\pi^i_h$ may seem to not be a composite operator, in terms of $v^i$, it is given by the composite expression \eqn{pihalodef}, and so must also be renormalized}.   Note that we choose to include the non-zero mean of $\delta_h$ in the renormalized $[\delta_h]_R$, so that $\mathcal{O}_{\delta_h} \supset - \langle \delta_h \rangle$.   The non-trivial step now is to understand how non-locally-contributing terms containing the differential operator $\partial_i \partial_j \partial_k / \partial^2$ in the renormalization of $\pi^i$  enter the renormalization for biased tracers.  In the dark-matter case, the form is dictated entirely by having a local stress tensor $\tau^{ij}$ and solving the equations of motion.  However, biased tracers do not have an explicit equation of motion from which to derive the form of all of the counterterms.

To proceed, we write the renormalization of the tracer momentum density in terms of the dark-matter momentum density, which we know from \secref{dmrenormrssec} and contains the new non-locally-contributing counterterm.   The term $\mathcal{O}_{\pi_h}^i$ in \eqn{rhovrenormbias} gives the counterterms that we explicitly add to the renormalized tracer density $[\pi_h^i]_R$, so we would like to write that in terms of dark-matter quantities.  By the {equivalence principle}, we know that {\cite{Senatore:2014eva, Mirbabayi:2014zca, Perko:2016puo}}
\be \label{diffvelren}
[ v^i_h  - v^i ]_R = \mathcal{O}^i_{\Delta v} \ , 
\ee
where $\mathcal{O}^i_{\Delta v}$ is a {higher-derivative Galilean scalar built of spatially-local products of $\partial_i \partial_j \Phi$ and $\partial_i v^j$}, {that vanishes when the wavenumber of $[v_h^i - v^i]_R$ goes to zero}.  Now, we recall that the renormalized velocity is given by \cite{Carrasco:2013mua}
\be \label{physveldef}
[v^i]_R = \frac{[\pi^i]_R}{[\rho]_R} + \mathcal{O}_{v,HD}^i  \ , 
\ee
where $\mathcal{O}_{v,HD}^i$ are the standard higher-derivative {Galilean scalar} counterterms used when defining the renormalized velocity of dark matter, which arise because $v^i$ is a contact operator in terms of $\pi^i$ and $\rho$  (for the same reason as described in {\secref{renormdmrsssec}}). Using an analogous expression to \eqn{physveldef} for $[ v^i_h]_R$, and plugging $[ v^i]_R$ and $[ v^i_h]_R$ into \eqn{diffvelren} gives
\be
[ \pi^i_h]_R = \frac{[\rho_h]_R}{[\rho]_R} [\pi^i]_R + [\rho_h]_R \left( \mathcal{O}^i_{\Delta v} - \mathcal{O}_{v_h,HD}^i + \mathcal{O}^i_{v,HD}  \right)  \ ,
\ee
which using \eqn{rhovrenorm} and \eqn{rhovrenormbias} to plug in the expressions for $[\pi^i]_R$ and $[\pi^i_h]_R$ implies
\begin{align}
\begin{split}
(\rho_h + \mathcal{O}_{\rho_h} ) v^i + \mathcal{O}_{\pi_h}^i = \frac{[\rho_h]_R}{[\rho]_R} \left( (\rho + \mathcal{O}_{\rho} ) v^i + \mathcal{O}_{\pi}^i  \right) +  [\rho_h]_R \left( \mathcal{O}^i_{\Delta v} - \mathcal{O}_{v_h,HD}^i + \mathcal{O}^i_{v,HD}  \right)         \ . 
\end{split}
\end{align}
Finally, using $\rho + \mathcal{O}_\rho = [\rho]_R$ and $\rho_h + \mathcal{O}_{\rho_h} = [\rho_h]_R$, we have
\be \label{ophiexpression2}
\mathcal{O}_{\pi_h}^ i = \bar \rho_h ( 1 + \delta_h + \mathcal{O}_{\delta_h}  ) \left( \frac{\mathcal{O}^i_\pi}{\bar \rho ( 1 + \delta + \mathcal{O}_\delta ) } + \mathcal{O}^i_{\Delta v} - \mathcal{O}_{v_h,HD}^i + \mathcal{O}^i_{v,HD} \right)  \ .
\ee
{On the right-hand side of the above expression, $\mathcal{O}_\pi^i$ is already the solution for the dark-matter momentum in terms of the local stress tensor (see \eqn{Opieq} and \eqn{piict2sol1}), and so contains the non-locally-contributing Green's function.  The other terms $\mathcal{O}^i_{\Delta v}$,  $\mathcal{O}_{v_h,HD}^i$, and $\mathcal{O}^i_{v,HD}$, are local functions of the dark-matter fields, so $\mathcal{O}_\pi^i$ is the only term with non-locally-contributing terms already at field level.  Thus, we have}
\be \label{nonlocalctseq}
\frac{\mathcal{O}_{\pi_h}^ i \big|_{\rm NLC}}{\bar \rho_h} = \frac{  1 + \delta_h + \mathcal{O}_{\delta_h}  }{1 + \delta + \mathcal{O}_\delta}   \frac{\mathcal{O}^i_\pi \big|_{\rm NLC} }{\bar \rho} \ ,
\ee
where $\mathcal{O} |_{\rm NLC}$ stands for the non-locally-contributing terms in $\mathcal{O}$.  Thus, we see that, because of the equivalence principle and locality of the renormalization of contact operators, the non-locally-contributing counterterms in $[\pi^i_h]_R$ are {determined} by those in $[\pi^i]_R$ {(and the ratio of renormalized densities $[\delta ]_R$ and $[\delta_h]_R$).}

We now specialize to the order relevant to this paper.  The non-locally-contributing counterterms start at second order in fields in  $\mathcal{O}^i_\pi$, so we have 
\be \label{ctrelation2}
\frac{\mathcal{O}_{\pi_h, (2) }^i \big|_{\rm NLC} }{\bar \rho_h } =    \frac{ \mathcal{O}^i_{\pi , (2)}  \big|_{\rm NLC} }{\bar \rho}  + \dots  \ , 
\ee
where the $\dots$ above stand for higher order terms (note that we have also left off the constant contribution $\langle \delta_h \rangle$ from  $\mathcal{O}_{\delta_h}$ since $\langle \delta_h \rangle$ counts as two powers of $\delta^{(1)}$, as can be seen from \eqn{deltahaverage}).  Then, since it is actually the ratios $\mathcal{O}^i_\pi / \bar \rho$ for dark matter and $\mathcal{O}^i_{\pi_h} / \bar \rho_h$ for tracers that enters the renormalization of the redshift-space overdensity (as can be seen from \eqn{rssrealsp} and \eqn{rsssbias}), this means that the relevant EFT coefficients for dark matter and tracers are actually equal at this order.  {This fact has some intriguing consequences.  First, measurement of this quantity in galaxy clustering data is a direct measurement of the underlying dark-matter properties.  In fact, one can also imagine using dark-matter simulations directly to set expectations for the size of this parameter, since it is unaffected by the bias.  {Any mismatch between the measured value and the value expected from dark matter would point to a violation of the equivalence principle or locality.}  Second, since this EFT coefficient is the same for all tracers, the common value should enter any analysis of multiple tracers or sky patches at the same redshift, thus reducing the number of overall parameters of the theory. Finally, the same parameter will enter any lensing analysis (which only depends on the overall mass distribution, dominated by dark matter), thus again reducing the number of parameters of the theory and providing interesting physical consistency checks.  We leave exploration of these exciting topics to future work.}

\subsection{Biased tracers in redshift space counterterms} \label{uvmatchingsecbias}

We can now write down the most general local-in-space counterterms in a derivative expansion of $\partial_i \partial_j \Phi$ and $\partial_i v^j$, up to second order, that obey the symmetries of the problem, which are rotation and Galilean invariance.  We focus on the first few orders in derivatives, which, because number and momentum of galaxies is not conserved, is $\mathcal{O}(k^2 P_{11})$ for $P_{13}^{{r, h}}$, $\mathcal{O}(k^2 P_{11}^2)$ for $B_{411}^{r,h}$ and $B_{321}^{r,h,(II)}$, $\mathcal{O}(k^0)$ and $\mathcal{O}(k^2)$ for $P^{r,h}_{22}$, $\mathcal{O} (k^0)$ and $\mathcal{O} (k^2)$ for $B_{222}^{r,h}$, and $\mathcal{O}(k^0 P_{11})$ and $\mathcal{O}(k^2 P_{11})$ for $B_{321}^{r,h,(I)}$.  The $\mathcal{O}(k^0 P_{11})$ in $P^{r,h}_{13}$ and $\mathcal{O}(k^0 P_{11}^2)$ in $B_{411}^{r,h}$ and $B_{321}^{r,h,(II)}$ are taken care of by shifts in the bias parameters in \secref{biasm}.  

For the response terms, we include the counterterms
\begin{align}
\begin{split}
\frac{\knl^2 \mathcal{O}_{\rho_h} }{\bar \rho_h}  & = D^3 c^h_{1} \partial^2 \tilde \delta^{(1)}    + D^4 \Bigg( c^h_{1} \partial_i \partial^2 \tilde \delta^{(1)} \frac{\partial_i \tilde \delta^{(1)} }{\partial^2} + c^h_{2} \partial^2 \tilde \delta^{(2)} - c^h_{2}  \partial_i \partial^2 \tilde \delta^{(1)} \frac{\partial_i \tilde \delta^{(1)} }{\partial^2} \\
&  \hspace{.7in} + c^h_{3} \partial^2 ( \tilde \delta^{(1)} \tilde \delta^{(1)} ) + c^h_{4} \partial^2 \left( \frac{\partial_i \partial_j \tilde \delta^{(1)}}{\partial^2}  \frac{\partial_i \partial_j \tilde \delta^{(1)}}{\partial^2}  \right)  + c^h_{5} \partial_i \tilde \delta^{(1)} \partial_i \tilde \delta^{(1)}  \Bigg)  \ , 
\end{split}
\end{align}
\begin{align} \label{opihi}
\frac{\knl^2 \mathcal{O}_{\pi_h}^i }{\bar \rho_h a H f }  & =   D^3 c^{\pi}_1 \partial_i \tilde \delta^{(1)}  + D^4 \Bigg( c^{\pi}_1 \partial_j \partial_i \tilde \delta^{(1)} \frac{\partial_j \tilde \delta^{(1)}}{\partial^2}  + c^{\pi}_2 \partial_i \tilde \delta^{(2)}   - c^{\pi}_2  \partial_j \partial_i \tilde \delta^{(1)} \frac{\partial_j \tilde \delta^{(1)}}{\partial^2}  \\
& + c^{\pi}_3 \partial_i ( \tilde \delta^{(1)} \tilde \delta^{(1)} ) + c^{\pi}_4 \partial_i \left( \frac{\partial_j \partial_k \tilde \delta^{(1)}}{\partial^2}  \frac{\partial_j \partial_k \tilde \delta^{(1)}}{\partial^2}  \right)  + c^{\pi}_5 \frac{\partial_i \partial_j \partial_k}{\partial^2} \left( \frac{\partial_j \partial_l \tilde \delta^{(1)} }{\partial^2 } \frac{\partial_l \partial_k \tilde \delta^{(1)} }{\partial^2 } \right)  \Bigg)  \ ,    \nonumber
\end{align}
and
\begin{align}
\begin{split} 
\frac{\knl^2 \mathcal{O}^{ij}_{\pi_h v} }{\bar \rho_h ( a H f)^2}   & =  D^2 \delta_{ij}   c_0^{\pi v} + D^3 \left( c_1^{\pi v} \frac{\partial_i \partial_j \tilde \delta^{(1)}}{\partial^2} + c_3^{\pi v } \delta_{ij}  \tilde \delta^{(1)} \right)   
+   D^4\Bigg( c_1^{\pi v} \partial_k \frac{\partial_i \partial_j \tilde \delta^{(1)}}{\partial^2} \frac{\partial_k \tilde \delta^{(1)}}{\partial^2}   \\
& + c_3^{\pi v} \delta_{ij} \partial_k \tilde \delta^{(1)} \frac{\partial_k \tilde \delta^{(1)} }{\partial^2 } + c_2^{\pi v } \frac{\partial_i \partial_j \tilde \delta^{(2)} }{\partial^2 }  - c_2^{\pi v }  \partial_k \frac{\partial_i \partial_j \tilde \delta^{(1)}}{\partial^2} \frac{\partial_k \tilde \delta^{(1)}}{\partial^2}  + c_4^{\pi v} \delta_{ij} \tilde \delta^{(2)}   \\
&  - c_4^{\pi v} \delta_{ij} \partial_k \tilde \delta^{(1)} \frac{\partial_k \tilde \delta^{(1)} }{\partial^2 }    + c_5^{\pi v } \tilde \delta^{(1)} \frac{\partial_i \partial_j \tilde \delta^{(1)} }{\partial^2 } + c_6^{\pi v }  \frac{\partial_i \partial_k \tilde \delta^{(1)} }{\partial^2 }  \frac{\partial_k \partial_j \tilde \delta^{(1)} }{\partial^2 }   + c_7^{\pi v } \delta_{ij} \tilde \delta^{(1)} \tilde \delta^{(1)}\Bigg)    \ . 
\end{split}
\end{align}
We cannot write down any terms in $\mathcal{O}^{ijk}_{\pi_h v^2}$ or $\mathcal{O}^{ijkl}_{\pi_h v^3}$ that are Galilean scalars and contribute at the order that we work.  In addition to the terms explicitly written above, we also need to include the terms in \eqn{rhovrenormbias} that are inherited from the Galilean transformation properties.  The ones that contribute to the order that we work are $v^i \mathcal{O}_{\rho_h}$, $v^i \mathcal{O}_{\pi_h }^j + v^j \mathcal{O}_{\pi_h }^i $, $( v^i \mathcal{O}_{ \pi_h v }^{jk} + \text{2 perms.} )$, and $(v^i v^j \mathcal{O}_{\pi_h v }^{kl} + \text{5 perms.} )$.  Also, notice that the last two terms just mentioned here are only present because we allow a constant piece in $\mathcal{O}_{\pi_h v}^{ij}$.

{We also point out the presence of the new non-locally-contributing counterterm proportional to $c_5^\pi$, which is included with a free coefficient following the argument in \secref{renormalization} and is indeed needed to match the UV limit of $B_{411}^{r,h}$, for example.\footnote{{Even though $c_5^\pi$ is determined by the dark-matter value, this is still an unknown number for the sake of galaxy-clustering data analysis, which justifies the choice in \cite{DAmico:2022osl} to treat it as a free parameter.}}   A nice check of \eqn{ctrelation2} is to consider the UV matching with SPT loops, and compare the matching of coefficients for the non-locally-contributing terms in $\mathcal{O}_{\pi_h}^i / \bar \rho_h$ for biased tracers and $\mathcal{O}_{\pi}^i / \bar \rho$ for dark matter.  For biased tracers this comes from $c_5^\pi$, and for dark matter this comes from $c_1$, $c_2$, $c_3$, $c_5$, and $c_6$, as mentioned under \eqn{newtermspiexp}.  Specifically, decomposing the dark-matter terms in \eqn{tauij2} into the biased tracer basis in \eqn{opihi}, we find that  $\mathcal{O}^i_{\pi,(2)} |_{\rm NLC} / \bar \rho = \mathcal{O}^i_{\pi_h,(2)} |_{\rm NLC} / \bar \rho_h$ implies $ c_5^\pi = (2 / 99) ( 2 c_1 - c_2 +c_3 -c_5 -c_6)$, which is indeed true for the UV matching that we found in \eqn{c5pimatching} and \eqn{dmuvmatchingresponse}.  }

For the stochastic terms, we include the counterterms
\begin{align}
\begin{split}
\frac{\knl^2 \mathcal{O}_{\rho_h}  }{\bar \rho_h}  = D^2 \epsilon_1 + D^3 \Bigg( \partial_i \epsilon_1 \frac{\partial_i \tilde \delta^{(1)}}{\partial^2 }  + \epsilon_3^{ij} \frac{\partial_i \partial_j \tilde \delta^{(1)}}{\partial^2 } +  \epsilon_4^{ijk} \frac{\partial_i \partial_j  \partial_k \tilde \delta^{(1)}}{\partial^2 } +  \epsilon_5^{ijkl} \frac{\partial_i \partial_j \partial_k \partial_l \tilde \delta^{(1)}}{\partial^2 }   \Bigg)  \ , 
\end{split}
\end{align}
\begin{align}
\begin{split} \label{opihistoch}
\frac{ \knl^2 \mathcal{O}_{\pi_h}^i }{\bar \rho_h a H f }  & =  D^2 \epsilon_6^i + D^3 \Bigg( \partial_j \epsilon_6^i \frac{\partial_j \tilde \delta^{(1)}}{\partial^2} + \epsilon_8^{ijk}  \frac{\partial_j \partial_k  \tilde \delta^{(1)}}{\partial^2 }  + \epsilon_9^{ijkl}  \frac{\partial_j \partial_k \partial_l \tilde \delta^{(1)}}{\partial^2 }   + \frac{\partial_i \partial_j \partial_k}{\partial^2 } \left(  \epsilon_{13}^{jl} \frac{\partial_l \partial_k \tilde \delta^{(1)}}{\partial^2}  \right)  \Bigg)     \ , 
\end{split}
\end{align}
and
\begin{align}
\begin{split}
\frac{\knl^2 \mathcal{O}_{\pi_h v}^{ij} }{\bar \rho_h ( a H f)^2}  = D^2 \epsilon_{10}^{ij} + D^3 \Bigg( \partial_k \epsilon_{10}^{ij} \frac{\partial_k \tilde \delta^{(1)}}{\partial^2} + \epsilon_{12}^{ijkl} \frac{\partial_k \partial_l \tilde \delta^{(1)}}{\partial^2}   \Bigg)    \ .
\end{split}
\end{align}
 In addition to the terms explicitly written above, we also need to include the terms in \eqn{rhovrenormbias} that are inherited from the Galilean transformation properties.  The ones that contribute to the order that we work are $v^i \mathcal{O}_{\rho_h}$, $v^i \mathcal{O}_{\pi_h }^j + v^j \mathcal{O}_{\pi_h }^i $, $( v^i \mathcal{O}_{ \pi_h v }^{jk} + \text{2 perms.} )$.  The correlations of the stochastic fields $\epsilon^{ij\dots}_n$ are computed analogously to those in \secref{explicitstresstensor}, but expanded to the appropriate order in $k$ to match the terms that we are renormalizing.  Again, we point out the presence of the new non-locally-contributing counterterm containing $\epsilon_{13}^{jl}$ which has the form discussed in \secref{renormalization} and is indeed needed to match the UV of $B_{321}^{r,h,(I)}$.   Note also that above we have included the flow terms so that we have the correct IR limit \eqn{galscalarirlim} for Galilean scalars.  These were obtained in the same way as in \secref{explicitstresstensor}, which is a generalization of \cite{Senatore:2014eva} (with the clarification in \appref{stochflowtermsec} for the stochastic terms) to different tensor structures.   {The above list of counterterms is the full set of independent terms with given tensor structures (at the order in fields and derivatives that we consider) that are Galilean scalars so that the theory satisfies the equivalence principle, and therefore it is the minimal complete set one can consider.

Putting this all together, we reach the form of the counterterms used in \cite{DAmico:2022osl}.  Quantities and contributions to observables are defined in \appref{ctbtrssapp}.  Explicitly, the above leads to the following forms of the kernels and contractions.  Starting with the response terms, we have
\be \label{k1rhctexpression}
K_1^{r,h,ct} ( \kvec ; \hat z)  = \frac{k^2}{\knl^2} \left(  - c^h_{1} + \left( c^{\pi}_{1} -\half c_3^{\pi v} f \right) f (\hat k \cdot \hat z)^2  - \half c_1^{\pi v } f^2 (\hat k \cdot \hat z)^4  \right) \ .
\ee
We expand $K_2^{r,h,ct}$ in the following way  
\begin{align}
\begin{split} \label{k2rhct}
K_2^{r,h,ct} ( \kvec_1 , \kvec_2 ; \hat z ) = \sum_{i=1}^{14} \alpha_i   \, e^{K_2}_i ( \kvec_1 , \kvec_2 ; \hat z)  \ , 
\end{split}
\end{align}
with
\be
\alpha_i = \{  c_1^{h} , c_2^{h} ,c_3^{h} ,c_4^{h} ,c_5^{h} , c_1^{\pi} , c_5^{\pi} , c_1^{\pi v } , c_2^{\pi v } , c_3^{\pi v } , c_4^{\pi v } , c_5^{\pi v } , c_6^{\pi v } , c_7^{\pi v } \} \ ,
\ee
and the basis elements $e_i^{K_2}$ and UV matching are given in \appref{responsematchapp}, {where one can see that the new non-locally-contributing counterterm enters through the term $c_5^\pi e_7^{K_2}$ in $K_2^{r,h,ct}$.} We have found that the EFT parameters $\{ c_2^{\pi}, c_3^{\pi}, c_4^{\pi}, c_0^{\pi v} \}$ are degenerate with other EFT parameters included above for the expressions for $K_1^{r,h,ct}$ and $K_2^{r,h,ct}$; this of course may not be true when including higher order kernels.

For the stochastic terms, we have  (where $\bar n$ is the tracer number density) 
\be
P_{22}^{r,h,\epsilon} ( k , \hat k \cdot \hat z  ) = \frac{1}{\bar n} \left( c_1^{\rm St}  +c^{\rm St}_2  \frac{k^2}{\knl^2}  + c^{\rm St}_3 \frac{k^2}{\knl^2} f (\hat k \cdot \hat z)^2    \right)  \ , 
\ee
and
\be \label{b321epdef1}
B_{321}^{r,h, (I), \epsilon}  = \bar B_{321}^{r,h, (I), \epsilon} ( \vec k_1 , \vec k_2 , \vec k_3 ; \hat z) +\bar B_{321}^{r,h, (I), \epsilon} ( \vec k_3 , \vec k_1 , \vec k_2 ; \hat z )  + \bar B_{321}^{r,h, (I), \epsilon} ( \vec k_2 , \vec k_3 , \vec k_1 ; \hat z)   \ ,
\ee
where we have defined
\begin{align}
\begin{split} \label{b321epdef2}
& \bar B_{321}^{r,h, (I), \epsilon} (\vec  k_1 ,\vec k_2 ,  \vec k_3 ; \hat z)    =   \langle  \tilde \delta_{r,h}^{(1)} ( \kvec_1 , \hat z  )  \tilde \delta^{(1)}_{r,h,\epsilon} ( \kvec_2 ,\hat z ) \tilde \delta^{(2)}_{r,h,\epsilon} ( \kvec_3  , \hat z ) \rangle'  + \langle  \tilde \delta_{r,h}^{(1)} ( \kvec_1 , \hat z  )  \tilde \delta^{(1)}_{r,h,\epsilon} ( \kvec_3  , \hat z )  \tilde \delta^{(2)}_{r,h,\epsilon} ( \kvec_2 , \hat z ) \rangle'    \ .
\end{split}
\end{align}
Then, we have
\be \label{b321barexpand}
\bar B_{321}^{r,h, (I), \epsilon} ( \vec k_1 ,  \vec k_2 ,  \vec k_3 ;  \hat z ) = \frac{ b_1 + f (\hat k_1 \cdot \hat z)^2}{\bar n} P_{11}(k_1) \sum_{i=1}^{13}   c^{\rm St}_i  e_i^{\rm St} (\vec k_1 ,   \vec k_2 , \vec  k_3 ; \hat z)  \ , 
\ee
with $e_3^{\rm St} = 0$; the basis elements $e_i^{\rm St}$ and UV matching is given in \appref{stochmatchapp}. {In particular, the new non-locally-contributing counterterm enters through $e_{13}^{\rm St}$.}\footnote{Note that \eqn{ctrelation2} also in principle implies a relationship between the non-locally-contributing stochastic terms of dark matter (in \eqn{piict2sol1ep}) and tracers ($\epsilon_{13}^{ij}$ in \eqn{opihistoch}).  However, since the dark-matter non-linear term is contracted with other dark-matter stochastic terms and the tracer non-linear term is contracted with other tracer stochastic terms, the resulting EFT coefficients after contractions will in general be different.  Non-trivial relationships may result at higher orders in perturbation theory, for example when the non-locally-contributing terms contract with themselves.}  Recall that the renormalization of $B_{222}^{r,h}$ involves three-point functions of the stochastic fields, which {can in general be independent} from the two-point functions {(but they can be related after assuming a Poisson distribution, which is a reasonable assumption)}.  We find 
\be \label{b222hstoch}
B_{222}^{r,h,\epsilon}  =\frac{1}{\bar n^2}  \left( c_1^{(222)} + \frac{1}{\knl^2} \left( c^{(222)}_2 (k_1^2 + k_2^2 + k_3^2 )   + c_5^{(222)} \hat z^i \hat z^j \left(k_1^i k_2^j + k_1^i k_3^j + k_2^i k_3^j  \right) \right) \right) \ ,
\ee
which indeed is the most general function up to $\mathcal{O}(k^2)$, symmetric in $\{\kvec_1 , \kvec_2 , \kvec_3\}$, that can be made out of contractions of these vectors, when momentum conservation $\kvec_1 + \kvec_2 + \kvec_3 = 0$ is considered.

For convenience, we quote here the dependencies of the counterterm quantities on biases and EFT parameters
\begin{align}
\begin{split}
& P_{13}^{r,h,ct} [ b_1 , c_1^{h} , c_1^{\pi} , c_1^{\pi v} , c_3^{\pi v } ] \ ,   \quad P_{22}^{r,h,\epsilon} [ c^{\rm St}_1 , c^{\rm St}_2 , c^{\rm St}_3]  \ , \\
& B_{321}^{r,h,(II),ct}  [ b_1 , b_2 , b_5 , c_1^{h} , c_1^{\pi } , c_1^{\pi v } , c_3^{\pi v } ]  \ , \quad B_{321}^{r,h,{(I), \epsilon}} [b_1  , c^{\rm St}_1, c^{\rm St}_2 , \{ c^{\rm St}_i \}_{i = 4 , \dots , 13} ]   \ ,  \\
& B_{411}^{r,h,ct} [b_1 , \{c_i^{h} \}_{i=1,\dots,5}  , c_1^{\pi  } , c_5^{\pi } , \{ c_j^{\pi v } \}_{j = 1, \dots , 7}   ]  \ , \quad  B_{222}^{r,h,\epsilon}[ c^{(222)}_1, c^{(222)}_2 , c^{(222)}_5 ] \ . 
\end{split}
\end{align}
Thus, overall, to the order that we work in this paper, we have 11 independent bias parameters, 14 independent response counterterms, and 16 independent stochastic counterterms.  For a conversion from the parameters used in this paper to those used in \texttt{PyBird}, see \appref{pybirdapp}.

\vspace{.4in}
\noindent{\bf Note Added:}    {While {the preprint of} this paper was being prepared {(which explains the theory model already used in \cite{DAmico:2022osl,DAmico:2022gki})}, {the preprint of} \cite{Philcox:2022frc} appeared which also computed the one-loop bispectrum for tracers in redshift space.  It seems to us that there are a few crucial differences in the theory model, though.  First, it does not appear that \cite{Philcox:2022frc} takes into account the non-locally-contributing counterterms.  For the response terms, for us, this shows up through $e_7^{K_2}$ as a contribution to $K_2^{r,h,ct}$ that has a dependence on $\kvec_3$ going like $k_3^0 \mu_3^2$, whereas \cite{Philcox:2022frc} seems to only have $k_3^2 \mu_3^4$ in their $Z^{\rm ctr}_2$ (which we also have from $e_9^{K_2}$).  For the stochastic terms, for us, this shows up in $\bar B_{321}^{r,h,(I),\epsilon}$ through $e_{13}^{\rm St}$ as a dependence going like $k_3^0 \mu_3^2$, whereas \cite{Philcox:2022frc} does not seem to have a term with this dependence in their $B_{\rm stoch}^{k^2P\bar n^{-1}}$.  Second, it does not seem to us that the renormalization of contact operators for redshift space in \cite{Philcox:2022frc} preserves the relevant Galilean transformation properties.  For example, assuming that the $\mathcal{O}$ operators in \cite{Philcox:2022frc} do not contain explicit factors of the velocity (which is what is said to be assumed there), this should imply that their $\mathcal{O}^{(1),i}_{u^2} = \mathcal{O}^i_u$.  However, it seems to us that their Eq. (C.4) implies a non-zero $\mathcal{O}^i_u$, but their Eq. (C.5) has $\mathcal{O}^{(1),i}_{u^2} = 0$. {We remind the reader that the terms mentioned here are non-trivially necessary and sufficient to renormalize the UV regions of the loops.} Upon comparison to simulations in \cite{Philcox:2022frc}, the model of \cite{Philcox:2022frc} seems {to have a tolerable theoretical error up to} $k_{\rm max} = 0.2 \unitsk$ for the BOSS survey, while \cite{DAmico:2022osl} (which uses the model presented in this paper) has {a tolerable} theoretical error only up to $k_{\rm max} = 0.23 \unitsk$ {(as found in \cite{DAmico:2022osl} by assessing the error from the truncation of the next-to-next-leading order term not included in our analysis, with further checks upon comparison to simulations).} 
{Since the relevance of the counterterms is larger at higher $k$'s, we speculate that the difference in the $k_{\rm reach}$ is a manifestation of the importance and advantage of using the fully correct physical model. However a detailed understanding of this point is left for future work.}

%%%%%%%%%%%%%%%
%
%
%
%
\section*{Acknowledgements}

{M.L. thanks N. Pavao, A. Seifi, and S. Zekio\u{g}lu for helpful Mathematica advice.} Y.D. acknowledges support from the STFC.  M.L. acknowledges the Northwestern University Amplitudes and Insight group, Department of Physics and Astronomy, and Weinberg College, and is also supported by the DOE under contract DE-SC0021485 and the Alfred P. Sloan Foundation.

%%%%%%%%%%%%%%%%%%%
%
%
%
%

\appendix

%%%%%%%%%%%%%%
%
%
\section{Details for dark matter} \label{rsdmapp}

\subsection{EdS Green's function} \label{edsgreensfunctionsec}

In this section we give some details about the growing-mode EdS Green's function for the overdensity.  The equation of motion for $\delta$  can be written as
\be \label{lineareom2}
a^2 \delta'' ( \kvec , a )  + \left( 2 + \frac{a \cH'}{\cH} \right) a \delta' (\kvec,a)- \frac{3}{2} \om (a) \delta (\kvec , a ) = S_\delta ( \kvec , a ) \  ,
\ee
where $S_{\delta} ( \kvec , a )$ is the non-linear source term.  For the $n$-th order perturbation, we take $\delta^{(n)} ( \kvec , a )  = D(a)^n \tilde \delta^{(n)} ( \kvec ) $, where $D(a)$ is the linear growth factor, which solves \eqn{growthfactor}.  Plugging this $\delta^{(n)} ( \kvec , a )$ into \eqn{lineareom2} and using \eqn{growthfactor} to replace $D''(a)$, we obtain
\be \label{perteds}
\half ( n-1 ) D(a)^{n}  \left( 3 \om ( a ) + 2  n  \frac{ a^2 D'(a)^2}{D(a)^2}  \right) \tilde \delta^{(n)} ( \kvec ) =   S^{(n)}_{\delta} ( \kvec , a ) \ .
\ee
Now we use that in SPT {we have two different time dependences for the source terms, which we write as $\tilde S^{(n)}_{\delta,1 } ( \kvec)$ and $\tilde S^{(n)}_{\delta,2 } ( \kvec)$, giving }
\be \label{sourcetimedep}
S^{(n)}_\delta ( \kvec , a ) = D(a)^n \left(  \om ( a) \tilde S^{(n)}_{\delta,1 } ( \kvec) + \left(\frac{a D'(a)}{D(a)} \right)^2 \tilde S^{(n)}_{\delta,2 } ( \kvec) \right) \ ,
\ee
which can be seen from \eqn{fulldiffeqdelta}.   We see that all terms in \eqn{perteds} and \eqn{sourcetimedep} have the same time dependence, {proportional to $D^n \om$}, if we assume the standard EdS condition \eqn{edscondition}, after which we obtain
\be \label{edsgffactor}
\tilde \delta^{(n)} ( \kvec )  =\frac {2}{(n-1)(3+2n)} \left( S^{(n)}_{\delta,1 } ( \kvec)+S^{(n)}_{\delta,2 } ( \kvec) \right)  \ . 
\ee
The numerical factor $2 / (n-1)(3 + 2n)$ comes from the EdS approximation of the Green's function.  

We can see how this relates to the actual Green's function of the linear equations in the EdS universe, where $\om = 1$ and $a \cH' / \cH = - 1/2$.  In that case, the linear equation of motion is
\be
a^2 \delta'' + \frac{3}{2} a \delta '  - \frac{3}{2} \delta = 0 \ ,
\ee
which has two solutions, $\delta ( a ) = a$ and $\delta ( a ) = a^{-3/2}$.  These can be combined to form the Green's function $G(a,a_1)$ satisfying the boundary conditions $G(a_1 , a_1) = 0$ and $\partial_a G(a , a_1) |_{a = a_1} = a_1^{-2}$
\be
G( a , a_1 ) = \frac{2}{5\, a_1} \left( \frac{a}{ a_1} - \frac{a^{-3/2}}{a_1^{-3/2}} \right)  \theta_H ( a - a_1) \ , 
\ee
where $\theta_H$ is the Heaviside step function. The connection with \eqn{edsgffactor} is that applying the above to a source term $\propto a^n$, we have
\be
\int^a_0 d a_1 G(a , a_1 ) \, a_1^n = \frac{2 \, a^n}{(n-1)(3+2n)}  \ ,
\ee
which gives exactly the same factor that we found.

%%%%%%%%%%%%%%
%
\subsection{Counterterm expressions in real space} \label{dmrealspctkernelssec}

The response terms are proportional only to powers of the linear field, and specifically we can write
\begin{align}
\begin{split}
 \tilde \delta^{(1)}_{ct}(\kvec ) =   F_1^{ct} ( \kvec ) \tilde \delta^{(1)}_{\kvec }  \andd \tilde \delta^{(2)}_{ct} ( \kvec  ) =  \int_{\qvec_1 , \qvec_2}^{\kvec} F_2^{ct} ( \qvec_1 , \qvec_2 ) \tilde \delta^{(1)}_{ \qvec_1 } \tilde  \delta^{(1)}_{\qvec_2 } \ ,
\end{split}
\end{align}
where definitions of the tilde fields are given in \eqn{cttimedep}.  We group all of the terms that are only proportional to the stochastic fields in $\delta_\epsilon^{(1)} ( \kvec,a) $ and terms that contain one stochastic field and one long-wavelength field (semi-stochastic) in $\delta_\epsilon^{(2)} ( \kvec,a)$. 

The response counterterms enter in
\begin{align}
\begin{split}
& P_{13}^{ct} ( k )  \equiv  2 F_1^{ct} ( \kvec ) P_{11} ( k ) \ , \\
& B_{411}^{ct} ( k_1 , k_2 , k_3 ) \equiv 2 P_{11} ( k_1 ) P_{11} ( k_2 ) F_2^{ct} ( - \kvec_1 , - \kvec_2 ) + \text{ 2 perms.} \ , \\
& B_{321}^{(II),ct} ( k_1 , k_2 , k_3 ) \equiv  2 P_{11} ( k_1 ) P_{11} ( k_2 ) F_{1}^{ct} ( \kvec_1 ) F_2 ( - \kvec_1 , - \kvec_2 )  + \text{ 5 perms.} \ ,
\end{split}
\end{align}
so that the combinations
\begin{align}
\begin{split}
P_{13} ( k ) & +P_{13}^{ct} ( k )  \ , \\
B_{411} ( k_1 , k_2 , k_3 ) & + B_{411}^{ct} ( k_1 , k_2 , k_3 ) \ , \\
B_{321}^{(II)} ( k_1 , k_2 , k_3 ) &+ B_{321}^{(II),ct} ( k_1 , k_2 , k_3 ) \ , 
\end{split}
\end{align}
are renormalized.  Furthermore, for the stochastic fields, we define
\begin{align}
\begin{split}\label{dmstochrenormeqs}
& \langle \tilde \delta_\epsilon^{(1)} ( \kvec  ) \tilde  \delta^{(1)}_{\epsilon} ( \kvec '  ) \rangle \equiv  ( 2 \pi)^3 \delta_D ( \kvec + \kvec' ) P_{22}^\epsilon ( k )  \ , \\
& \langle \tilde \delta_\epsilon^{(1)} ( \kvec_1  ) \tilde  \delta^{(1)}_{\epsilon} ( \kvec_2  ) \tilde  \delta^{(1)}_{\epsilon} ( \kvec_3  )  \rangle \equiv  ( 2 \pi)^3 \delta_D ( \kvec_1 + \kvec_2 + \kvec_3 ) B_{222}^\epsilon ( k_1 , k_2 , k_3)  \ ,  \\
& \langle \tilde  \delta^{(1)} ( \kvec_1  ) \tilde  \delta^{(1)}_\epsilon ( \kvec_2  ) \tilde \delta^{(2)}_\epsilon ( \kvec_3  ) \rangle + \text{5 perms.}    \equiv  ( 2 \pi)^3 \delta_D ( \kvec_1 + \kvec_2 + \kvec_3 ) B_{321}^{ (I) , \epsilon} ( k_1 , k_2 , k_3 )  \ , 
\end{split}
\end{align}
so that  the combinations
\begin{align}
\begin{split} \label{dmstochrenormeqs2}
 P_{22} ( k ) & + P_{22}^\epsilon ( k ) \ ,  \\
 B_{222} ( k_1 , k_2 , k_3 ) & + B_{222}^\epsilon ( k_1 , k_2 , k_3 ) \ ,  \\
 B_{321}^{(I)} ( k_1 , k_2 , k_3 ) & + B_{321}^{ (I), \epsilon} ( k_1 , k_2 , k_3 ) \ , 
\end{split}
\end{align}
are renormalized.

%%%%%%%%%%%%%%
%
%
%
\subsection{Response terms} \label{responsefieldsols}

Here we show some details for the results given in \secref{tauijdmsec} for the response terms.  First we write the response stress tensor $\tau^{ij}_{ct} $ as a sum of first- and second-order terms
\be
\tau^{ij}_{ct}( a ) = \Om ( a )  \left(  \hat \tau^{ij}_{ct, (1)} ( a ) +  \hat \tau^{ij}_{ct, (2)} ( a ) \right)  \ , 
\ee
where, in order to cancel the UV terms coming from the SPT loop expansion, the time dependence must be
\be
\hat \tau^{ij}_{ct,(1)} ( a ) = \frac{a^2 H(a)^2 \bar \rho ( a)  D(a)^3 }{\knl^2} \tilde \tau^{ij}_{ct, (1)} \andd  \hat \tau^{ij}_{ct, (2)} ( a ) = \frac{a^2 H(a)^2 \bar \rho (a) D(a)^4}{\knl^2} \tilde \tau^{ij}_{ct, (2)}  \ ,
\ee
where $\tilde \tau^{ij}_{ct, (1)}$ and $\tilde \tau^{ij}_{ct, (2)}$ are time independent {(the factor of $\om(a)$ can be seen from \appref{edsgreensfunctionsec})}.  We suppress spatial dependence of all fields in this section to remove clutter.

 Being careful to keep track of the {time dependence $D(a)^n$} and the EdS Green's functions (which are simply numerical factors coming from the linear equations and the time dependence $D(a)^n$, see \appref{edsgreensfunctionsec}), we have
\begin{align} 
\begin{split}\label{delta1ct}
\delta^{(1)}_{ct} ( a ) & = \frac{1}{9 a^2 H^2 \bar \rho } \partial_i \partial_j  \hat \tau^{ij}_{ct,(1)} ( a )  \ , \quad \pi_{S,ct}^{(1)} ( a ) = \frac{-f}{3 a H} \partial_i \partial_j \hat  \tau^{ij}_{ct,(1)} ( a )  \ , \\
 \pi^{i }_{V,ct, (1)} ( a ) & = \frac{-2 f }{ 7 a H } \epsilon^{ijk} \partial_j \partial_l \hat \tau^{kl}_{ct,(1)} (a) \andd 
  \pi^i_{ct,(1)} ( a )  =  - \frac{f}{3 aH}  \frac{\partial_i \partial_j \partial_k}{\partial^2} \hat \tau^{jk}_{ct,(1)} ( a )  \ 
  \end{split}
\end{align}
where, as in \secref{tauijdmsec}, we assume that $\partial_i \partial_j \partial_k \tilde \tau^{jk}_{ct,(1)} = \partial^2 \partial_j \tilde  \tau^{ij}_{ct,(1)}$.  {Note that in the above, and in all instances, we solve for $\pi_S$ and $\pi_V^i$ directly through the equations of motion \eqn{greateom}, and then we form $\pi^i$ using the definition \eqn{piidef}.}\footnote{A useful manipulation to remember is that for any vector $V^i$ satisfying $V^i = \epsilon^{ijk} \partial_j \partial_l A_{kl}$, we have
\be
\epsilon^{ijk} \frac{\partial_j}{\partial^2 } V^k = \frac{\partial_i \partial_j \partial_k}{\partial^2} A_{jk} - \partial_j A_{ij} \  . 
\ee  } 

The second-order expressions are 
\begin{align}
\begin{split}
\delta_{ct}^{(2)} ( a ) & = \frac{2}{33a^2 H^2 \bar \rho} \partial_i \partial_j \hat \tau^{ij}_{ct,(2)} ( a )  \\
& + \frac{2}{33 a^2 H^2 \bar \rho f } \partial_i \partial_j \Bigg[  4 \mpl^2 a^{-2} \left( \partial_i \Phi^{(1)} ( a ) \partial_j \Phi^{(1)}_{ct} ( a )  - \half \delta_{ij} \left(  \partial_k \Phi^{(1)}(a) \partial_k \Phi^{(1)}_{ct} ( a ) \right) \right)    \\
& \hspace{1.5in} + \frac{2}{\bar \rho} \pi^i_{(1)} ( a ) \pi^{j}_{ct,(1)} ( a ) \Bigg] \ , \\
\pi^{(2)}_{S,ct} ( a ) & = - 4 a \bar \rho H  f \delta^{(2)}_{ct} ( a )  \ ,  \\
\pi^{i}_{V,ct, (2)} ( a ) & = \frac{-2   }{9 a H f } \epsilon^{ijk} \partial_j \partial_l \Bigg[   2 \mpl^2 a^{-2} \left( \partial_k \Phi^{(1)}(a) \partial_l \Phi^{(1)}_{ct} ( a )  + \partial_k \Phi^{(1)}_{ct}(a) \partial_l \Phi^{(1)} ( a )   \right)  \\
& \hspace{1.1in}  + \frac{1}{\bar \rho}  \left( \pi^k_{(1)} ( a )  \pi^l_{ct,(1)} ( a )  + \pi^k_{ct,(1)} ( a ) \pi^l_{(1)} ( a ) \right) + \tau^{kl}_{ct,(2)} ( a )    \Bigg] \  ,
\end{split}
\end{align}
and some extra terms that we need to plug in to the above are
\be \label{extraexpressions}
\partial_i \Phi^{(1)} ( a ) = \frac{3}{2} \Om a^2 H^2 \frac{\partial_i }{\partial^2 } \delta^{(1)} ( a ) \ , \quad \pi^i_{(1)}(a) = - a H \bar \rho  f \frac{\partial_i }{\partial^2 } \delta^{(1)} ( a ) \ , \quad \partial_i \Phi^{(1)}_{ct} ( a ) = \frac{ \Om }{6 \bar \rho} \frac{\partial_i \partial_j \partial_k }{\partial^2} \hat  \tau^{jk}_{ct,(1)} ( a ) \ . 
\ee

We then {factorize the time dependence by} defining $\tilde \pi_{S,ct}^{(1)}$, $\tilde \pi_{S,ct}^{(2)}$, $\tilde \pi_{V,ct,(1)}^{i}$, and $\tilde \pi_{V,ct,(2)}^{i}$ in the same way as $\tilde \pi_{ct,(1)}^{i}$ and $\tilde \pi_{ct,(2)}^{i}$ in \eqn{cttimedeppi}, which leads to the linear solutions
\begin{align}
\begin{split}  \label{linsol11}
\tilde \pi_{S,ct}^{(1)} = \frac{1}{3 \knl^2} \partial_i \partial_j  \tilde \tau^{ij}_{ct, (1)}  \andd
\tilde  \pi^{i }_{V,ct,(1)}  = 0  \ ,
\end{split}
\end{align}
along with $\tilde \delta^{(1)}_{ct}$ and $\tilde \pi^i_{ct,(1)}$ given in \eqn{linsol1}.  
The second order expressions are $ \tilde \pi^{(2)}_{S,ct} = 4 \tilde \delta^{(2)}_{ct} $ and
\begin{align}
\begin{split} \label{piict2}
\tilde \pi^{i}_{V,ct,(2)} &  = \frac{ 1}{ 9 \knl^2} \epsilon^{ijk} \partial_j \partial_l \left[   \frac{\partial_k \tilde \delta^{(1)}}{\partial^2} \frac{\partial_l \partial_m \partial_n}{\partial^2} \tilde \tau^{mn}_{ct,(1)}  +  \frac{\partial_l \tilde \delta^{(1)}}{\partial^2} \frac{\partial_k \partial_m \partial_n}{\partial^2} \tilde \tau^{mn}_{ct,(1)}  + 2 \tilde \tau^{kl}_{ct,(2)}   \right]  \ , 
\end{split}
\end{align}
with $\tilde \delta^{(2)}_{ct}$ and $\tilde \pi^i_{ct,(2)}$ given in \eqn{delta2ct} and \eqn{piict2sol1} respectively.

%%%%%%%%%%%%%
%
%
\subsection{Stochastic terms}

Here we show some details for the results given in \secref{tauijdmsec} for the stochastic terms.  First we write the stochastic stress tensor $\tau^{ij}_{\epsilon} $ as a sum of first- and second-order terms
\be
\tau^{ij}_{\epsilon}( a ) = \Om ( a )  \left(  \hat \tau^{ij}_{\epsilon, (1)} ( a ) +  \hat \tau^{ij}_{\epsilon, (2)} ( a ) \right)  \ , 
\ee
where, in order to cancel the UV terms coming from the SPT loop expansion the time dependence must be
\be
\hat \tau^{ij}_{\epsilon, (1)} ( a ) = \frac{a^2 H(a)^2 \bar \rho(a)  D(a)^2}{\knl^2} \tilde \tau^{ij}_{\epsilon, (1)} \andd  \hat \tau^{ij}_{\epsilon,(2)} ( a ) = \frac{a^2 H(a)^2 \bar \rho (a) D(a)^3}{\knl^2} \tilde \tau^{ij}_{\epsilon, (2)}  \ ,
\ee
where $\tilde \tau^{ij}_{\epsilon, (1)}$ and $\tilde \tau^{ij}_{\epsilon, (2)}$ are time independent.  We suppress spatial dependence of all fields in this section to remove clutter.

 Being careful to keep track of the {time dependence $D(a)^n$} and the EdS Green's functions (which are simply numerical factors coming from the linear equations and the time dependence $D(a)^n$, see \appref{edsgreensfunctionsec}), we have 
\begin{align}
\begin{split} \label{stoch1sols}
\delta^{(1)}_{\epsilon} ( a ) & = \frac{2}{7 a^2 H^2 \bar \rho} \partial_i \partial_j \hat \tau^{ij}_{\epsilon,(1)} ( a )  \ , \quad \pi_{S,\epsilon}^{(1)} ( a ) = \frac{-4 f}{7 a H} \partial_i \partial_j \hat \tau^{ij}_{\epsilon,(1)} ( a )  \ , \\
 \pi^{i (1)}_{V,\epsilon} ( a ) &= \frac{-2 f  }{ 5 a H } \epsilon^{ijk} \partial_j \partial_l \hat\tau^{kl}_{\epsilon,(1)} (a) \ , \\
  \pi^i_{\epsilon,(1)} ( a ) & =  - \frac{f}{aH} \left[  \frac{4}{7} \frac{\partial_i \partial_j \partial_k}{\partial^2} \hat \tau^{jk}_{\epsilon, (1)} ( a ) - \frac{2}{5} \left( \frac{\partial_i \partial_j \partial_k}{\partial^2} \hat \tau^{jk}_{\epsilon, (1)} ( a ) - \partial_j  \hat \tau^{ij}_{\epsilon,(1)}(a) \right) \right]  \ . 
\end{split}
\end{align}

The second-order expressions are
\begin{align}
\begin{split}
\delta_{\epsilon}^{(2)} ( a ) & = \frac{1}{9a^2 H^2 \bar \rho} \partial_i \partial_j  \hat \tau^{ij}_{\epsilon,(2)} ( a )  \\
& + \frac{1}{9 a^2 H^2 \bar \rho f } \partial_i \partial_j \Bigg[  4 \mpl^2 a^{-2} \left( \partial_i \Phi^{(1)} ( a ) \partial_j \Phi^{(1)}_{\epsilon} ( a )  - \half \delta_{ij} \left(  \partial_k \Phi^{(1)}(a) \partial_k \Phi^{(1)}_{\epsilon} ( a ) \right) \right)    \\
& \hspace{1.5in} + \frac{2}{\bar \rho} \pi^i_{(1)} ( a ) \pi^{j}_{\epsilon,(1)} ( a ) \Bigg] \ , \\
\pi^{(2)}_{S,\epsilon} ( a ) & = - 3 a \bar \rho H  f \delta^{(2)}_{\epsilon} ( a )  \ ,  \\
\pi^{i}_{V,\epsilon, (2)} ( a ) & = \frac{-2}{7 a H f} \epsilon^{ijk} \partial_j \partial_l \Bigg[   2 \mpl^2 a^{-2} \left( \partial_k \Phi^{(1)}(a) \partial_l \Phi^{(1)}_{\epsilon} ( a )  + \partial_k \Phi^{(1)}_{\epsilon}(a) \partial_l \Phi^{(1)} ( a )   \right)  \\
& \hspace{1.1in}  + \frac{1}{\bar \rho}  \left( \pi^k_{(1)} ( a )  \pi^l_{\epsilon,(1)} ( a )  + \pi^k_{\epsilon,(1)} ( a ) \pi^l_{(1)} ( a ) \right) + \tau^{kl}_{\epsilon, (2)} ( a )    \Bigg] \ .
\end{split}
\end{align}
 Some extra terms that we need to plug in to the above are given in \eqn{extraexpressions} and 
\be
\partial_i \Phi^{(1)}_{\epsilon} ( a ) = \frac{3 \Om }{7 \bar \rho} \frac{\partial_i \partial_j \partial_k }{\partial^2} \hat \tau^{jk}_{\epsilon,(1)} ( a ) \ . 
\ee

We then factorize the time dependence by defining $\tilde \pi_{S,\epsilon}^{(1)}$, $\tilde \pi_{S,\epsilon}^{(2)}$, $\tilde \pi_{V,\epsilon,(1)}^{i}$, and $\tilde \pi_{V,\epsilon,(2)}^{i}$ in the same way as $\tilde \pi_{\epsilon,(1)}^{i}$ and $\tilde \pi_{\epsilon,(2)}^{i}$ in \eqn{cttimedeppi}, which leads to the linear solutions
\begin{align}
\begin{split}  \label{linsol1st1}
\tilde \pi_{S,\epsilon}^{(1)}  = \frac{4 }{7 \knl^2} \partial_i \partial_j  \tilde \tau^{ij}_{(1)}  \andd  \tilde \pi^{i }_{V,\epsilon, (1) }  = \frac{2  }{ 5 \knl^2 } \epsilon^{ijk} \partial_j \partial_l \tilde \tau^{kl}_{\epsilon, (1)}  \ , 
\end{split}
\end{align}
along with $\tilde \delta_{\epsilon}^{(1)}$ and $\tilde \pi_{\epsilon,(1)}^i$ given in \eqn{linsol1st}.   The second order are $\tilde \pi^{(2)}_{S , \epsilon} = 3 \tilde \delta^{(2)}_{\epsilon}$ and
\begin{align}
\tilde \pi^{i}_{V,\epsilon,(2)} &  =  \frac{  2}{ 7 \knl^2} \epsilon^{ijk} \partial_j \partial_l \left[   \frac{\partial_k \tilde \delta^{(1)}}{\partial^2} \frac{\partial_l \partial_m \partial_n}{\partial^2} \tilde \tau^{mn}_{\epsilon, (1)}  +  \frac{\partial_l \tilde \delta^{(1)}}{\partial^2} \frac{\partial_k \partial_m \partial_n}{\partial^2} \tilde \tau^{mn}_{\epsilon, (1)}  +  \tilde \tau^{kl}_{\epsilon, (2)}   \right] \\
& -  \frac{4 }{35 \knl^2} \epsilon^{ijk} \partial_j \partial_l \left[  \frac{\partial_k \tilde \delta^{(1)}}{\partial^2} \left( \frac{\partial_l \partial_m \partial_n}{\partial^2} \tilde \tau^{mn}_{\epsilon, (1)} - \partial_m \tilde \tau^{lm}_{\epsilon, (1)}  \right) +  \frac{\partial_l \tilde \delta^{(1)}}{\partial^2} \left( \frac{\partial_k \partial_m \partial_n}{\partial^2} \tilde \tau^{mn}_{\epsilon, (1)} - \partial_m \tilde \tau^{km}_{\epsilon, (1)}  \right)   \right]  \ , \nonumber 
\end{align}
with $\tilde \delta^{(2)}_{\epsilon}$ and $\tilde \pi^i_{\epsilon,(2)}$ given in \eqn{delta2ctep} and \eqn{piict2sol1ep} respectively.

%%%%%%%%%%%%%
%
%
%
\subsection{Stochastic flow terms} \label{stochflowtermsec}

Here we derive the expression given in \eqn{stochstress} for the stochastic counterterms.  The only subtlety is in the way that the flow term enters, and here we sharpen and clarify the expressions in \cite{Senatore:2014eva,Angulo:2015eqa}.  Let us start with two stochastic fields $e_1^{ij} ( \xvec , a)$ and $e_3^{ijkl} ( \xvec , a )$ and write the general non-local in time expression
\begin{align}
\begin{split}
\tau_\epsilon^{ij} ( \xvec , a ) = & \int^a \frac{d a'}{a'} \Bigg( \kappa_1 ( a , a' ) e_1^{ij} ( \xfl ( \xvec , a , a' ) , a' )  \\
& \hspace{1in} + \kappa_3 ( a , a' ) e^{ijkl}_{3} ( \xfl ( \xvec , a , a' ) , a' )  \frac{\partial_k \partial_l \delta ( \xfl ( \xvec  , a , a' ) , a' ) }{\partial^2} \Bigg)  \ ,
\end{split}
\end{align}
for some non-local kernels $\kappa_1$ and $\kappa_3$.  Next, we expand this expression up to second order to get
\begin{align}
\tau_\epsilon^{ij} ( \xvec , a ) \approx  & \int^a \frac{d a'}{a'} \Bigg( \kappa_1 ( a , a' ) \left[ e_{1,(1)}^{ij} ( \xvec , a' )   + \frac{\partial_k \delta^{(1)} ( \xvec , a)}{\partial^2}  \partial_k e^{ij}_{1,(1)} ( \xvec , a' )  \left( 1 - \frac{D(a')}{D(a)} \right) + e_{1,(2)}^{ij} ( \xvec , a' ) \right] \nonumber \\
& \hspace{1in} + \kappa_3 ( a , a' ) e^{ijkl}_{3,(1)} ( \xvec, a' )  \frac{\partial_k \partial_l \delta^{(1)} (  \xvec  , a' ) }{\partial^2} \Bigg)  \ . \label{tauepsexpand}
\end{align}
This expression for $\tau_{\epsilon}^{ij} ( \xvec , a)$ is a Galilean scalar as long as both $e_1^{ij} ( \xvec , a)$ and $e_3^{ijkl} ( \xvec , a )$ are Galilean scalars, which at the order that we work, means
\be \label{secondorderep}
e_{1,(2)}^{ij} ( \xvec , a' )  = \partial_k e_{1,(1)}^{ij} ( \xvec , a' ) \frac{\partial_k \delta^{(1)} ( \xvec , a' )}{\partial^2} + e^{ijkl}_{2,(1)} ( \xvec, a' )  \frac{\partial_k \partial_l \delta^{(1)} (  \xvec  , a' ) }{\partial^2} \ ,
\ee
for some new field $e^{ijkl}_{2,(1)}$.  The first term above is fixed by the assumption that $e_1^{ij}$ is a Galilean scalar, while the second term is the most general second-order term that can be included.  We assume that $e_{1,(1)}^{ij}$ contains all of the purely stochastic terms, so we do not include any of those in the second-order expression.  Plugging \eqn{secondorderep} into \eqn{tauepsexpand},  we get
\begin{align}
\begin{split}
\tau_\epsilon^{ij} ( \xvec , a ) \approx  & \int^a \frac{d a'}{a'} \Bigg( \kappa_1 ( a , a' ) \left[ e_{1,(1)}^{ij} ( \xvec , a' )   + \frac{\partial_k \delta^{(1)} ( \xvec , a)}{\partial^2}  \partial_k e^{ij}_{1,(1)} ( \xvec , a' )  \right]  \\
& \hspace{1in} + \left[ \kappa_1 ( a , a' ) e^{ijkl}_{2,(1)} ( \xvec, a' )+ \kappa_3 ( a , a' ) e^{ijkl}_{3,(1)} ( \xvec, a' )  \right]  \frac{\partial_k \partial_l \delta^{(1)} (  \xvec  , a' ) }{\partial^2} \Bigg)  \ ,\label{tauepsexpand2}
\end{split}
\end{align}
where there is now crucially only one flow term, with $\partial_k \delta^{(1)} / \partial^2$ evaluated at the external time $a$.\footnote{As a piece of intuition, one could take the quadratic bias $\delta^2$ and inspect the resulting flow terms.  One sees that there is only one free coefficient for the flow terms.  The second flow term that would naively come from integration over time has a fixed coefficient that cancels the IR limit of the third order expression of $\delta^2$.  This matches what we find here because one can think of a stochastic term as arising from the limit of $\delta(\xvec)^2$ where each $\delta(\xvec)$ is taken to be as short wavelength as possible.}
  We finally get the form \eqn{stochstress} by setting 
\begin{align}
 \epsilon_1^{ij} ( \xvec ) & = \frac{\knl^2} {\Om (a) \cH(a)^2 \bar \rho ( a ) D(a)^2  }\int^a \frac{d a'}{a'}  \kappa_1 ( a , a' )  e_{1,(1)}^{ij} ( \xvec , a' )    \ , \\ 
  \epsilon_3^{ijkl} ( \xvec ) & =  \frac{\knl^2}{\Om (a) \cH(a)^2 \bar \rho ( a ) D(a)^2 }  \int^a \frac{d a'}{a'} \left[ \kappa_1 ( a , a' ) e^{ijkl}_{2,(1)} ( \xvec, a' )+ \kappa_3 ( a , a' ) e^{ijkl}_{3,(1)} ( \xvec, a' )  \right] \frac{ D(a')}{D(a)}  \ , \nonumber
\end{align}
where above, to match \eqn{stochstress}, we have assumed EdS time dependence.  So we see how the final $\epsilon_3^{ijkl}$ is made up of contributions from the originally included $e_3^{ijkl}$ and the second order piece $e_{2,(1)}^{ijkl}$, both integrated over past times.  Since they are both unknown non-linear stochastic fields, we combine them into the single field $\epsilon_3^{ijkl}$.  As always, these integrals can be formally done to define the final coefficients used in \secref{explicitstresstensor}.

%%%%%%%%%%%%%
%
%
%
\subsection{UV matching in real space} \label{uvmatchingdm}

The UV limits of the loops {that are renormalized by response terms} are
\be
P_{13}( k ) \rightarrow - \frac{61}{630 \pi^2} k^2 P_{11}(k) \int \, dq \, P_{11} ( q )  \ ,
\ee
\begin{align}
\begin{split} \label{b411dmuvlimit}
& B_{411} ( k_1 , k_2 , k_3) \rightarrow  -  \frac{ P_{11}(k_1 ) P_{11} ( k_2 )}{1358280\, \pi^2 k_1^2 k_2^2 }  \Bigg[  12409 k_3^6 +   20085 k_3^4 ( k_1^2 + k_2^2) \\
& \hspace{1.5in} + k_3^2 ( -44518 k_1^4 +76684 k_1^2 k_2^2 -44518 k_2^4  ) \\
& \hspace{1.5in} + 12024 ( k_1^2 - k_2^2)^2 (k_1^2 + k_2^2 )      \Bigg] \int \, dq \, P_{11}(q) + \text{ 2 perms.} \ , 
\end{split}
\end{align}
and
\begin{align}
\begin{split}
B_{321}^{(II)} ( k_1 , k_2 , k_3 ) \rightarrow  P_{11}(k_2) F_2 ( \kvec_1 , \kvec_2) \left( \frac{-61 k_1^2 P_{11}(k_1)}{630 \pi^2}  \int \, dq \, P_{11} ( q ) \right)  + \text{ 5 perms.} \ .
\end{split} 
\end{align}
An explicit solution for the EFT coefficients that absorbs the UV contributions above is  
\begin{align}
\begin{split} \label{dmuvmatchingresponse}
& c_3 =  - \frac{61 \knl^2 }{140 \pi^2 } \int  d q P_{11}( q ) \ , \quad  c_4  = -\frac{12409 \knl^2 }{11760 \pi^2 }  \int \, dq \, P_{11}(q)  \ , \\
& c_5  = -\frac{6997 \knl^2 }{6860 \pi^2 }  \int \, dq \, P_{11}(q) \ , \quad  c_7  = \frac{63149 \knl^2 }{82320 \pi^2 }  \int \, dq \, P_{11}(q)  \ , 
\end{split}
\end{align}
with all of the other coefficients zero (they are degenerate for these observables).

The UV limits of the loops that are renormalized by stochastic terms are
\be
P_{22} ( k ) \rightarrow \frac{9}{196 \pi^2 } k^4 \int d q \frac{P_{11}(q)^2}{q^2} \ ,
\ee
\begin{align}
\begin{split} \label{b321IUV}
B_{321}^{(I)} ( k_1 , k_2 , k_3 ) & \rightarrow \frac{ P_{11}(k_1) }{35280 \pi^2 k_1^2} \Bigg[1060 k_1^6 - 2337 k_1^4 (k_2^2 + k_3^2 ) - 217 (k_2^2 - k_3^2 )^2 (k_2^2 + k_3^2)  \\
& + 2 k_1^2 (747 k_2^4 + 512 k_2^2 k_3^2 + 747 k_3^4 )  \Bigg] \int dq \frac{P_{11}(q)^2}{q^2}  + \text{ 2 cyclic perms.}\ . 
\end{split}
\end{align}
and
\begin{align}
\begin{split} \label{b222UV}
B_{222}(k_1,k_2,k_3) & \rightarrow - \frac{15}{2401 \pi^2} \Bigg[   k_1^6 -  k_1^4 ( k_2^2 + k_3^2 ) +(k_2^2 -k_3^2 )^2 (k_2^2 +k_3^2) \\
&\hspace{1.5in}  - k_1^2 \left( k_2^4 - \frac{k_2^2 k_3^2 }{30} + k_3^4 \right)  \Bigg] \int d q \frac{P_{11}(q)^3}{q^4} \ . 
\end{split}
\end{align}
If we write the final forms of the stochastic counterterms  as 
\be
P_{22}^\epsilon (k) =\frac{ c_{{\rm DM},1}^{\rm St} }{\bar n_{\rm DM}} \frac{k^4}{\knl^4} \ , 
\ee
\begin{align}
\begin{split}
B_{321}^{(I),\epsilon} ( k_1 , k_2 , k_3 ) & = \frac{P_{11}(k_1)}{\bar n_{\rm DM} \, \knl^4} \Bigg[  c_{\rm DM,1}^{\rm St} \frac{-7 k_1^6 -14 (k_2^2 - k_3^2)^2 (k_2^2 + k_3^2) + k_1^2 (21 k_2^4 -2 k_2^2 k_3^2 + 21 k_3^4 )}{ 72 k_1^2} \\
& + c_{\rm DM, 2}^{\rm St} (k_1  -k_2 - k_3)(k_1 - k_2 + k_3 )(k_1 + k_2 - k_3) (k_1 + k_2 + k_3) + c_{\rm DM,3}^{\rm St} k_2^2 k_3^2 \\
& + c_{\rm DM,4}^{\rm St} \frac{-2 k_1^2 (k_2^2 - k_3^2)^2 + ( k_1^4 +(k_2^2 -k_3^2)^2) ( k_2^2 + k_3^2) }{k_1^2} + c_{\rm DM,5}^{\rm St} (k_1^2 - k_2^2 -k_3^2)^2 \\
& - c_{\rm DM,6}^{\rm St}  \frac{( -k_1^2 + k_2^2 + k_3^2) ( k_1^2 - k_2^2 + k_3^2) (k_1^2 + k_2^2 - k_3^2)}{k_1^2} \Bigg]  + \text{ 2 cyclic perms.}  \ , 
\end{split}
\end{align}
and 
\begin{align}
\begin{split}
B_{222}^\epsilon ( k_1 , k_2 , k_3 ) & = \frac{1}{\bar n_{\rm DM}^2 \,\knl^6} \Bigg(   c_{\rm DM,1}^{(222)} k_1^2 k_2^2 k_3^2  + c_{\rm DM,2}^{(222)}(k_1^6 + k_2^6 + k_3^6  - ( k_1^4 k_2^2 + k_1^2 k_2^4 + \text{ 2 perms.} ) )      \Bigg)  \ , 
\end{split}
\end{align}
then an explicit solution to the UV matching is
\begin{align}
\begin{split}
c_{\rm DM,1}^{\rm St} & = - \frac{9\, \bar n_{\rm DM} \, \knl^4}{196 \pi^2 } \int dq \frac{P_{11}(q)^2}{q^2} \ ,  \quad
c_{\rm DM,4}^{\rm St}   = \frac{683\, \bar n_{\rm DM} \, \knl^4}{ 70560 \pi^2} \int dq \frac{P_{11}(q)^2}{q^2} \  , \\
c_{\rm DM,5}^{\rm St}  & = - \frac{389\, \bar n_{\rm DM} \, \knl^4}{17640 \pi^2} \int dq \frac{P_{11}(q)^2}{q^2} \ , \quad
c_{\rm DM,6}^{\rm St}  = - \frac{293 \, \bar n_{\rm DM} \, \knl^4}{23520 \pi^2} \int dq \frac{P_{11}(q)^2}{q^2}  \ , 
\end{split}
\end{align}
and
\be
c_{\rm DM,1}^{(222)} =   \frac{ \bar n_{\rm DM}^2   \knl^6 }{4802 \pi^2} \int dq \frac{P_{11} ( q )^3}{q^4}  \ , \quad  
c_{\rm DM,2}^{(222)} =  \frac{15\, \bar n_{\rm DM}^2  \knl^6 }{2401 \pi^2} \int dq \frac{P_{11} ( q )^3}{q^4} \ ,
\ee
with other coefficients zero (they are degenerate for these observables). 
%%%%%%%%%%%%%%%%
%
%
%

\subsection{Counterterm expressions in redshift space} \label{dmrssctkernelssec}

The response terms are proportional only to powers of the linear field, and specifically we can write
\begin{align}
\begin{split}
 \tilde \delta^{(1)}_{r, ct}(\kvec , \hat z  ) =   F_1^{r, ct} ( \kvec  ; \hat z ) \tilde \delta^{(1)}_{\kvec }  \andd \tilde \delta^{(2)}_{r, ct} ( \kvec  , \hat z  ) =  \int_{\qvec_1 , \qvec_2}^{\kvec} F_2^{r, ct} ( \qvec_1 , \qvec_2 ; \hat z  ) \tilde \delta^{(1)}_{ \qvec_1 } \tilde  \delta^{(1)}_{\qvec_2 } \ .
\end{split}
\end{align}
where definitions of the tilde fields are given in \eqn{cttimedeprss}.  We group all of the terms that are only proportional to the stochastic fields in $\delta_{r,\epsilon}^{(1)} ( \kvec, \hat z , a) $ and terms that contain one stochastic field and one long-wavelength field (semi-stochastic) in $\delta_{r,\epsilon}^{(2)} ( \kvec, \hat z ,  a)$.  

The response counterterms enter in 
\begin{align}
\begin{split}
& P_{13}^{r, ct} ( k , \hat k \cdot \hat z  )  \equiv  2 F_1^r ( \kvec ; \hat z)  F_1^{r, ct} ( -  \kvec ; \hat z ) P_{11} ( k ) \ , \\
& B_{411}^{r, ct}  \equiv 2 P_{11} ( k_1 ) P_{11} ( k_2 ) F_1^{r} ( \kvec_1 ; \hat z)  F_1^r ( \kvec_2 ;\hat  z)  F_2^{r, ct} ( - \kvec_1 , - \kvec_2 ; \hat z) + \text{ 2 perms.} \ , \\
& B_{321}^{r, (II),ct}  \equiv  2 P_{11} ( k_1 ) P_{11} ( k_2 ) F_{1}^{r, ct} ( \kvec_1 ; \hat z )  F_1^r ( \kvec_2 ; \hat z ) F^r_2 ( - \kvec_1 , - \kvec_2 ; \hat z )  + \text{ 5 perms.} \ ,
\end{split}
\end{align}
(we have suppressed the argument $(k_1, k_2, k_3, \hat k_1 \cdot \hat z , \hat k_2 \cdot \hat z )$ of the bispectra terms to remove clutter) so that the combinations
\begin{align}
\begin{split}
P_{13}^r ( k , \hat k \cdot \hat z ) & +P_{13}^{r , ct} ( k , \hat k \cdot \hat z )  \ , \\
B_{411}^r  ( k_1 , k_2 , k_3, \hat k_1 \cdot \hat z , \hat k_2 \cdot \hat z  ) & + B_{411}^{r , ct} ( k_1 , k_2 , k_3 , \hat k_1 \cdot \hat z , \hat k_2 \cdot \hat z ) \ , \\
B_{321}^{r, (II)} ( k_1 , k_2 , k_3 , \hat k_1 \cdot \hat z , \hat k_2 \cdot \hat z  ) &+ B_{321}^{r , (II),ct} ( k_1 , k_2 , k_3 , \hat k_1 \cdot \hat z , \hat k_2 \cdot \hat z   ) \ , 
\end{split}
\end{align}
are renormalized.  Furthermore, for the stochastic fields, we define
\begin{align}
& \langle \tilde \delta_{r,\epsilon}^{(1)} ( \kvec , \hat z  ) \tilde  \delta^{(1)}_{r , \epsilon} ( \kvec '  , \hat z ) \rangle \equiv  ( 2 \pi)^3 \delta_D ( \kvec + \kvec' ) P_{22}^{r,\epsilon} ( k  , \hat k \cdot \hat z)  \ ,  \\
& \langle \tilde \delta_{r, \epsilon}^{(1)} ( \kvec_1 , \hat z  ) \tilde  \delta^{(1)}_{r , \epsilon} ( \kvec_2  , \hat z ) \tilde  \delta^{(1)}_{r, \epsilon} ( \kvec_3 , \hat z  )  \rangle \equiv  ( 2 \pi)^3 \delta_D ( \kvec_1 + \kvec_2 + \kvec_3 ) B_{222}^{r, \epsilon} ( k_1 , k_2 , k_3 , \hat k_1 \cdot \hat z , \hat k_2 \cdot \hat z )  \ ,  \nonumber \\
& \langle \tilde  \delta^{(1)}_r ( \kvec_1  , \hat z ) \tilde  \delta^{(1)}_{r,\epsilon} ( \kvec_2  , \hat z ) \tilde \delta^{(2)}_{r, \epsilon} ( \kvec_3 , \hat z ) \rangle + \text{5 perms.}    \equiv  ( 2 \pi)^3 \delta_D ( \kvec_1 + \kvec_2 + \kvec_3 ) B_{321}^{ r , (I) , \epsilon} ( k_1 , k_2 , k_3 , \hat k_1 \cdot \hat z , \hat k_2 \cdot \hat z  )   \ ,\nonumber 
\end{align}
so that  the combinations
\begin{align}
\begin{split}
 P_{22}^r ( k , \hat k \cdot \hat z ) & + P_{22}^{r,\epsilon } ( k , \hat k \cdot \hat z ) \ ,  \\
 B_{222}^r ( k_1 , k_2 , k_3 , \hat k_1 \cdot \hat z , \hat k_2 \cdot \hat z  ) & + B_{222}^{r,\epsilon} ( k_1 , k_2 , k_3, \hat k_1 \cdot \hat z , \hat k_2 \cdot \hat z   ) \ ,  \\
 B_{321}^{r, (I)} ( k_1 , k_2 , k_3, \hat k_1 \cdot \hat z , \hat k_2 \cdot \hat z   ) & + B_{321}^{r ,  (I), \epsilon} ( k_1 , k_2 , k_3, \hat k_1 \cdot \hat z , \hat k_2 \cdot \hat z   ) \ , 
\end{split}
\end{align}
are renormalized.

The basis elements $e^{F_2}_i$ used to define $F_2^{r,ct}$ in \eqn{f2rct} are related to the functions that we use for biased tracers in \appref{responsematchapp} by 
\begin{align}
e^{F_2}_1&  = \frac{1}{9} e^{K_2}_1 + \frac{8}{99} e^{K_2}_2 - \frac{19}{693} e^{K_2}_3 - \frac{16}{693} e^{K_2}_4 - \frac{1}{33} e^{K_2}_5 +\frac{2}{99} e^{K_2}_7 + \frac{2}{3 f} e^{K_2}_{10}
 + \frac{7}{99 f} e^{K_2}_{11} - \frac{2}{3 } e^{K_2}_{12} +\frac{8}{99 f} e^{K_2}_{14}  \ , \nonumber \\
 e^{F_2}_2 & = \frac{10}{231} e^{K_2}_{3} +\frac{4}{231} e^{K_2}_{4}  + \frac{16}{33 f} e^{K_2}_{11}   \nonumber \ ,  \\
 e^{F_2}_3 & = \frac{1}{33} e^{K_2}_{2} + \frac{2}{231} e^{K_2}_{3} - \frac{2}{231} e^{K_2}_{4} + \frac{1}{33} e^{K_2}_{5}+ \frac{2}{9} e^{K_2}_{6} - \frac{2}{99} e^{K_2}_{7} + \frac{4}{9 f} e^{K_2}_{10} - \frac{7}{99 f} e^{K_2}_{11} - \frac{4}{9} e^{K_2}_{12} + \frac{29}{99 f} e^{K_2}_{14} \nonumber \ ,  \\
  e^{F_2}_4 & = \frac{2}{33} e^{K_2}_{3} + \frac{16}{33 f} e^{K_2}_{14}\nonumber  \ ,  \\
e^{F_2}_n &  = e^{K_2}_{n+3} \quad \text{for} \quad n = 5, \dots , 11 \ .  \label{dmrssbasisfns}
 \end{align}
 As discussed in more detail in \secref{rssresponsesec}, we note that the new non-locally-contributing counterterm comes from the function $e_7^{K_2}$, and so enters the expression for $e_1^{F_2}$ and $e_3^{F_2}$ above.  This means that in the basis that we have chosen, $c_3$ and $c_5$ contribute the non-locally-contributing counterterm.

Next, we give the UV matching.  The values for the dark-matter parameters $c_3$, $c_4$, $c_5$, and $c_7$ are given in \eqn{dmuvmatchingresponse}, and the rest are  
\begin{align}
\begin{split} \label{dmuvmatchingresponse2}
& c^{\pi v}_{\rm DM,1} =  - \frac{(46 + 35 f) \knl^2 }{210 \pi^2 } \int  d q P_{11}( q ) \ , \quad  c^{\pi v}_{\rm DM,2}  = -\frac{(11+15 f) \knl^2 }{150 \pi^2 }  \int \, dq \, P_{11}(q)  \ , \\
& c^{\pi v}_{\rm DM,3}  = -\frac{83 \knl^2 }{210 \pi^2 }  \int \, dq \, P_{11}(q) \ , \quad  c^{\pi v}_{\rm DM,4}  =  - \frac{172 \knl^2 }{735 \pi^2 }  \int \, dq \, P_{11}(q)  \ , \\ 
& c^{\pi v}_{\rm DM,5}  = -\frac{2683  \knl^2 }{5145 \pi^2 }  \int \, dq \, P_{11}(q) \ , \quad  c^{\pi v}_{\rm DM,6}  =  - \frac{(4626 + 1715 f) \knl^2 }{25725 \pi^2 }  \int \, dq \, P_{11}(q)  \ , \\ 
& c^{\pi v}_{\rm DM,7}  = -\frac{269  \knl^2 }{686 \pi^2 }  \int \, dq \, P_{11}(q)  \ . 
\end{split}
\end{align}

%%%%%%%%%%%%%%%%%
%
%
%

\section{Bispectrum loop integrals in redshift space}\label{bisploopapp}

In this appendix, we give some information on how to evaluate the one-loop bispectrum integrals in redshift space in \eqn{bispexpressionsrssbias}. The most straightforward way to evaluate the one-loop bispectrum integrals is by choosing a coordinate system and integrating numerically.   Because of rotation invariance, a generic bispectrum one-loop term $B$ is given by
\be \label{genericb1loop}
B(k_1 , k_2 , k_3 , \mu_1 , \mu_2 ) = \int_{\qvec} \mathcal{K} ( k_1 , k_2 , k_3 , \mu_1 , \mu_2, q, \hat k_1 \cdot \hat q , \hat k_2 \cdot \hat q, \hat q \cdot \hat z ) \ ,
\ee
where we have used momentum conservation $\kvec_3 = - \kvec_1 - \kvec_2$, and as always $\mu_i \equiv \hat k_i \cdot \hat z $.  One choice of coordinate system is 
\begin{align}
\begin{split} \label{coordsys}
&\vec k_1 =k_1 \left(0,0,1\right) \  , \\ 
&\vec k_2 = k_2 \left(0,\sqrt{1-y^2},y\right) \ ,  \\ 
&\vec q = q \left(\cos(\beta)\sqrt{1-x^2},\sin(\beta)\sqrt{1-x^2},x\right) \ , \\ 
&\hat z =   \left(\cos(\phi)\sqrt{1-\mu_1^2},\sin(\phi)\sqrt{1-\mu_1^2},\mu_1\right) \ , 
\end{split}
\end{align}
where $y \equiv ( k_3^2 - k_1^2 - k_2^2) / ( 2 k_1 k_2)$, $x\in [-1,1]$, $ \mu_1 \in [-1,1]$, $\phi \in [0,2\pi)$, and $\beta \in [0,2\pi)$. In this coordinate system the measure is simply
\be
\int_{\qvec} = \int \frac{dq \, q^2}{(2 \pi)^3} \int_{-1}^1 dx \int_{0}^{2 \pi} d \beta  \ . 
\ee

Next we move to a tensor reduction method that is better suited for analytic integration.  For this, we write a generic bispectrum one-loop term $B$  as
\be \label{genericb2}
B(k_1 , k_2 , k_3, \mu_1 , \mu_2 ) = \sum_a f_a ( k_1 , k_2 , k_3 , \mu_1 , \mu_2 ) \int_{\qvec} \mathcal{K}_a ( q , |\kvec_1 + \qvec | , | \kvec_2 - \qvec| ; \hat q \cdot \hat z) \ ,
\ee
where the functions $f_a$ do not depend on the loop momentum $\qvec$ and the kernels $\mathcal{K}_a$ depend on the scalar combinations $q$, $|\vec{k}_1 + \vec{q}|$, $|\vec{k}_2 - \vec{q}|$ because of momentum conservation, and on the projection along $\hat{z}$ of the integrated momentum, $\hat{q} \cdot \hat{z}$.   For reasons that we will comment on later, we choose to parameterize $\hat k_1 \cdot \hat q$ and $\hat k_2 \cdot \hat q$ by $|\kvec_1 + \qvec|$ and $| \kvec_2 - \qvec|$ respectively.  {The specific form in \eqn{genericb2} is possible because the dependence on $\mu_1$ and $\mu_2$ is polynomial, and we choose to include any $k_i$ dependence that does not come through $|\kvec_1 + \qvec|$ or $|\kvec_2 - \qvec|$ in the functions $f_a$ to reduce the number of terms in the sum.} Note that both $f_a$ and $\mathcal{K}_a$ can depend on the linear power spectrum $P_{11}$.  If there were no dependence on $\hat{q} \cdot \hat{z}$, the integrals over $\mathcal{K}_a$ could be done using analytic techniques, see for example {\cite{Simonovic:2017mhp, Anastasiou:2022udy}}.
Luckily, the dependence on $\hat{q} \cdot \hat{z}$ is very simple: the $\mathcal{K}_a$ are polynomials in $\hat{q} \cdot \hat{z}$ (which simply comes from the redshift space expression \eqn{rsssbias}), of order up to six for $B_{222}$, up to four for $B_{321}^I$, and up to two for $B_{321}^{II}$ and $B_{411}$.  Thus, each $\mathcal{K}_a$ can be written as
\be \label{calka}
\mathcal{K}_a ( q , |\kvec_1 + \qvec | , | \kvec_2 - \qvec| ; \hat q \cdot \hat z)  =  \sum_n \mathcal{T}_{a,(n)} ( q , |\kvec_1 + \qvec | , | \kvec_2 - \qvec| ) ( \hat q \cdot \hat z )^n \ ,
\ee
where the sum over $n$ is over a finite number of terms as stated above.  

We would now like to write \eqn{calka} in such a way that allows us to express the integral over $\qvec$ as a sum over integrals of functions of $( q , |\kvec_1 + \qvec | , | \kvec_2 - \qvec| )$, which is a form conducive to analytic integration.  To do that, we first define
\begin{equation} 
    I^{i_1 \dots i_n}_{a,(n)} (\vec{k}_1, \vec{k}_2) = \int_{\vec{q}} \mathcal{T}_{a,(n)} (q, |\vec{k}_1 + \vec{q}|, |\vec{k}_2 - \vec{q}|) \, \hat{q}^{i_1} \cdots \hat{q}^{i_n} \, ,
    \label{eq:vector}
\end{equation}
and from \eqn{calka}, we are interested in computing $\hat z^{i_1} \dots \hat z^{i_n}     I^{i_1 \dots i_n}_{a,(n)} (\vec{k}_1, \vec{k}_2) $.  Now, because of rotation invariance, the function $ I^{i_1 \dots i_n}_{a,(n)} (\vec{k}_1, \vec{k}_2) $ can be generally written as
\be \label{ianfn}
I^{i_1 \dots i_n}_{a,(n)} (\vec{k}_1, \vec{k}_2)  = \sum_{\alpha=1}^{N_n} c_{a,(n),\alpha} ( k_1 , k_2 , \hat k_1 \cdot \hat k_2 ) T^{i_1 \dots i_n}_{a , (n),\alpha} ( \hat k_1 , \hat k_2, \delta_{ij} ) \ , 
\ee
where the functions $c_{a,(n),\alpha}$ depend on scalar products of $\kvec_1$ and $\kvec_2$, and the functions $T^{i_1 \dots i_n}_{a , (n),\alpha} $ are all of the $N_n$ symmetric tensors with indices $i_1 \dots i_n$ made up of products of $\hat k_1$, $\hat k_2$, and $\delta_{ij}$.  To see how this has helped us, we go back to the expression that we are interested in, which now becomes
\be \label{finalloopform}
\hat z^{i_1} \dots \hat z^{i_n}     I^{i_1 \dots i_n}_{a,(n)} (\vec{k}_1, \vec{k}_2)  = \sum_\alpha c_{a,(n),\alpha} ( k_1 , k_2 , \hat k_1 \cdot \hat k_2 ) t_{a,(n),\alpha} ( \mu_1 , \mu_2) \ ,
\ee
where we have defined $ t_{a,(n),\alpha} ( \mu_1 , \mu_2)  \equiv \hat z^{i_1} \dots \hat z^{i_n}  T^{i_1 \dots i_n}_{a , (n),\alpha} ( \hat k_1 , \hat k_2, \delta_{ij} )$.  This is now exactly the form that we wanted: all of the loop integrals are contained in the $c_{a,(n),\alpha}$ functions, which is over functions of $( q , |\kvec_1 + \qvec | , | \kvec_2 - \qvec| )$, and all of the $\hat z$ dependence has been transferred to the external momenta in the $t_{a,(n),\alpha} $ functions.  

Let us now determine the $c_{a,(n),\alpha}$ functions explicitly.  To do that, we contract \eqn{ianfn} with the $N_n$ symmetric tensors $T^{i_1 \dots i_n}_{a , (n),\beta} $, giving
\begin{align}
\begin{split} \label{cafneqs}
& \int_{\vec{q}} \mathcal{T}_{a,(n)} (q, |\vec{k}_1 + \vec{q}|, |\vec{k}_2 - \vec{q}|) \,  T^{i_1 \dots i_n}_{a , (n),\beta} ( \hat k_1 , \hat k_2, \delta_{ij} ) \hat{q}^{i_1} \cdots \hat{q}^{i_n}  \\
  &\hspace{1.5in}  = \sum_{\alpha=1}^{N_n} c_{a,(n),\alpha} ( k_1 , k_2 , \hat k_1 \cdot \hat k_2 ) T^{i_1 \dots i_n}_{a , (n),\alpha} ( \hat k_1 , \hat k_2, \delta_{ij} ) T^{i_1 \dots i_n}_{a , (n),\beta} ( \hat k_1 , \hat k_2, \delta_{ij} ) \ ,
\end{split}
\end{align}
which is a system of $N_n$ equations that can be used to solve for the $N_n$ functions $c_{a,(n),\alpha}$; on the left-hand side, using 
\be \label{qdotk1ork2}
\hat{q} \cdot \hat{k}_1 = \frac{1}{2 q k_1} \left(|\vec{k}_1 + \vec{q}|^2 - k_1^2 - q^2\right) \andd  \hat{q} \cdot \hat{k}_2 = \frac{1}{2 q k_2} \left (k_2^2 + q^2 - |\vec{k}_2 - \vec{q}|^2 \right) \ , 
\ee
 we have our desired form of loop integrals over $\qvec$ of functions of $( q , |\kvec_1 + \qvec | , | \kvec_2 - \qvec| )$, while on the right-hand side, we have the $c_{a,(n),\alpha}$ functions multiplied by scalar products between $\hat k_1 $ and $\hat k_2$.  In particular, defining the matrix
 \be
 M^{\alpha \beta}_{a,(n)} (\hat k_1 \cdot \hat k_2) \equiv T^{i_1 \dots i_n}_{a , (n),\alpha} ( \hat k_1 , \hat k_2, \delta_{ij} ) T^{i_1 \dots i_n}_{a , (n),\beta} ( \hat k_1 , \hat k_2, \delta_{ij} ) \ , 
 \ee
 we have
 \begin{align}
\begin{split} \label{cafnsol}
&  c_{a,(n),\alpha} ( k_1 , k_2 , \hat k_1 \cdot \hat k_2 )  =\\
 &\hspace{.5in}  \sum_{\beta =1}^{N_n}  [M^{-1}_{a,(n)}]^{\alpha \beta} (\hat k_1 \cdot \hat k_2)  \int_{\vec{q}} \mathcal{T}_{a,(n)} (q, |\vec{k}_1 + \vec{q}|, |\vec{k}_2 - \vec{q}|) \,  T^{i_1 \dots i_n}_{a , (n),\beta} ( \hat k_1 , \hat k_2, \delta_{ij} ) \hat{q}^{i_1} \cdots \hat{q}^{i_n}  \ . 
 \end{split}
 \end{align}
  Thus, this solution for the $c_{a,(n),\alpha}$ functions, plugged into \eqn{finalloopform}, gives our final result.  
  
  The above manipulations can also be presented in a slightly different way.  Again, we are interested in computing the right-hand side of \eqn{eq:vector} contracted with $\hat z^{i_1} \cdots \hat z^{i_n}$, which, using \eqn{eq:vector} and \eqn{ianfn}, is given by
\be
 \int_{\vec{q}} \mathcal{T}_{a,(n)} (q, |\vec{k}_1 + \vec{q}|, |\vec{k}_2 - \vec{q}|) \, (\hat q \cdot \hat z)^n = \sum_\alpha c_{a,(n),\alpha} ( k_1 , k_2 , \hat k_1 \cdot \hat k_2 ) t_{a,(n),\alpha} ( \mu_1 , \mu_2) \ . 
\ee
Now, given the solution \eqn{cafnsol} for the $c_{a,(n),\alpha}$ functions, we can rewrite this as
\begin{align}
& \int_{\vec{q}} \mathcal{T}_{a,(n)} (q, |\vec{k}_1 + \vec{q}|, |\vec{k}_2 - \vec{q}|) \, (\hat q \cdot \hat z)^n = \\
& \int_{\vec{q}} \mathcal{T}_{a,(n)} (q, |\vec{k}_1 + \vec{q}|, |\vec{k}_2 - \vec{q}|)  \sum_{\alpha = 1}^{N_n} \sum_{\beta =1}^{N_n}  [M^{-1}_{a,(n)}]^{\alpha \beta} (\hat k_1 \cdot \hat k_2)  \,  T^{i_1 \dots i_n}_{a , (n),\beta} ( \hat k_1 , \hat k_2, \delta_{ij} ) \hat{q}^{i_1} \cdots \hat{q}^{i_n}  t_{a,(n),\alpha} ( \mu_1 , \mu_2) \ , \nonumber
\end{align}
which implies that, under the integrals that we are interested in, we can simply replace  
\be \label{tensorreduxqz}
(\hat q \cdot \hat z)^n \rightarrow  \sum_{\alpha = 1}^{N_n} \sum_{\beta =1}^{N_n}  [M^{-1}_{a,(n)}]^{\alpha \beta} (\hat k_1 \cdot \hat k_2)  \,  T^{i_1 \dots i_n}_{a , (n),\beta} ( \hat k_1 , \hat k_2, \delta_{ij} ) \hat{q}^{i_1} \cdots \hat{q}^{i_n}  t_{a,(n),\alpha} ( \mu_1 , \mu_2)  \ , 
\ee
which again, using \eqn{qdotk1ork2}, is in the form that we want.

At this point, a number of comments are in order.  First, the above form of the loop integrals is well suited for situations where integrals over functions of $( q , |\kvec_1 + \qvec | , | \kvec_2 - \qvec| )$ can be done analytically, see for example \cite{Simonovic:2017mhp, Anastasiou:2022udy}.  If this is not the case, one can simply use an explicit coordinate system to perform the integrals, as described above.  Second, it turns out that for the bispectrum, it can be quite slow to invert the matrix $M^{\alpha \beta}_{a , (n)}$ for values of $n$ greater than four (which is relevant in particular for the $B_{222}$ diagram).

Thus, in practice, it is sometimes easier to use the following more straightforward way to arrive at the above result, again with an eye towards analytic integration.  We start with the form of the loop in \eqn{genericb2}, and using the above definitions, we recall that we need to evaluate terms of the form
\be \label{caltanq}
 \int_{\vec{q}} \mathcal{T}_{a,(n)} (q, |\vec{k}_1 + \vec{q}|, |\vec{k}_2 - \vec{q}|) \, (\hat q \cdot \hat z)^n  \ , 
\ee
so we would like to find a replacement for $(\hat q \cdot \hat z)^n$ in terms of $q$, $|\vec{k}_1 + \vec{q}|$, and $|\vec{k}_2 - \vec{q}|$.  For this, we can simply use the coordinate system \eqn{coordsys}.  Without loss of generality we can take the first component of $\hat z$ to be positive, and use $\mu_2 = \hat k_2 \cdot \hat z$ to get
\be
\hat z = \left(\sqrt{\frac{-\mu _1^2-\mu _2^2-y^2+2 \mu _1 \mu _2 y+1}{1-y^2}},\frac{\mu _2-\mu
   _1 y}{\sqrt{1-y^2}},\mu _1\right) \ . 
\ee
{To see why the sign of the first component of $\hat z$ does not matter, consider \eqn{caltanq} and send $\hat z_1 \rightarrow - \hat z_1$, where the subscript ${}_1$ denotes the first component.  This can be compensated by sending $\hat q_1 \rightarrow - \hat q_1$, and since none of $q$, $|\kvec_1 + \qvec|$, or $|\kvec_2 - \qvec|$ depend on $\hat q_1$ for the parameterization \eqn{coordsys}, the integral is invariant.}
Dotting this with $\hat q$ from \eqn{coordsys}, we get
\be
\hat q \cdot \hat z= \cos (\beta) \sqrt{1-x^2}  \sqrt{\frac{-\mu _1^2-\mu _2^2-y^2+2 \mu _1 \mu _2
   y+1}{1-y^2}}+\frac{ \sin (\beta ) \sqrt{1-x^2}  \left(\mu _2-\mu _1 
   y\right)}{\sqrt{1-y^2}}+\mu _1 x \ . 
\ee 
Now, we raise this to the $n$-th power, and plug it into \eqn{caltanq}.  The resulting expression can be simplified by noting the following.  First, $|\kvec_1 + \qvec|$ does not depend on $\beta$, and $| \kvec_2 - \qvec|$ only depends on $\sin(\beta)$.  Thus, in terms of $\beta$ dependence, the integral in \eqn{caltanq} is equal to a sum over terms of the form
\be
\int_0^{2\pi} d \beta f ( \sin ( \beta) ) \cos ( \beta)^m
\ee 
for some functions $f$ and integer powers $m$.  {This integral is clearly zero when $m$ is odd, since in that case $\cos(\beta)^m$ is an odd function around $\beta = \pi / 2$ and $\sin ( \beta)$ is an even function around $\beta = \pi / 2$, and both are periodic with periods $2 \pi$. } 
Thus, we see that when we expand out $(\hat q \cdot \hat z)^n$, we can immediately set to zero any terms with an odd power of $\cos (\beta)$, and any time that we encounter an even power of $\cos (\beta)$ we can replace it with $\cos(\beta)^2 = 1 - \sin(\beta)^2$, so our result only depends on $\sin(\beta)$.  

It now remains to express $x$ and $\sin(\beta)$ in terms of the desired $(q, |\vec{k}_1 + \vec{q}|, |\vec{k}_2 - \vec{q}|)$.  This is straightforward using \eqn{coordsys}, and we find
\be \label{xdef}
x = \frac{-k_1^2+ |\vec{k}_1 + \vec{q}|^2-q^2}{2 k_1 q} \ , 
\ee
and
\be \label{sinbetadef}
\sin(\beta) = \frac{-2 k_2 \, q\,  x\,  y+k_2^2- |\vec{k}_2 - \vec{q}|^2+q^2}{2 k_2 q \sqrt{1-x^2} \sqrt{1-y^2}} \ . 
\ee
Taking into account that only even powers of $\cos(\beta)$ contribute, {and defining
\be
\nu \equiv \sqrt{1-x^2} \sin(\beta) \ , 
\ee
} this means that we can replace $(\hat q \cdot \hat z)^n$ in \eqn{caltanq} with (using the multinomial theorem)
\begin{align}
\begin{split} \label{repqz}
(\hat q \cdot \hat z)^n \rightarrow & \sum_{i+j+2 k=n}\frac{n! \left(\mu _1 x\right){}^i}{i!j!(2k)!} 
   \left[\nu \left(\mu _2-\mu _1 y\right)\right]^j  \left(1-y^2\right)^{-\frac{1}{2} (j+2 k)}   \\
   & \hspace{1in} \times \left[\left(1-x^2-\nu^2\right) \left(-\mu _1^2-\mu _2^2-y^2+2 \mu _1 \mu _2 y+1\right)\right]^k\ 
\end{split}
\end{align}
which, using \eqn{xdef} and \eqn{sinbetadef}, again puts the loop integral in \eqn{caltanq} in the desired form {(notice that since \eqn{repqz} depends on $\sin ( \beta )$ only through $\nu$, the possible non-analytic dependence $\sqrt{1-x^2}$ cancels and does not appear in the final answer)}.  One can check that the right-hand sides of the expressions in \eqn{tensorreduxqz} and \eqn{repqz} are equal.

%%%%%%%%%%%%%%%%%
%
%
%

\section{Details for biased tracers to fourth order}\label{bias4}

In this appendix, we give the explicit calculation to obtain \eqn{finalbias}.  We do this in three steps. First, we expand the bias expansion in terms of all operators allowed by the equivalence principle. This gives the list of operators that we schematically wrote in \eqn{bias_raw}. Next we do the explicit Taylor expansion of the operators evaluated at $\xfl$ around $\vx$, and define the explicit $\mathbb{C}^{(n)}_{\mathcal{O},i}$ that are in \eqn{opt}. In a last step we put them together, and remove degeneracies, to obtain \eqn{finalbias}. In this whole appendix, we focus on the fourth order calculation and refer to \cite{Senatore:2014eva,Angulo:2015eqa,Fujita:2016dne} for the calculation up to third order.  {The dark-matter kernels, in real space and redshift space, can be obtained from the expressions in this appendix by setting $b_1 = b_2 = b_3 = b_4 = 1$ and $b_5 , \dots , b_{15} = 0$.  }

\subsection{Bias expansion} \label{biasexpandsubapp}
As mentioned in \secref{biasm}, the tracer overdensity can only depend on second derivatives of the gravitational potential, and first derivatives of the velocity field (and higher spatial derivatives of these). To look at all possible operators, we define the building blocks 
		\be
		r_{ij} = \frac{2}{3 \om \cH^2}\pd_i\pd_j \Phi \andd  p_{ij} = -  \frac{1}{f a H } \pd_i v^j  \ ,
		\ee
 so that $\delta^{ij} r_{ij}=\delta$ and $\delta^{ij} p_{ij}=\theta$, with $\theta \equiv - \partial_i v^i / (f a H)$.\footnote{{Notice that $p_{ij}$ is symmetric for a velocity with vanishing vorticity.  For the biases that we discuss in this section, this is true.}} For notational convenience, we further define
		\begin{align}
		\begin{split}
		&r^2 = r_{ij}r_{ij}  \ ,\qquad rp = r_{ij}p_{ij} \ , \qquad p^2 = p_{ij}p_{ij} \ , \qquad r^3 = r_{ij}r_{jl } r_{li} \ , \\ 
		&r^2p = r_{ij}r_{jl}p_{li} \ , \qquad rp^2 = r_{ij}p_{jl }p_ {il} \ , \qquad p^3 = p_{ij}p_{jl} p_{il} \ , \qquad r^4 = r_{ij}r_{jl} r_{lk} r_{ki} \ .\\ 
		\end{split}
		\end{align} 
		We can now write down the full expansion for the tracer overdensity up to fourth order, which we only schematically gave in \eqn{bias_raw}.  {To determine the operators, we write down all contractions of $r_{ij}$ and $p_{ij}$ up to fourth order, and obtain}   
		\begin{align}
		\begin{split} \label{fullexp}
		\delta_{h}(\vx,t) &= \int^t dt' H(t') \big[c_{\delta}(t,t')\delta(\xfl,t')+c_{\theta}(t,t')\theta(\xfl,t')\\ 
		&+c_{\delta^2}(t,t')\delta^2(\xfl,t')+c_{\delta\theta}(t,t')\delta\theta(\xfl,t')+c_{\theta^2}(t,t')\theta^2(\xfl,t')\\ 
		&+c_{r^2}(t,t')r^2(\xfl,t')+c_{rp}(t,t')rp(\xfl,t')+c_{p^2}(t,t')p^2(\xfl,t')\\ 
		&+c_{\delta^3}(t,t')\delta^3(\xfl,t')+c_{\delta^2\theta}(t,t')\delta^2\theta(\xfl,t')+c_{\delta\theta^2}(t,t')\delta\theta^2(\xfl,t')+c_{\theta^3}(t,t')\theta^3(\xfl,t')\\ 
		&+c_{r^3}(t,t')r^3(\xfl,t')+c_{r^2p}(t,t')r^2p(\xfl,t')+c_{rp^2}(t,t')rp^2(\xfl,t')+c_{p^3}(t,t')p^3(\xfl,t')\\ 
		&+c_{r^2\delta}(t,t')r^2\delta(\xfl,t')+c_{rp\delta}(t,t')rp\delta(\xfl,t')+c_{p^2\delta}(t,t')p^2\delta(\xfl,t')\\ 
		&+c_{r^2\theta}(t,t')r^2\theta(\xfl,t')+c_{rp\theta}(t,t')rp\theta(\xfl,t')+c_{p^2\theta}(t,t')p^2\theta(\xfl,t')\\ 
		&+c_{\delta^4}(t,t')\delta^4(\xfl,t')+c_{\delta r^3}(t,t')\delta r^3(\xfl,t')+c_{\delta^2r^2}(t,t')\delta^2r^2(\xfl,t')\\ 
		&+c_{\left(r^2\right)^2}(t,t')\left(r^2\right)^2(\xfl,t')+c_{ r^4}(t,t')r^4(\xfl,t')\big] \big|_{\xfl = \xfl( \xvec , t , t')} \ .
		\end{split}
		\end{align}
		{Since we only go up to fourth order in perturbations, terms that explicitly start at fourth order above are evaluated on the linear fields, and for those terms we have used $\theta^{(1)} = \delta^{(1)}$.}	
		
		\subsection{Expansion in fluid element} \label{fluidelementapp}
		The operators in \eqn{fullexp} are all evaluated at $\xfl$, where $\xfl$ is given implicitly in \eqn{xfl}. In this section, we Taylor expand all the fields evaluated at $\xfl$ around $\vx$. Going up to products of four fields, for a generic operator $\mathcal{O}$ we have	 (following \cite{Senatore:2014eva})
 \bea\label{eq:xfl_expansion}
		&&\mathcal{O}(\xfl(\xvec, t,t'),t')\approx \mathcal{O}(\vec x,t')+ \partial_i\mathcal{O}(\xvec ,t')\int_{t}^{t'} \frac{dt_1}{a(t_1)} \; v^i(\vec x,t_1)\\ \nonumber
		&&\quad \quad\quad\quad\quad\quad+\frac{1}{2}\partial_i\partial_j \mathcal{O}(\xvec,t')\int_{t}^{t'} \frac{dt_1}{a(t_1)}\; v^i(\vec x,t_1)\int_{t}^{t'} \frac{dt_2}{a(t_2)}\;v^j(\vec x,t_2)\\ \nonumber
		&&\quad \quad\quad\quad\quad\quad+\partial_i\mathcal{O}(\xvec,t')\int_{t}^{t'} \frac{dt_1}{a(t_1)} \;\partial_jv^i(\xvec,t_1)\int_{t}^{t_1} \frac{dt_2}{a(t_2)} \; v^j(\vec x,t_2)\\ \nonumber
		&&\quad \quad\quad\quad\quad\quad+\frac{1}{6}\partial_i\partial_j\partial_k \mathcal{O}(\xvec,t')\int_{t}^{t'} \frac{dt_1}{a(t_1)}\; v^i(\vec x,t_1)\int_{t}^{t'} \frac{dt_2}{a(t_2)}\;v^j(\vec x,t_2)\int_{t}^{t'} \frac{dt_3}{a(t_3)}\;v^k(\vec x,t_3) \\ \nonumber
		&&\quad \quad\quad\quad\quad\quad +\frac{1}{2}\partial_i\mathcal{O}(\xvec,t')\int_{t}^{t'} \frac{dt_1}{a(t_1)} \;\partial_j\partial_k v^i(\xvec,t_1)\int_{t}^{t_1} \frac{dt_2}{a(t_2)}\; v^j(\vec x,t_2)\int_{t}^{t_1} \frac{dt_3}{a(t_3)}\;v^k(\vec x,t_3)\\ \nonumber
		&&\quad \quad\quad \quad\quad \quad + \partial_i\mathcal{O}(\xvec,t')\int_{t}^{t'} \frac{dt_1}{a(t_1)} \;\partial_jv^i(\xvec,t_1)\int_{t}^{t_1} \frac{dt_2}{a(t_2)}\; \partial_k v^j(\vec x,t_2) \int_t^{t_2} \frac{ d t_3}{a(t_3)}\;v^k(\vec x,t_3)\\ \nonumber
		&&\quad \quad\quad\quad\quad\quad + \partial_i\partial_j \mathcal{O}(\xvec,t')\int_{t}^{t'} \frac{dt_1}{a(t_1)}\; v^i(\vec x,t_1)\int_{t}^{t'} \frac{dt_2}{a(t_2)}\;\partial_kv^j(\xvec,t_2)\int_{t}^{t_2} \frac{dt_3}{a(t_3)} \; v^k(\vec x,t_3) \ . 
		\eea	
		We now perturbatively expand and rewrite the velocity in terms of the divergence only	
		\bea
		v^{i}_{(n)}(\vx,t ') = - \frac{a(t')  \dot D(t')}{D(t')} \frac{D(t')^n}{D(t)^n} \fr{\pd_i}{\pd^2}\theta^{(n)}(\vx, t) \ . 
		\eea
		With the EdS approximation as done above (and potentially also without it \cite{Donath:2020abv}), it is always possible to solve the time integrals in \eqn{eq:xfl_expansion} analytically, since all integrals can be reduced to 
\bea\label{fulltaylorexp}
		\int_{t}^{t' } d t_1 \;\fr{ \dot D(t_1)D( t_1 )^{n-1}}{D(t )^n} = \fr{1}{n}\left[\fr{D(t ')^n}{D( t )^n} - 1 \right] \ . 
		\eea
		This then gives the expression for a Taylor expanded operator at fourth order in perturbations 
\begin{align}
\begin{split}\label{eq:xfl_expansion4}
		&\mathcal{O}^{(4)}(\xfl(\xvec, t,t'),t')=\mathcal{O}^{(4)}\fr{D(t')^4}{D(t)^4}       +\pd_i\mathcal{O}^{(1)} \fr{\pd_i \theta^{(3)}}{\pd^2} \left[\fr{1}{3}\fr{D(t')}{D(t)}-\fr{1}{3}\fr{D(t')^4}{D(t)^4}\right]\\ 
		&       +\pd_i\mathcal{O}^{(2)} \fr{\pd_i \theta^{(2)}}{\pd^2}\left[\frac{1}{2}\fr{D(t')^2}{D(t)^2}-\frac{1}{2}\fr{D(t')^4}{D(t)^4}\right] + \pd_i\mathcal{O}^{(3)} \fr{\pd_i \theta^{(1)}}{\pd^2} \left[\fr{D(t')^3}{D(t)^3}-\fr{D(t')^4}{D(t)^4}\right]\\ 
		&       +\pd_i\pd_j \mathcal{O}^{(1)}  \fr{\pd_i \theta^{(1)}}{\pd^2} \fr{\pd_j \theta^{(2)}}{\pd^2} \left[\frac{1}{2}\fr{D(t')}{D(t)}-\frac{1}{2}\fr{D(t')^2}{D(t)^2}-\frac{1}{2}\fr{D(t')^3}{D(t)^3}+\frac{1}{2}\fr{D(t')^4}{D(t)^4}\right]\\ 
		&       +\pd_i\pd_j \mathcal{O}^{(2)} \fr{\pd_i \theta^{(1)}}{\pd^2}  \fr{\pd_j \theta^{(1)} }{\pd^2}\left[\frac{1}{2}\fr{D(t')^2}{D(t)^2}-\fr{D(t')^3}{D(t)^3}+\frac{1}{2}\fr{D(t')^4}{D(t)^4}\right]\\ 
		&       +\pd_i\mathcal{O}^{(1)}  \fr{\pd_j\pd_i \theta^{(1)}}{\pd^2}  \fr{\pd_j \theta^{(2)}}{\pd^2}\left[\fr{1}{3}\fr{D(t')}{D(t)}-\frac{1}{2}\fr{D(t')^2}{D(t)^2}+\fr{1}{6}\fr{D(t')^4}{D(t)^4}\right]\\ 
		&       +\pd_i\mathcal{O}^{(1)}  \fr{\pd_i\pd_j \theta^{(2)}}{\pd^2} \fr{\pd_j \theta^{(1)} }{\pd^2}\left[\fr{1}{6}\fr{D(t')}{D(t)}-\frac{1}{2}\fr{D(t')^3}{D(t)^3}+\fr{1}{3}\fr{D(t')^4}{D(t)^4}\right]\\ 
		&       +\pd_i\mathcal{O}^{(2)}  \fr{\pd_i\pd_j \theta^{(1)}}{\pd^2}\fr{\pd_j \theta^{(1)}}{\pd^2}\left[\frac{1}{2}\fr{D(t')^2}{D(t)^2}-\fr{D(t')^3}{D(t)^3}+\frac{1}{2}\fr{D(t')^4}{D(t)^4}\right]\\ 
		&       +\pd_i\pd_j\pd_k \mathcal{O}^{(1)} \fr{\pd_i \theta^{(1)}}{\pd^2} \fr{\pd_j \theta^{(1)}}{\pd^2} \fr{\pd_k \theta^{(1)}}{\pd^2} \left[\frac{1}{6}\fr{D(t')}{D(t)}-\frac{1}{2}\fr{D(t')^2}{D(t)^2}+\frac{1}{2}\fr{D(t')^3}{D(t)^3}-\frac{1}{6}\fr{D(t')^4}{D(t)^4}\right]  \\ 
		&       +\pd_i\mathcal{O}^{(1)} \fr{\pd_i \pd_j \pd_k \theta^{(1)}}{\pd^2} \fr{\pd_j \theta^{(1)}}{\pd^2} \fr{\pd_k \theta^{(1)}}{\pd^2} \bigg[\fr{1}{6}\fr{D(t')}{D(t)}-\frac{1}{2}\fr{D(t')^2}{D(t)^2}+\frac{1}{2}\fr{D(t')^3}{D(t)^3}-\fr{1}{6}\fr{D(t')^4}{D(t)^4}\bigg]\\ 
		&         +\pd_i\mathcal{O}^{(1)} \fr{\pd_i\pd_j \theta^{(1)}}{\pd^2}  \fr{\pd_j\pd_k \theta^{(1)}}{\pd^2} \fr{\pd_k \theta^{(1)}}{\pd^2} \bigg[\fr{1}{6}\fr{D(t')}{D(t)}-\frac{1}{2}\fr{D(t')^2}{D(t)^2}+\frac{1}{2}\fr{D(t')^3}{D(t)^3}-\fr{1}{6}\fr{D(t')^4}{D(t)^4}]\bigg]\\ 
		&       +\;\pd_i\pd_j \mathcal{O}^{(1)} \fr{\pd_i \theta^{(1)}}{\pd^2} \fr{\pd_j\pd_k \theta^{(1)}}{\pd^2} \fr{\pd_k \theta^{(1)}}{\pd^2}  \left[\fr{1}{2}\fr{D(t')}{D(t)} -\frac{3}{2}\fr{D(t')^2}{D(t)^2} + \frac{3}{2}\fr{D(t')^3}{D(t)^3} - \fr{1}{2}\fr{D(t')^4}{D(t)^4}\right] \ , 
		\end{split}
		\end{align}
where all fields $\mathcal{O}^{(n)}$ and $\theta^{(n)}$ are evaluated at $(\xvec , t)$, and we refer the reader to \cite{Angulo:2015eqa, Fujita:2016dne} for expressions up to third order.  

From the above expansion, we can now define the $\mathbb{C}_{\mathcal{O},i}$ operators that appear in  \eqn{opt} and later in \eqn{finalbias}. Note that for an operator $\mathcal{O}_m$ that starts at order $m$, such as a product operator of $m$ fields, {$\mathcal{O}_m^{(n)} = 0$ for $n < m$}. Finally, for an operator $\mathcal{O}_m$ starting at order $m$, we collect the terms that multiply the same power of $D(t') / D(t)$, which gives the $n$-th order terms in the expansion 
\bea \label{biasfnbasis}
\mathcal{O}_{m}^{(n)} ( \xvec_{\rm fl} ( \xvec, t , t') , t')  =   \sum_{\alpha = 1}^{n-m+1}  \left( \frac{D(t' )}{D(t)} \right)^{\alpha+m-1}  \mathbb{C}^{(n)}_{\mathcal{O}_m , \alpha}  ( \xvec,t )  \ ,
\eea	
which allows us to read off the $\mathbb{C}_{\mathcal{O},i}$ from \eqn{eq:xfl_expansion4}. Summing up all the orders gives \eqn{opt}.

		After also performing the time integrals and defining coefficients following \eqn{eq:param},  we obtain the Taylor expanded and time integrated version of \eqn{fullexp} at fourth order,
		\begin{align}
		\begin{split} \label{schematic}
\hspace{-.1in} \delta_{h}^{(4)}(\vx,t) &= c_{\delta,1} \mathbb{C}_{\delta,1}^{(4)} +c_{\delta,2} \mathbb{C}_{\delta,2}^{(4)} +c_{\delta,3} \mathbb{C}_{\delta,3}^{(4)} +c_{\delta,4} \mathbb{C}_{\delta,4}^{(4)} +c_{\theta,1} \mathbb{C}_{\theta,1}^{(4)} +c_{\theta,2} \mathbb{C}_{\theta,2}^{(4)} +c_{\theta,3} \mathbb{C}_{\theta,3}^{(4)}  \\  
		&+c_{\theta,4} \mathbb{C}_{\theta,4}^{(4)} +c_{\delta^2,1} \mathbb{C}_{\delta^2,1}^{(4)} +c_{\delta^2,2} \mathbb{C}_{\delta^2,2}^{(4)} +c_{\delta^2,3} \mathbb{C}_{\delta^2,3}^{(4)}  +c_{\delta\theta,1} \mathbb{C}_{\delta\theta,1}^{(4)} +c_{\delta\theta,2} \mathbb{C}_{{\delta\theta},2}^{(4)}  \\  
		& +c_{\delta\theta,3} \mathbb{C}_{\delta\theta,3}^{(4)} +c_{\theta^2,1} \mathbb{C}_{\theta^2,1}^{(4)} +c_{\theta^2,2} \mathbb{C}_{\theta^2,2}^{(4)} +c_{\theta^2,3} \mathbb{C}_{\theta^2,3}^{(4)} +c_{r^2,1} \mathbb{C}_{r^2,1}^{(4)} +c_{r^2,2} \mathbb{C}_{r^2,2}^{(4)} \\  
		& +c_{r^2,3} \mathbb{C}_{r^2,3}^{(4)} +c_{rp,1} \mathbb{C}_{rp,1}^{(4)} +c_{rp,2} \mathbb{C}_{rp,2}^{(4)}  +c_{rp,3} \mathbb{C}_{rp,3}^{(4)} +c_{p^2,1} \mathbb{C}_{p^2,1}^{(4)} +c_{p^2,2} \mathbb{C}_{p^2,2}^{(4)} \\
		& +c_{p^2,3} \mathbb{C}_{p^2,3}^{(4)} +c_{\delta^3,1} \mathbb{C}_{\delta^3,1}^{(4)} +c_{\delta^3,2} \mathbb{C}_{\delta^3,2}^{(4)} +c_{\theta^3,1} \mathbb{C}_{\theta^3,1}^{(4)} +c_{\theta^3,2} \mathbb{C}_{\theta^3,2}^{(4)} +c_{\delta^2\theta,1} \mathbb{C}_{\delta^2\theta,1}^{(4)} \\  
		& +c_{\delta^2\theta,2} \mathbb{C}_{\delta^2\theta,2}^{(4)} +c_{\delta\theta^2,1} \mathbb{C}_{\delta\theta^2,1}^{(4)} +c_{\delta\theta^2,2} \mathbb{C}_{\delta\theta^2,2}^{(4)} +c_{r^3,1} \mathbb{C}_{r^3,1}^{(4)} +c_{r^3,2} \mathbb{C}_{r^3,2}^{(4)}+c_{p^3,1} \mathbb{C}_{p^3,1}^{(4)}   \\  
		& +c_{p^3,2} \mathbb{C}_{p^3,2}^{(4)} +c_{r^2p,1} \mathbb{C}_{r^2p,1}^{(4)} +c_{r^2p,2} \mathbb{C}_{r^2p,2}^{(4)} +c_{rp^2,1} \mathbb{C}_{rp^2,1}^{(4)} +c_{rp^2,2} \mathbb{C}_{rp^2,2}^{(4)} +c_{r^2\delta,1} \mathbb{C}_{r^2\delta,1}^{(4)} \\  
		& +c_{r^2\delta,2} \mathbb{C}_{r^2\delta,2}^{(4)} +c_{rp\delta,1} \mathbb{C}_{rp\delta,1}^{(4)} +c_{rp\delta,2} \mathbb{C}_{rp\delta,2}^{(4)}  +c_{p^2\delta,1} \mathbb{C}_{p^2\delta,1}^{(4)} +c_{p^2\delta,2} \mathbb{C}_{p^2\delta,2}^{(4)}  +c_{r^2\theta,1} \mathbb{C}_{r^2\theta,1}^{(4)} \\  
		& +c_{r^2\theta,2} \mathbb{C}_{r^2\theta,2}^{(4)} +c_{rp\theta,1} \mathbb{C}_{rp\theta,1}^{(4)} +c_{rp\theta,2} \mathbb{C}_{rp\theta,2}^{(4)} +c_{p^2\theta,1} \mathbb{C}_{p^2\theta,1}^{(4)}  +c_{p^2\theta,2} \mathbb{C}_{p^2\theta,2}^{(4)} \\  
		&  +c_{\delta^4,1} \mathbb{C}_{\delta^4,1}^{(4)} +c_{\delta r^3,1} \mathbb{C}_{\delta r^3,1}^{(4)} +c_{\delta^2r^2,1} \mathbb{C}^{(4)}_{\delta^2r^2,1}+c_{\left(r^2\right)^2,1} \mathbb{C}_{\left(r^2\right)^2,1}^{(4)} +c_{ r^4,1} \mathbb{C}_{r^4,1}^{(4)}  \ ,
		\end{split}
		\end{align}
		where all of the $c_{\mathcal{O},i}$ functions are evaluated at $t$, and all of the $\mathbb{C}^{(4)}_{\mathcal{O},i}$ functions are evaluated at $(\xvec , t)$.
		 As mentioned in the main text, the expansion above is not yet irreducible, for instance $\mathbb{C}_{r^2,1}^{(4)} =  \sfrac{7}{2}\mathbb{C}_{\delta,2}^{(4)} -  \sfrac{5}{2}\mathbb{C}_{\delta^2,1}^{(4)}$. Therefore by a redefinition of coefficients, we can reduce the number of coefficients needed. A full list of degeneracies is given in \eqn{eq:degen} in the next section.
					
\subsection{Degeneracies, local basis, {and comparison to literature}}	\label{degenandlocalapp}

We find that at fourth order, the number of independent operators is fifteen, and we choose the basis 
\be \label{cker4basis}
\{\mathbb{C}_{\delta,1}^{(4)},\mathbb{C}_{\delta,2}^{(4)},\mathbb{C}_{\delta,3}^{(4)},\mathbb{C}_{\delta,4}^{(4)},\mathbb{C}_{\delta^2,1}^{(4)},\mathbb{C}_{\delta^2,2}^{(4)},\mathbb{C}_{\delta^2,3}^{(4)},\mathbb{C}_{r^2,2}^{(4)},\mathbb{C}_{r^2,3}^{(4)},\mathbb{C}_{\delta^3,1}^{(4)},\mathbb{C}_{\delta^3,2}^{(4)},\mathbb{C}_{r^3,2}^{(4)} ,\mathbb{C}_{r^2\delta,2}^{(4)} ,\mathbb{C}_{\delta^4,1}^{(4)} ,\mathbb{C}_{\delta r^3,1}^{(4)} \} \ .
\ee We give these fifteen functions explicitly in \appref{explicit}.
The other operators are given in terms of these by the following relationships 
		\begingroup
	\allowdisplaybreaks
	\bea\label{eq:degen}
	&&\mathbb{C}_{r^2,1}^{(4)} =  \sfrac{7}{2}\mathbb{C}_{\delta,2}^{(4)} -  \sfrac{5}{2}\mathbb{C}_{\delta^2,1}^{(4)} \ , \quad	\mathbb{C}_{r^3,1}^{(4)} = \sfrac{45}{4} \mathbb{C}_{\delta,3}^{(4)} - \sfrac{105}{16}\mathbb{C}_{\delta^2,2}^{(4)} -
	\sfrac{3}{4} \mathbb{C}_{r^2,2}^{(4)} + \sfrac{35}{8} \mathbb{C}_{\delta^3,1}^{(4)}\ , \\ \nonumber 
	&&	\mathbb{C}_{r^2\delta,1}^{(4)} = \sfrac{7}{4} \mathbb{C}_{\delta^2,2}^{(4)} - \sfrac{5}{2} \mathbb{C}_{\delta^3,1}^{(4)}\ , \quad
	\mathbb{C}_{r^2\delta^2}^{(4)} = \sfrac{7}{6}\mathbb{C}_{\delta^3,2}^{(4)} - \sfrac{5}{2} \mathbb{C}_{\delta^4,1}^{(4)}\ , \\ \nonumber &&
	\mathbb{C}_{\left(r^2\right)^2,1}^{(4)} = \sfrac{735}{32}\mathbb{C}_{\delta^2,3}^{(4)} - \sfrac{105}{4} \mathbb{C}_{\delta^3,2}^{(4)} -\sfrac{49}{16} \mathbb{C}_{r^2\delta,2}^{(4)} + \sfrac{2315}{96} \mathbb{C}_{\delta^4,1}^{(4)} - \sfrac{49}{12} \mathbb{C}_{\delta r^3,1}^{(4)} \ , \\ \nonumber &&
	\mathbb{C}_{r^4,1}^{(4)} = \sfrac{735}{64} \mathbb{C}_{\delta^2,3}^{(4)} - \sfrac{343}{24} \mathbb{C}_{\delta^3,2}^{(4)} -\sfrac{49}{32} \mathbb{C}_{r^2\delta,2}^{(4)} + \sfrac{2827}{192} \mathbb{C}_{\delta^4,1}^{(4)} - \sfrac{17}{24} \mathbb{C}_{\delta r^3,1}^{(4)} \ , \\ \nonumber &&
	\mathbb{C}_{\theta,1}^{(4)} = \mathbb{C}_{\delta,1}^{(4)}\ , \quad
	\mathbb{C}_{\theta,2}^{(4)} = 2 \mathbb{C}_{\delta,2}^{(4)} - \mathbb{C}_{\delta^2,1}^{(4)}\ , \quad
	\mathbb{C}_{\theta,3}^{(4)} = 
	3 \mathbb{C}_{\delta,3}^{(4)} - \sfrac{3}{2} \mathbb{C}_{\delta^2,2}^{(4)} + \mathbb{C}_{\delta^3,1}^{(4)}\ , \\ \nonumber &&
	\mathbb{C}_{\theta,4}^{(4)} = 
	4 \mathbb{C}_{\delta,4}^{(4)} - 2 \mathbb{C}_{\delta^2,3}^{(4)} + \sfrac{4}{3}\mathbb{C}_{\delta^3,2}^{(4)} - 
	\mathbb{C}_{\delta^4,1}^{(4)}\ , \quad
	\mathbb{C}_{\delta\theta,1}^{(4)} = \mathbb{C}_{\delta^2,1}^{(4)}\ ,\\ \nonumber &&
	\mathbb{C}_{\delta\theta,2}^{(4)} =  \sfrac{3}{2}\mathbb{C}_{\delta^2,2}^{(4)} - \mathbb{C}_{\delta^3,1}^{(4)}\ , \quad
	\mathbb{C}_{\delta\theta,3}^{(4)} = 
	2 \mathbb{C}_{\delta^2,3}^{(4)} -  \sfrac{4}{3}\mathbb{C}_{\delta^3,2}^{(4)} + \mathbb{C}_{\delta^4,1}^{(4)}\ , \quad
	\mathbb{C}_{\theta^2,1}^{(4)} = \mathbb{C}_{\delta^2,1}^{(4)}\ , \\ \nonumber &&
	\mathbb{C}_{\theta^2,2}^{(4)} = 2 \mathbb{C}_{\delta^2,2}^{(4)} - 2 \mathbb{C}_{\delta^3,1}^{(4)}\ ,\quad
	\mathbb{C}_{\theta^2,3}^{(4)} = \sfrac{39}{8} \mathbb{C}_{\delta^2,3}^{(4)} - 5 \mathbb{C}_{\delta^3,2}^{(4)} -\sfrac{1}{4} \mathbb{C}_{r^2\delta,2}^{(4)} + \sfrac{107}{24} \mathbb{C}_{\delta^4,1}^{(4)} - \sfrac{1}{3}\mathbb{C}_{\delta r^3,1}^{(4)} \ , \\ \nonumber && 
	\mathbb{C}_{rp,1}^{(4)} = \sfrac{7}{2}  \mathbb{C}_{\delta,2}^{(4)}- \sfrac{5}{2} \mathbb{C}_{\delta^2,1}^{(4)}\ , \quad
	\mathbb{C}_{rp,2}^{(4)} =  \sfrac{9}{2}\mathbb{C}_{\delta,3}^{(4)} - \sfrac{21}{8} \mathbb{C}_{\delta^2,2}^{(4)} + \sfrac{1}{2} \mathbb{C}_{r^2,2}^{(4)}+ \sfrac{7}{4}\mathbb{C}_{\delta^3,1}^{(4)}\ , \\ \nonumber &&
	\mathbb{C}_{rp,3}^{(4)} = \sfrac{55}{2} \mathbb{C}_{\delta,4}^{(4)} - \sfrac{387}{16}\mathbb{C}_{\delta^2,3}^{(4)} + \sfrac{643}{30} \mathbb{C}_{\delta^3,2}^{(4)} -  \sfrac{2}{5}\mathbb{C}_{r^3,2}^{(4)} + \sfrac{49}{40}\mathbb{C}_{r^2\delta,2}^{(4)} - \sfrac{4627}{240}\mathbb{C}_{\delta^4,1}^{(4)} + \sfrac{17}{30} \mathbb{C}_{\delta r^3,1}^{(4)} \ , \\ \nonumber &&
	\mathbb{C}_{p^2,1}^{(4)} = \sfrac{7}{2} \mathbb{C}_{\delta,2}^{(4)} - \sfrac{5}{2} \mathbb{C}_{\delta^2,1}^{(4)}\ , \quad	\mathbb{C}_{p^2\theta,1}^{(4)} = \sfrac{7}{4} \mathbb{C}_{\delta^2,2}^{(4)} - \sfrac{5}{2} \mathbb{C}_{\delta^3,1}^{(4)}\ , \\ \nonumber &&
	\mathbb{C}_{p^2,2}^{(4)} = 
	9 \mathbb{C}_{\delta,3}^{(4)} - \sfrac{21}{4} \mathbb{C}_{\delta^2,2}^{(4)} + \sfrac{7}{2}\mathbb{C}_{\delta^3,1}^{(4)}, \quad
	\mathbb{C}_{p^2,3}^{(4)} = \sfrac{33}{2} \mathbb{C}_{\delta,4}^{(4)} - 9 \mathbb{C}_{\delta^2,3}^{(4)} + 
	6 \mathbb{C}_{\delta^3,2}^{(4)} - \sfrac{9}{2} \mathbb{C}_{\delta^4,1}^{(4)}\ , \\ \nonumber &&
	\mathbb{C}_{\delta^2\theta,1}^{(4)} = \mathbb{C}_{\delta^3,1}^{(4)}\ , \quad
	\mathbb{C}_{\delta^2\theta,2}^{(4)} = \sfrac{4}{3} \mathbb{C}_{\delta^3,2}^{(4)} - \mathbb{C}_{\delta^4,1}^{(4)}\ ,\quad
	\mathbb{C}_{\delta\theta^2,1}^{(4)} = \mathbb{C}_{\delta^3,1}^{(4)}\ , \\ \nonumber &&
	\mathbb{C}_{\delta\theta^2,2}^{(4)} = \sfrac{5}{3}\mathbb{C}_{\delta^3,2}^{(4)} - 2 \mathbb{C}_{\delta^4,1}^{(4)}\ , \quad
		\mathbb{C}_{\theta^3,1}^{(4)} = \mathbb{C}_{\delta^3,1}^{(4)}\ , \quad
	\mathbb{C}_{\theta^3,2}^{(4)} = 2 \mathbb{C}_{\delta^3,2}^{(4)} - 3 \mathbb{C}_{\delta^4,1}^{(4)}\ , \\ \nonumber &&
	\mathbb{C}_{pr^2,1}^{(4)} = \sfrac{45}{4}\mathbb{C}_{\delta,3}^{(4)} -  \sfrac{105}{16}\mathbb{C}_{\delta^2,2}^{(4)} - 
	\sfrac{3}{4}\mathbb{C}_{r^2,2}^{(4)} +  \sfrac{35}{8}\mathbb{C}_{\delta^3,1}^{(4)}\ , \\ \nonumber &&
	\mathbb{C}_{pr^2,2}^{(4)} =  \sfrac{147}{32}\mathbb{C}_{\delta^2,3}^{(4)} - \sfrac{343}{60} \mathbb{C}_{\delta^3,2}^{(4)} + \sfrac{13}{15} \mathbb{C}_{r^3,2}^{(4)} - \sfrac{49}{80} \mathbb{C}_{r^2\delta,2}^{(4)} + 
	\sfrac{2827}{480} \mathbb{C}_{\delta^4,1}^{(4)} - \sfrac{17}{60} \mathbb{C}_{\delta r^3,1}^{(4)} \ , \\ \nonumber &&
	\mathbb{C}_{rp^2,1}^{(4)} = \sfrac{45}{4} \mathbb{C}_{\delta,3}^{(4)} - \sfrac{105}{16} \mathbb{C}_{\delta^2,2}^{(4)} - \sfrac{3}{4}\mathbb{C}_{r^2,2}^{(4)} + \sfrac{35}{8} \mathbb{C}_{\delta^3,1}^{(4)}\ , \\ \nonumber &&
	\mathbb{C}_{rp^2,2}^{(4)} = \sfrac{147}{16} \mathbb{C}_{\delta^2,3}^{(4)} - \sfrac{343}{30} \mathbb{C}_{\delta^3,2}^{(4)} + 
	\sfrac{11}{15} \mathbb{C}_{r^3,2}^{(4)} - \sfrac{49}{40} \mathbb{C}_{r^2\delta,2}^{(4)} + \sfrac{2827}{240} \mathbb{C}_{\delta^4,1}^{(4)} - \sfrac{17}{30}\mathbb{C}_{\delta r^3,1}^{(4)} \ , \\ \nonumber &&
	\mathbb{C}_{p^3,1}^{(4)} =  \sfrac{45}{4}\mathbb{C}_{\delta,3}^{(4)} - \sfrac{105}{16}\mathbb{C}_{\delta^2,2}^{(4)} - \sfrac{3}{4} \mathbb{C}_{r^2,2}^{(4)} + \sfrac{35}{8} \mathbb{C}_{\delta^3,1}^{(4)}\ , \\ \nonumber &&
	\mathbb{C}_{p^3,2}^{(4)} = \sfrac{441}{32} \mathbb{C}_{\delta^2,3}^{(4)} - \sfrac{343}{20}  \mathbb{C}_{\delta^3,2}^{(4)}+ \sfrac{3}{5}\mathbb{C}_{r^3,2}^{(4)} - \sfrac{147}{80}  \mathbb{C}_{r^2\delta,2}^{(4)}+ \sfrac{2827}{160} \mathbb{C}_{\delta^4,1}^{(4)} - \sfrac{17}{20} \mathbb{C}_{\delta r^3,1}^{(4)} \ , \\ \nonumber &&
	\mathbb{C}_{p^2\delta,1}^{(4)} = \sfrac{7}{4} \mathbb{C}_{\delta^2,2}^{(4)} - \sfrac{5}{2} \mathbb{C}_{\delta^3,1}^{(4)}\ , \quad
	\mathbb{C}_{rp\delta,1}^{(4)} = \sfrac{7}{4}\mathbb{C}_{\delta^2,2}^{(4)} - \sfrac{5}{2} \mathbb{C}_{\delta^3,1}^{(4)}\ , \\ \nonumber &&
	\mathbb{C}_{r^2\theta,1}^{(4)} = \sfrac{7}{4}  \mathbb{C}_{\delta^2,2}^{(4)}- \sfrac{5}{2} \mathbb{C}_{\delta^3,1}^{(4)}\ , \quad
	\mathbb{C}_{rp\theta,1}^{(4)} = \sfrac{7}{4}  \mathbb{C}_{\delta^2,2}^{(4)}- \sfrac{5}{2} \mathbb{C}_{\delta^3,1}^{(4)}\ , \\ \nonumber &&
	\mathbb{C}_{rp\delta,2}^{(4)} = \sfrac{21}{16} \mathbb{C}_{\delta^2,3}^{(4)} - \sfrac{4}{3} \mathbb{C}_{\delta^3,2}^{(4)} + 
	\sfrac{5}{8} \mathbb{C}_{r^2\delta,2}^{(4)} + \sfrac{49}{48}  \mathbb{C}_{\delta^4,1}^{(4)}+ \sfrac{1}{6}\mathbb{C}_{\delta r^3,1}^{(4)}\ , \\ \nonumber &&
		\mathbb{C}_{p^2\delta,2}^{(4)} = \sfrac{21}{8} \mathbb{C}_{\delta^2,3}^{(4)} - \sfrac{8}{3} \mathbb{C}_{\delta^3,2}^{(4)} +\sfrac{1}{4} \mathbb{C}_{r^2\delta,2}^{(4)} + \sfrac{49}{24}\mathbb{C}_{\delta^4,1}^{(4)} +\sfrac{1}{3} \mathbb{C}_{\delta r^3,1}^{(4)} \ , \\ \nonumber &&
	\mathbb{C}_{r^2\theta,2}^{(4)} = \sfrac{105}{16} \mathbb{C}_{\delta^2,3}^{(4)}- \sfrac{47}{6} \mathbb{C}_{\delta^3,2}^{(4)} + \sfrac{1}{8}
	\mathbb{C}_{r^2\delta,2}^{(4)} + \sfrac{365}{48} \mathbb{C}_{\delta^4,1}^{(4)} - \sfrac{7}{6} \mathbb{C}_{\delta r^3,1}^{(4)} \ , \\ \nonumber &&
		\mathbb{C}_{rp\theta,2}^{(4)} = \sfrac{63}{8}  \mathbb{C}_{\delta^2,3}^{(4)}- \sfrac{55}{6} \mathbb{C}_{\delta^3,2}^{(4)} -\sfrac{1}{4}\mathbb{C}_{r^2\delta,2}^{(4)} + \sfrac{69}{8} \mathbb{C}_{\delta^4,1}^{(4)} - \mathbb{C}_{\delta r^3,1}^{(4)} \ , \\ \nonumber &&
		\mathbb{C}_{p^2\theta,2}^{(4)} = \sfrac{147}{16} \mathbb{C}_{\delta^2,3}^{(4)} - \sfrac{21}{2} \mathbb{C}_{\delta^3,2}^{(4)} - \sfrac{5}{8} \mathbb{C}_{r^2\delta,2}^{(4)} + \sfrac{463}{48} \mathbb{C}_{\delta^4,1}^{(4)} - \sfrac{5}{6}\mathbb{C}_{\delta r^3,1}^{(4)} 	 \ . 
	\eea
		\endgroup

 The local-in-time limit is obtained by setting $c_{\mathcal{O}_m } ( t , t') = c_{\mathcal{O}_m} ( t ) \delta_D ( t - t' )  / H(t)$ in \eqn{eq:param}.  Using that, along with \eqn{opt} with $t = t'$, in \eqn{fin}, we have the local-in-time expansion
 \be
 \delta_{h, \text{ loc}}^{(n)} ( \xvec , t) = \sum_{\mathcal{O}_m} c_{\mathcal{O}_m} ( t ) \mathcal{O}_m^{(n)} ( \xvec , t) \ , 
 \ee
where $\mathcal{O}_m^{(n)} ( \xvec , t)$ is the normal expression for the operator $\mathcal{O}_m$ at $n$-th order in perturbations.  For $n=4$, we find that a basis for all of the $\mathcal{O}_m^{(4)} ( \xvec , t)$ is $ \vec B_{\rm loc} = \{ \mathcal{O}_m^{(4)}   \}$ where $\mathcal{O}_m$ is given by
\be
\{ p^2 , r^2 , r^2 r^2 , r^2 p , r^3 , r^4 , r p , \delta , r^2 \delta , r^3 \delta , rp \delta , \delta^2 , r^2 \delta^2 , \delta^3 , \theta \} \ .
\ee
Then, calling $\vec B$ the basis given in \eqn{cker4basis}, we find that
\be
\vec B_{\rm loc} = M \cdot \vec B \ ,
\ee
where the change of basis matrix $M$ is given by  	
\be
M =\left(
\begin{array}{ccccccccccccccc}
 0 & \frac{7}{2} & 9 & \frac{33}{2} & -\frac{5}{2} & -\frac{21}{4} & -9 & 0 & 0 & \frac{7}{2} & 6 & 0 & 0 & -\frac{9}{2} & 0 \\
 0 & \frac{7}{2} & 0 & 0 & -\frac{5}{2} & 0 & 0 & 1 & 1 & 0 & 0 & 0 & 0 & 0 & 0 \\
 0 & 0 & 0 & 0 & 0 & 0 & \frac{735}{32} & 0 & 0 & 0 & -\frac{105}{4} & 0 & -\frac{49}{16} & \frac{2315}{96} & -\frac{49}{12} \\
 0 & 0 & \frac{45}{4} & 0 & 0 & -\frac{105}{16} & \frac{147}{32} & -\frac{3}{4} & 0 & \frac{35}{8} & -\frac{343}{60} & \frac{13}{15} & -\frac{49}{80} &
   \frac{2827}{480} & -\frac{17}{60} \\
 0 & 0 & \frac{45}{4} & 0 & 0 & -\frac{105}{16} & 0 & -\frac{3}{4} & 0 & \frac{35}{8} & 0 & 1 & 0 & 0 & 0 \\
 0 & 0 & 0 & 0 & 0 & 0 & \frac{735}{64} & 0 & 0 & 0 & -\frac{343}{24} & 0 & -\frac{49}{32} & \frac{2827}{192} & -\frac{17}{24} \\
 0 & \frac{7}{2} & \frac{9}{2} & \frac{55}{2} & -\frac{5}{2} & -\frac{21}{8} & -\frac{387}{16} & \frac{1}{2} & 0 & \frac{7}{4} & \frac{643}{30} & -\frac{2}{5} &
   \frac{49}{40} & -\frac{4627}{240} & \frac{17}{30} \\
 1 & 1 & 1 & 1 & 0 & 0 & 0 & 0 & 0 & 0 & 0 & 0 & 0 & 0 & 0 \\
 0 & 0 & 0 & 0 & 0 & \frac{7}{4} & 0 & 0 & 0 & -\frac{5}{2} & 0 & 0 & 1 & 0 & 0 \\
 0 & 0 & 0 & 0 & 0 & 0 & 0 & 0 & 0 & 0 & 0 & 0 & 0 & 0 & 1 \\
 0 & 0 & 0 & 0 & 0 & \frac{7}{4} & \frac{21}{16} & 0 & 0 & -\frac{5}{2} & -\frac{4}{3} & 0 & \frac{5}{8} & \frac{49}{48} & \frac{1}{6} \\
 0 & 0 & 0 & 0 & 1 & 1 & 1 & 0 & 0 & 0 & 0 & 0 & 0 & 0 & 0 \\
 0 & 0 & 0 & 0 & 0 & 0 & 0 & 0 & 0 & 0 & \frac{7}{6} & 0 & 0 & -\frac{5}{2} & 0 \\
 0 & 0 & 0 & 0 & 0 & 0 & 0 & 0 & 0 & 1 & 1 & 0 & 0 & 0 & 0 \\
 1 & 2 & 3 & 4 & -1 & -\frac{3}{2} & -2 & 0 & 0 & 1 & \frac{4}{3} & 0 & 0 & -1 & 0 \\
\end{array}
\right) \ .
\ee	
Since $\det M \neq 0$, this means that $M$ is invertible and thus the local-in-time basis is equivalent to the non-local-in-time basis.

{In \cite{Senatore:2014eva}, the formalism to construct a complete bias basis was originally presented.  Then, other procedures, which are expected to be equivalent, were developed and carried out to fourth order.}  For reference, we provide the explicit transformation connecting our real-space fourth-order basis to the one presented in \cite{Desjacques:2016bnm}, and the connection between the basis used in \cite{Desjacques:2016bnm} and the basis used in \cite{Eggemeier:2018qae} can be found in the latter reference.  The basis elements at fourth order in \cite{Desjacques:2016bnm} are given by the fourth-order terms in the perturbative expansion of the following operators
\begin{align}
\begin{split}
\vec B' =  \Big\{  & \tr [ \Pi^{[1]} ] , \tr [ (\Pi^{[1]} )^2 ]  , ( \tr [ \Pi^{[1]} ] )^2 ,  \tr [ (\Pi^{[1]})^3 ] \  , \tr[( \Pi^{[1]})^2 ] \tr [ \Pi^{[1]} ]  , ( \tr [ \Pi^{[1]} ] )^3 , \tr [ \Pi^{[1]} \Pi^{[2]}]  , \\
& \tr [( \Pi^{[1]})^4 ] , \tr [ (\Pi^{[1]})^3 ] \tr[ \Pi^{[1]}] , \left( \tr [ ( \Pi^{[1]} )^2 ] \right)^2 , ( \tr [ \Pi^{[1]}] )^4 , \\
&  \tr [ \Pi^{[1]} ] \tr [ \Pi^{[1]} \Pi^{[2]} ] , \tr [ \Pi^{[1]} \Pi^{[1]} \Pi^{[2]} ] , \tr [ \Pi^{[1]} \Pi^{[3]} ] , \tr[ \Pi^{[2]} \Pi^{[2]} ] \Big\} \ , 
\end{split}
\end{align}
and the relevant notation is defined in the same reference.  Then, defining the change of basis matrix $M'$ by  $\vec B '  = M ' \cdot \vec B$, we have
\be
M' = \left(
\begin{array}{ccccccccccccccc}
 1 & 1 & 1 & 1 & 0 & 0 & 0 & 0 & 0 & 0 & 0 & 0 & 0 & 0 & 0 \\
 0 & \frac{7}{2} & 0 & 0 & -\frac{5}{2} & 0 & 0 & 1 & 1 & 0 & 0 & 0 & 0 & 0 & 0 \\
 0 & 0 & 0 & 0 & 1 & 1 & 1 & 0 & 0 & 0 & 0 & 0 & 0 & 0 & 0 \\
 0 & 0 & \frac{45}{4} & 0 & 0 & -\frac{105}{16} & 0 & -\frac{3}{4} & 0 & \frac{35}{8} & 0 & 1 & 0 & 0 & 0 \\
 0 & 0 & 0 & 0 & 0 & \frac{7}{4} & 0 & 0 & 0 & -\frac{5}{2} & 0 & 0 & 1 & 0 & 0 \\
 0 & 0 & 0 & 0 & 0 & 0 & 0 & 0 & 0 & 1 & 1 & 0 & 0 & 0 & 0 \\
 0 & 0 & 0 & 0 & 0 & 0 & 0 & \frac{1}{2} & 1 & 0 & 0 & 0 & 0 & 0 & 0 \\
 0 & 0 & 0 & 0 & 0 & 0 & \frac{735}{64} & 0 & 0 & 0 & -\frac{343}{24} & 0 & -\frac{49}{32} & \frac{2827}{192} & -\frac{17}{24} \\
 0 & 0 & 0 & 0 & 0 & 0 & 0 & 0 & 0 & 0 & 0 & 0 & 0 & 0 & 1 \\
 0 & 0 & 0 & 0 & 0 & 0 & \frac{735}{32} & 0 & 0 & 0 & -\frac{105}{4} & 0 & -\frac{49}{16} & \frac{2315}{96} & -\frac{49}{12} \\
 0 & 0 & 0 & 0 & 0 & 0 & 0 & 0 & 0 & 0 & 0 & 0 & 0 & 1 & 0 \\
 0 & 0 & 0 & 0 & 0 & 0 & -\frac{105}{32} & 0 & 0 & 0 & \frac{10}{3} & 0 & \frac{15}{16} & -\frac{245}{96} & \frac{7}{12} \\
 0 & 0 & 0 & 0 & 0 & 0 & 0 & 0 & 0 & 0 & 0 & \frac{1}{3} & 0 & 0 & 0 \\
 0 & 0 & 0 & \frac{1925}{16} & 0 & 0 & -\frac{15015}{128} & 0 & -\frac{21}{8} & 0 & \frac{1729}{16} & -\frac{17}{6} & \frac{441}{64} &
   -\frac{12681}{128} & \frac{51}{16} \\
 0 & 0 & 0 & -\frac{1925}{8} & 0 & 0 & \frac{15015}{64} & 0 & \frac{25}{4} & 0 & -\frac{1729}{8} & \frac{17}{3} & -\frac{441}{32} & \frac{12681}{64} &
   -\frac{51}{8} \\
\end{array}
\right) \ . 
\ee
Since $\det M' \neq 0$, this shows that the two bases are equivalent.

\subsection{Explicit expressions for fourth order kernels}\label{explicit}

We here give the final $\mathbb{C}_{\mathcal{O},i}$ used in \eqn{finalbias} that are linearly independent, as Fourier space kernels. We use the notation 
\bea
	\mathbb{C}^{(n)}_{\mathcal{O},i}(\vk, t )= D ( t )^n \int_{\vq_1,...,\vq_n}^{\kvec}  K^{\mathcal{O},i}_{n}(\vq_1,..., \vq_n) \tilde \delta^{(1)}_{\vq_1} \cdots \tilde \delta^{(1)}_{\vq_n} \ . 
	\eea
Dropping the $(\vq_1 , \vq_2, \vq_3,\vq_4)$ dependence on the left-hand sides to avoid clutter, we explicitly have 
\begingroup
\allowdisplaybreaks
{\begin{align}
K_{4}^{\delta ,1}= & \frac{\left(\vq_1\cdot\left(\vq_3+\vq_4\right) \vq_2\cdot\left(\vq_3+\vq_4\right)+\vq_1\cdot\vq_2 \left(3 \vq_1+2 \vq_2\right)\cdot\left(\vq_3+\vq_4\right)\right) G_2\left(\vq_3,\vq_4\right)}{6 q_2^2
   \left(\vq_3+\vq_4\right){}^2}\\ \nonumber
   &+\frac{\vq_1\cdot\left(\vq_2+\vq_3+\vq_4\right) G_3\left(\vq_2,\vq_3,\vq_4\right)}{3 \left(\vq_2+\vq_3+\vq_4\right){}^2} +\frac{\vq_1\cdot\vq_2 \left(\vq_2\cdot\vq_3 \vq_4\cdot\left(\vq_2+\vq_3\right)+\vq_1\cdot\vq_3
   \vq_4\cdot\left(\vq_1+3 \vq_3\right)\right)}{6 q_2^2 q_3^2 q_4^2}\ , \\ \nonumber
   K_{4}^{\delta ,2} =&\frac{\left(\vq_1+\vq_2\right)\cdot\left(\vq_3+\vq_4\right) \left(q_2^2 F_2\left(\vq_1,\vq_2\right)-\vq_1\cdot\vq_2\right) G_2\left(\vq_3,\vq_4\right)}{2 q_2^2
   \left(\vq_3+\vq_4\right){}^2}  + \frac{\vq_3\cdot\left(\vq_1+\vq_2\right) \vq_4\cdot\left(\vq_1+\vq_2+\vq_3\right) F_2\left(\vq_1,\vq_2\right)}{2 q_3^2 q_4^2} \\ \nonumber 
   &-\frac{\vq_1\cdot\vq_2 \left(\vq_2\cdot\vq_3
   \vq_4\cdot\left(\vq_2+\vq_3\right)+\vq_1\cdot\vq_3 \vq_4\cdot\left(\vq_1+3 \vq_3\right)\right)}{2 q_2^2 q_3^2 q_4^2}\ , \\ \nonumber
K_{4}^{\delta ,3}=&\frac{\vq_4\cdot\left(\vq_1+\vq_2+\vq_3\right)F_3\left(\vq_1,\vq_2,\vq_3\right)}{q_4^2} +\frac{ \vq_1\cdot\vq_2 \left(\vq_2\cdot\vq_3 \vq_4\cdot\left(\vq_2+\vq_3\right)+\vq_1\cdot\vq_3 \vq_4\cdot\left(\vq_1+3 \vq_3\right)\right)}{2 q_2^2 q_3^2
   q_4^2}\\ \nonumber
   &- \frac{\vq_4\cdot\left(\vq_1+\vq_2+\vq_3\right)\vq_3\cdot\left(\vq_1+\vq_2\right) }{q_4^2}  \left(\frac{ F_2\left(\vq_1,\vq_2\right)}{q_3^2}+\frac{
   G_2\left(\vq_1,\vq_2\right)}{2 \left(\vq_1+\vq_2\right){}^2}\right)\ , \\ \nonumber
  K_{4}^{\delta ,4}=& F_4\left(\vq_1,\vq_2,\vq_3,\vq_4\right)-\sum_{i=1}^3 K_{4}^{\delta ,i}\left(\vq_1,\vq_2,\vq_3,\vq_4\right)\ , \\ \nonumber
  K_{4}^{\delta^2 ,1}=& \frac{\vq_2\cdot\left(\vq_3+\vq_4\right) G_2\left(\vq_3,\vq_4\right)}{\left(\vq_3+\vq_4\right){}^2}+\frac{
    \vq_1\cdot\vq_3 \vq_2\cdot\vq_4+\vq_2\cdot\vq_3 \vq_4\cdot\left(\vq_2+\vq_3\right)}{q_3^2 q_4^2} \ , \\ \nonumber
    K_{4}^{\delta^2 ,2}=& \frac{2 \vq_4\cdot\left(\vq_1+\vq_2+\vq_3\right) F_2\left(\vq_1,\vq_2\right)}{q_4^2}-\frac{2 \left(\vq_1\cdot\vq_3 \vq_2\cdot\vq_4+\vq_2\cdot\vq_3 \vq_4\cdot\left(\vq_2+\vq_3\right)\right)}{q_3^2 q_4^2}\ , \\ \nonumber
    K_{4}^{\delta^2 ,3}=&2 F_3\left(\vq_1,\vq_2,\vq_3\right)+F_2\left(\vq_1,\vq_2\right)F_2\left(\vq_3,\vq_4\right)-\sum_{i=1}^2 K_{4}^{\delta^2 ,i}\left(\vq_1,\vq_2,\vq_3,\vq_4\right)\ , \\ \nonumber
    K_{4}^{r^2 ,2}=&  \frac{2 \left(\vq_3\cdot\left(\vq_1+\vq_2\right)\right){}^2 \vq_4\cdot\left(\vq_1+\vq_2+\vq_3\right) F_2\left(\vq_1,\vq_2\right)}{\left(\vq_1+\vq_2\right){}^2 q_3^2 q_4^2}  -\frac{2 \left(\vq_1\cdot\vq_2\right){}^2
   \left(\vq_1\cdot\vq_3 \vq_2\cdot\vq_4+\vq_2\cdot\vq_3 \vq_4\cdot\left(\vq_2+\vq_3\right)\right)}{q_1^2 q_2^2 q_3^2 q_4^2}\ , \\ \nonumber
    K_{4}^{r^2 ,3}=&   \frac{\left(\left(\vq_1+\vq_2\right)\cdot\left(\vq_3+\vq_4\right)\right){}^2 F_2\left(\vq_1,\vq_2\right)F_2\left(\vq_3,\vq_4\right)}{\left(\vq_1+\vq_2\right){}^2\left(\vq_3+\vq_4\right){}^2} - \frac{2F_2\left(\vq_1,\vq_2\right)
   \left(\vq_3\cdot\left(\vq_1+\vq_2\right)\right){}^2 \vq_4\cdot\left(\vq_1+\vq_2+\vq_3\right)}{\left(\vq_1+\vq_2\right){}^2 q_3^2 q_4^2} \\ \nonumber
   &+\frac{2 \left(\vq_4\cdot\left(\vq_1+\vq_2+\vq_3\right)\right){}^2
   F_3\left(\vq_1,\vq_2,\vq_3\right)}{\left(\vq_1+\vq_2+\vq_3\right){}^2 q_4^2} -\frac{\vq_3\cdot\left(\vq_1+\vq_2\right) \left(\vq_3\cdot\vq_4\right){}^2 G_2\left(\vq_1,\vq_2\right)}{\left(\vq_1+\vq_2\right){}^2 q_3^2
   q_4^2}\\ \nonumber
   &+\frac{\left(\vq_1\cdot\vq_3 \vq_2\cdot\vq_4+\vq_2\cdot\vq_3 \left(\vq_2\cdot\vq_4+\vq_3\cdot\vq_4\right)\right) \left(\vq_1\cdot\vq_2\right){}^2}{q_1^2 q_2^2 q_3^2 q_4^2}  \ , \\ \nonumber
   K_{4}^{\delta^3 ,1} =&\frac{3 \vq_4\cdot \vq_3}{q_4^2} \ ,\quad
	K_{4}^{\delta^3 ,2}=3 F_2\left(\vq_1,\vq_2\right)-\frac{3 \vq_4\cdot \vq_3}{q_4^2} \ , \\ \nonumber
	K_{4}^{r^3 ,2}=& \frac{3 \vq_3\cdot\left(\vq_1+\vq_2\right) \vq_3\cdot\vq_4 \vq_4\cdot\left(\vq_1+\vq_2\right) F_2\left(\vq_1,\vq_2\right)}{\left(\vq_1+\vq_2\right){}^2 q_3^2 q_4^2} -\frac{3 \vq_1\cdot\vq_2 \vq_1\cdot\vq_3 \vq_2\cdot\vq_3 \vq_3\cdot\vq_4}{q_1^2 q_2^2
   q_3^2 q_4^2}\ , \\ \nonumber
   K_{4}^{r^2\delta ,2}=& \frac{2 \left(\vq_3\cdot\left(\vq_1+\vq_2\right)\right){}^2 F_2\left(\vq_1,\vq_2\right)}{\left(\vq_1+\vq_2\right){}^2 q_3^2}+\frac{
    \left(\vq_3\cdot\vq_4\right){}^2 F_2\left(\vq_1,\vq_2\right)}{q_3^2
   q_4^2} -\frac{\vq_3\cdot\vq_4 \left(2 \left(\vq_2\cdot\vq_3\right){}^2 q_1^2+\left(\vq_1\cdot\vq_2\right){}^2 q_3^2\right)}{q_1^2 q_2^2 q_3^2 q_4^2}\ , \\ \nonumber
   K_{4}^{\delta^4 ,1}=&1 \ , \quad	K_{4}^{\delta r^3 ,1} = \frac{\vq_1\cdot \vq_2 \vq_2\cdot \vq_3 \vq_3\cdot \vq_1}{q_1^2 q_2^2 q_3^2} \ , \nonumber
   \end{align}}   \endgroup 
\hspace{-.06in}The above, combined with \eqn{finalbias}, defines the final biased tracer kernels $K_n^{h}$ up to $n = 4$.  The conversion to redshift space to get $K_4^{r,h}(\vq_1,\vq_2,\vq_3,\vq_4;\hat{z})$ is then given by
\begingroup
\allowdisplaybreaks
{\begin{align}
	K_{4}^{r,h}  =& K_{4}^{h}(\vq_1,\vq_2,\vq_3,\vq_4)+f\mu^2 G_4(\vq_1,\vq_2,\vq_3,\vq_4) +k \mu f\frac{(\vq_1+\vq_2+\vq_3)\cdot \hat{z}}{(\vq_1+\vq_2+\vq_3)^2}G_3(\vq_1,\vq_2,\vq_3)K_{1}^{h}(\vq_4)\nonumber \\ 
	&+k\mu f\frac{(\vq_1+\vq_2)\cdot \hat{z}}{(\vq_1+\vq_2)^2}G_2(\vq_1,\vq_2)K_{2}^{h}(\vq_3,\vq_4)+k\mu f\frac{\vq_1\cdot \hat{z}}{q_1^2}K_{3}^{h}(\vq_2,\vq_3,\vq_4)\\ \nonumber
	&+k^2 \mu^2f^2\frac{(\vq_1+\vq_2+\vq_3)\cdot \hat{z}}{(\vq_1+\vq_2+\vq_3)^2}\frac{\vq_4\cdot \hat{z}}{q_4^2}G_3(\vq_1,\vq_2,\vq_3) +k^2 \mu^2f^2\frac{(\vq_1+\vq_2)\cdot \hat{z}}{(\vq_1+\vq_2)^2}\frac{\vq_3\cdot \hat{z}}{q_3^2}G_2(\vq_1,\vq_2)K_{1}^{h}(\vq_4) \\ \nonumber
	&+\frac{1}{2}k^2\mu^2f^2\frac{(\vq_1+\vq_2)\cdot \hat{z}}{(\vq_1+\vq_2)^2}\frac{(\vq_3+\vq_4)\cdot \hat{z}}{(\vq_3+\vq_4)^2}G_2(\vq_1,\vq_2)G_2(\vq_3,\vq_4) +\frac{1}{2}k^2 \mu^2f^2\frac{\vq_1\cdot \hat{z}}{q_1^2}\frac{\vq_2\cdot \hat{z}}{q_2^2}K_{2}^{h}(\vq_3,\vq_4) \\ \nonumber
	&+\frac{1}{2}k^3\mu^3f^3\frac{(\vq_1+\vq_2)\cdot \hat{z}}{(\vq_1+\vq_2)^2}\frac{\vq_3\cdot \hat{z}}{q_3^2}\frac{\vq_4\cdot \hat{z}}{q_4^2}G_2(\vq_1,\vq_2) +\frac{1}{6}k^3\mu^3f^3\frac{\vq_1\cdot \hat{z}}{q_1^2}\frac{\vq_2\cdot \hat{z}}{q_2^2}\frac{\vq_3\cdot \hat{z}}{q_3^2}K_{1}^{h}(\vq_4) \\ \nonumber
	&+\frac{1}{24}k^4\mu^4f^4\frac{\vq_1\cdot \hat{z}}{q_1^2}\frac{\vq_2\cdot \hat{z}}{q_2^2}\frac{\vq_3\cdot \hat{z}}{q_3^2}\frac{\vq_4\cdot \hat{z}}{q_4^2} \ , 
	\end{align}}
	\endgroup	
\hspace{-0.06in}where $\mu=\hat{k}\cdot\hat{z}$ and {$\kvec \equiv \qvec_1 + \qvec_2 + \qvec_3 + \qvec_4$}, and see \cite{Perko:2016puo} for the analogous redshift space expression up to third order.

%%%%%%%%%%%%%%%%%%
%
%
%

\section{Details for biased tracers in redshift space renormalization} \label{fullmatchingapp}

%%%%%%%%%%%
%
%

\subsection{Counterterm expressions for biased tracers in redshift space} \label{ctbtrssapp}

 For the response counterterms, we define the kernels $K_1^{r,h,ct}$ and $K_2^{r,h,ct}$ from 
\begin{align}
\begin{split} \label{biasctexp}
& \tilde \delta^{(1)}_{r,h,ct}(\kvec , \hat z  ) =   K_1^{r,h,ct} ( \kvec ; \hat z) \deltaone ( \kvec )  \ ,  \quad  \tilde \delta^{(2)}_{r,h,ct} ( \kvec , \hat z  ) =  \int_{\qvec_1 , \qvec_2}^{\kvec} K_2^{r,h,ct} ( \qvec_1 , \qvec_2 ; \hat z  ) \deltaone ( \qvec_1 ) \deltaone ( \qvec_2 ) \ ,
\end{split}
\end{align}
where the tilde fields are defined analogously to \eqn{cttimedeprss}, just with $r \rightarrow r,h$.  The response counterterms enter in 
\begin{align}
\begin{split}
& P_{13}^{r, h, ct} ( k , \hat k \cdot \hat z  )  \equiv  2 K_1^{r,h} ( \kvec ; \hat z)  K_1^{r,h, ct} ( -  \kvec ; \hat z ) P_{11} ( k ) \ , \\
& B_{411}^{r,h, ct}  \equiv 2 P_{11} ( k_1 ) P_{11} ( k_2 ) K_1^{r,h} ( \kvec_1 ; \hat z)  K_1^{r,h} ( \kvec_2 ;\hat  z)  K_2^{r,h, ct} ( - \kvec_1 , - \kvec_2 ; \hat z) + \text{ 2 perms.} \ , \\
& B_{321}^{r, h, (II),ct}  \equiv  2 P_{11} ( k_1 ) P_{11} ( k_2 ) K_{1}^{r,h, ct} ( \kvec_1 ; \hat z )  K_1^{r,h} ( \kvec_2 ; \hat z ) K^{r,h}_2 ( - \kvec_1 , - \kvec_2 ; \hat z )  + \text{ 5 perms.} \ ,
\end{split}
\end{align}
(we have suppressed the argument $(k_1, k_2, k_3, \hat k_1 \cdot \hat z , \hat k_2 \cdot \hat z )$ of the bispectra terms to remove clutter) so that the combinations
\begin{align}
\begin{split}
P_{13}^{r,h} ( k , \hat k \cdot \hat z ) & +P_{13}^{r,h , ct} ( k , \hat k \cdot \hat z )  \ , \\
B_{411}^{r,h}  ( k_1 , k_2 , k_3, \hat k_1 \cdot \hat z , \hat k_2 \cdot \hat z  ) & + B_{411}^{r , h, ct} ( k_1 , k_2 , k_3 , \hat k_1 \cdot \hat z , \hat k_2 \cdot \hat z ) \ , \\
B_{321}^{r,h, (II)} ( k_1 , k_2 , k_3 , \hat k_1 \cdot \hat z , \hat k_2 \cdot \hat z  ) &+ B_{321}^{r ,h,  (II),ct} ( k_1 , k_2 , k_3 , \hat k_1 \cdot \hat z , \hat k_2 \cdot \hat z   ) \ , 
\end{split}
\end{align}
are renormalized. 

For the stochastic terms, we write the first order solution as $  \delta^{(1)}_{r,h,\epsilon} ( \kvec , \hat z , a )  = D(a)^2  \tilde \delta^{(1)}_{r,h,\epsilon} ( \kvec , \hat z  )$ and the second order as
\begin{align}
\begin{split}
\tilde  \delta^{(2)}_{r,h,\epsilon} ( \kvec , \hat z) =  \int_{\qvec_1 , \qvec_2}^{\kvec} \delta_2^{r,h,\epsilon} ( \qvec_1 , \qvec_2; \hat z) \tilde \delta^{(1)}_{ \qvec_2 } \ ,
\end{split}
\end{align}
The term that renormalizes $P^{r,h}_{22}$ is
\be
P_{22}^{r,h,\epsilon} ( k  , \hat k \cdot \hat z )  \equiv \langle  \tilde \delta^{(1)}_{r,h,\epsilon} ( \kvec , \hat z  )   \tilde \delta^{(1)}_{r,h,\epsilon} ( \kvec ' , \hat z  )\rangle ' \ ,
\ee
the term that renormalizes $B^{r,h}_{222}$ is
\be
 B_{222}^{r,h,\epsilon} ( k_1 , k_2 , k_3 , \hat k_1 \cdot \hat z , \hat k_2 \cdot \hat z) = \langle  \tilde \delta^{(1)}_{r,h,\epsilon} ( \kvec_1 , \hat z  )     \tilde \delta^{(1)}_{r,h,\epsilon} ( \kvec_2 , \hat z  )     \tilde \delta^{(1)}_{r,h,\epsilon} ( \kvec_3 , \hat z  )    \rangle '  \ , 
\ee
and the term that renormalizes $B_{321}^{r,h,(I)}$ is defined in \eqn{b321epdef1} and \eqn{b321epdef2}.  
In this way, 
\begin{align}
\begin{split}
 P_{22}^{r,h} ( k , \hat k \cdot \hat z ) & + P_{22}^{r,h,\epsilon} ( k , \hat k \cdot \hat z )  \ , \\
 B_{222}^{r,h}  ( k_1 , k_2 , k_3 , \hat k_1 \cdot \hat z , \hat k_2 \cdot \hat z)   & +  B_{222}^{r,h,\epsilon} ( k_1 , k_2 , k_3 , \hat k_1 \cdot \hat z , \hat k_2 \cdot \hat z)  \ , \\
 B_{321}^{r,h,(I)}( k_1 , k_2 , k_3 , \hat k_1 \cdot \hat z , \hat k_2 \cdot \hat z)  & + B_{321}^{r,h, (I), \epsilon}  ( k_1 , k_2 , k_3 , \hat k_1 \cdot \hat z , \hat k_2 \cdot \hat z) \ , 
\end{split}
\end{align}
are renormalized.

\subsection{Response terms} \label{responsematchapp}

The functions that enter $K_2^{r,h,ct}$ in \eqn{k2rhct} are given by   
\begingroup
\allowdisplaybreaks
{\begin{align}
 \label{basisfns1}
e_1^{K_2} & = - \frac{ \kvec_1 \cdot \kvec_2  k_2^2 }{2\knl^2 k_1^2} +  \frac{f \kvec_3 \cdot \hat z \kvec_1 \cdot \hat z k_2^2}{2 \knl^2 k_1^2 }  + ( 1 \leftrightarrow 2)    \ , \quad e_2^{K_2} = - \frac{ k_3^2}{\knl^2}  F_2 ( \kvec_1 , \kvec_2 ) + \frac{\kvec_1 \cdot \kvec_2}{2 \knl^2 }\left( \frac{k_1^2}{k_2^2} + \frac{k_2^2}{k_1^2}   \right)  \nonumber  \\
   e_3^{K_2} & = - \frac{k_3^2}{\knl^2}  \ , \quad     e_4^{K_2}  = - \frac{(\kvec_1 \cdot \kvec_2)^2 k_3^2}{\knl^2 k_1^2 k_2^2}  \ , \quad           e_5^{K_2}  =   - \frac{\kvec_1 \cdot \kvec_2}{\knl^2} \nonumber  \\
   e_6^{K_2} & =  - \frac{f \kvec_3 \cdot \hat z \kvec_1 \cdot \hat z \kvec_1 \cdot \kvec_2}{2 \knl^2 k_2^2} + \frac{f^2 (\kvec_3 \cdot \hat z)^2 \kvec_2 \cdot \hat z \kvec_1 \cdot \hat z}{2 \knl^2 k_2^2}  + ( 1 \leftrightarrow 2) \ ,  \nonumber  \\
     e_7^{K_2} &  = \frac{f (\kvec_3 \cdot \hat z)^2 \kvec_1 \cdot \kvec_2 \kvec_2 \cdot \kvec_3 \kvec_3 \cdot \kvec_1}{ \knl^2 k_1^2 k_2^2 k_3^2 } \\
        e_8^{K_2} & =  - \frac{f^2 (\kvec_3 \cdot \hat z)^2 \kvec_1 \cdot \kvec_2 (\kvec_1 \cdot \hat z)^2}{4 \knl^2 k_1^2 k_2^2} + \frac{f^3 ( \kvec_3 \cdot \hat z)^3 \kvec_1 \cdot \hat z (\kvec_2 \cdot \hat z)^2}{4 \knl^2 k_1^2 k_2^2 }  + ( 1 \leftrightarrow 2) \nonumber \\
        e_9^{K_2} & = - \frac{f^2  (\kvec_3 \cdot \hat z)^2}{2 \knl^2} \left( \frac{ (\kvec_3 \cdot \hat z)^2}{k_3^2} F_2 (\kvec_1 , \kvec_2 ) - \frac{\kvec_1 \cdot \kvec_2 [ (\kvec_1 \cdot \hat z)^2 + (\kvec_2 \cdot \hat z)^2  ]}{2 k_1^2 k_2^2}  \right) \nonumber  \\
e_{10}^{K_2} & =   - \frac{f^2 (\kvec_3 \cdot \hat z)^2 \kvec_1 \cdot \kvec_2 }{4 \knl^2 k_2^2} + \frac{f^3 (\kvec_3 \cdot \hat z)^3 \kvec_2 \cdot \hat z}{4 \knl^2 k_2^2}  + ( 1 \leftrightarrow 2) \nonumber \\
   e_{11}^{K_2} & = - \frac{f^2 (\kvec_3 \cdot \hat z)^2}{2 \knl^2} \left( F_2 ( \kvec_1 , \kvec_2 ) - \frac{\kvec_1 \cdot \kvec_2}{2} \left( \frac{1}{k_1^2 } + \frac{1}{k_2^2} \right) \right)  \ , \quad  e_{12}^{K_2}  = - \frac{f^2 (\kvec_3 \cdot \hat z)^2}{4 \knl^2} \left( \frac{(\kvec_1 \cdot \hat z)^2}{k_1^2}  + \frac{(\kvec_2 \cdot \hat z)^2 }{k_2^2} \right)  \nonumber  \\
      e_{13}^{K_2} & =  - \frac{f^2 (\kvec_3 \cdot \hat z)^2 \kvec_1 \cdot \hat z \kvec_2 \cdot \hat z \kvec_1 \cdot \kvec_2}{2 \knl^2 k_1^2 k_2^2 }  \ , \quad e_{14}^{K_2}  =  - \frac{f^2 (\kvec_3 \cdot \hat z)^2}{2 \knl^2} \ , \nonumber 
\end{align}}
\endgroup
\hspace{-.06in}where we have defined $\kvec_3 \equiv - \kvec_1 - \kvec_2$.  {The new non-locally-contributing counterterm enters $K_2^{r,h,ct}$ through the term $e_7^{K_2}$.}

 The UV matching for the response terms is given by 
 \begingroup
\allowdisplaybreaks
{\begin{align}     \label{c5pimatching}
& c_1^{h}=\frac{ \sigma ^2 k_{\text{NL}}^2 \left(3 b_1-64 \left(b_3+15 b_8\right)\right) }{1260 \pi ^2} \ ,  \\
 & c_2^{h} = \frac{\sigma ^2 k_{\text{NL}}^2}{679140 \pi ^2}  \big(8 \big(-792 b_2-2926 b_3+1522 b_4+462 \left(6 b_7-95 b_8\right) \nonumber  \\
 & \hspace{1in} -6215 b_9+6930 b_{11}\big)-27335 b_1\big) \ , \nonumber \\
   & c_3^{h}  =\frac{ \sigma ^2 k_{\text{NL}}^2}{4753980 \pi ^2} \Big(194348 b_1+95205 b_2+160160 b_3-431040 b_4 -6529248 b_9 \nonumber  \\
   &\hspace{1in} +77 \left(147 b_5-5248 b_7+31200 b_8\right)  -2217600 b_{11} \Big)  \ ,\nonumber  \\
   & c_4^{h}  =\frac{\sigma
   ^2 k_{\text{NL}}^2}{1584660 \pi ^2} \big( 20559 b_1-2 \big(17677 b_2-13552 b_3+5664 b_4+7392 b_7 \nonumber  \\
   & \hspace{1in} -203280 b_8+55088 b_9+18480 b_{11}\big)  \big) \ ,  \nonumber \\
   & c_5^{h}  =- \frac{2 \sigma ^2
   k_{\text{NL}}^2}{169785 \pi ^2}  \big(6083 b_1+792 b_2+1078 b_3-6450 b_4-7700 b_7 \nonumber \\
   & \hspace{1in} +16170 b_8-35673 b_9-62370 b_{11}\big)  \ , \nonumber \\
        & c_1^{\pi}=-\frac{\sigma ^2 k_{\text{NL}}^2\left(4725 b_1+32 \left(36 b_2+35 \left(b_3+15 b_8\right)\right)\right) }{30870 \pi ^2} \ , \nonumber \\
   & c_5^{\pi} =\frac{668 \, \sigma ^2 k_{\text{NL}}^2}{56595 \pi ^2}   \ , \nonumber  \\
& c_1^{\pi v} =-\frac{  \sigma^2 k_{\text{NL}}^2(35 f+46)}{210 \pi ^2}  \ , \quad c_2^{\pi v} =-\frac{\sigma^2 k_{\text{NL}}^2 (15 f+11) }{150 \pi ^2} \ , \nonumber \\
& c_3^{\pi v} =-\frac{\sigma^2  k_{\text{NL}}^2\left(147 \left(35 b_1+48\right) f+9156 b_1+2304 b_2+2240 b_3+33600 b_8+9261\right) }{30870 \pi
   ^2 f}  \ ,\nonumber  \\
   & c_4^{\pi v} =\frac{\sigma ^2 k_{\text{NL}}^2  \left(-6 b_2 (245 f+514)+273 b_1+448 b_3+6720 b_8-594 f-1785\right) }{8820 \pi ^2 f}  \ , \nonumber \\
   & c_5^{\pi v}  =\frac{ \sigma ^2 k_{\text{NL}}^2 \left(2394 b_1+2304 b_2+2240 b_3+33600 b_8-75\right)}{30870 \pi ^2}  \ ,\nonumber  \\
   & c_6^{\pi v} =-\frac{ \sigma ^2 k_{\text{NL}}^2 (1715 f+4626) }{25725 \pi ^2}  \ ,\nonumber \\
   & c_7^{\pi v} =-\frac{\sigma ^2  k_{\text{NL}}^2}{679140 \pi ^2 f}  \big(924 b_1 (168 f+97)+3234 b_5 (35 f-2)-205920 b_2 \nonumber  \\
   & \hspace{1in} +19712 b_3+295680 b_8+111078 f+70809 \big) \ , \nonumber 
\end{align}}
\endgroup
\hspace{-0.06in}with all other coefficients set to zero ({\it i.e.} they are degenerate for the observables that we consider).  Above, $\sigma^2 = \int dq \, P_{11}(q)$.   {Notice that, as expected, the $c^{\pi v}_{{\rm DM}, i}$ from \eqn{dmuvmatchingresponse2} are related to the $c_i^{\pi v}$ above when the biases are evaluated on the dark-matter values $b_1 = b_2 = b_3 = b_4 = 1$ and $b_5 , \dots , b_{15} = 0$.  Using the expression for $F_1^{r,ct}$ in \eqn{f1rctexpression}, the expression for $K_1^{r,h,ct}$ in \eqn{k1rhctexpression}, the expression for $F_2^{r,ct}$ in \eqn{f2rct}, the expression for $K_{2}^{r,h,ct}$ in \eqn{k2rhct}, and the basis relation \eqn{dmrssbasisfns}, we find $c_j^{\pi v} \big|_{\text{DM bias}}  = c_{\text{DM},j}^{\pi v} $ for $j = 1 , 2 , 6$, and 
\begin{align}
\begin{split}
c_3^{\pi v} \big|_{\text{DM bias}} & = c_{\rm DM,3}^{\pi v} + \frac{2}{3 f} c_3 + \frac{4}{9 f} c_5 \ ,  \\
c_4^{\pi v}\big|_{\text{DM bias}} & = c_{\rm DM,4} + \frac{7}{99 f} c_3 + \frac{16}{33 f} c_4 - \frac{7}{99 f} c_5  \ , \\
c_5^{\pi v} \big|_{\text{DM bias}} & = c_{\rm DM,5}^{\pi v} - \frac{2}{3} c_3 - \frac{4}{9} c_5 \ , \\
c_7^{\pi v} \big|_{\text{DM bias}} & = c_{\rm DM,7}^{\pi v} + \frac{8}{99 f} c_3 + \frac{29}{99 f} c_5 + \frac{16}{33 f} c_7 \ , 
\end{split}
\end{align}
where $c |_{\text{DM bias}}$ means to evaluate $c$ on the dark-matter values for the bias parameters, which one can indeed confirm is true for the UV matching that we found in \eqn{dmuvmatchingresponse}, \eqn{dmuvmatchingresponse2}, and  \eqn{c5pimatching}.}

%%%%%%%%%%%%%5
\subsection{Stochastic terms} \label{stochmatchapp}

The functions that enter the stochastic counterterm $\bar B_{321}^{r,h,(I),\epsilon}$ in \eqn{b321barexpand} are 
\begingroup
\allowdisplaybreaks
{\begin{align}
& e_1^{\rm St} =   f \mu _1^2-1 \ ,     \\
& e_2^{\rm St} = -\frac{k_1^2 \left(k_2^2 \left(1-2 f \mu _1^2\right)+k_3^2\right)+2 f  \left(k_3^2-k_2^2\right) k_1 \mu _1 k_2
   \mu _2+\left(k_2^2-k_3^2\right){}^2}{2 k_1^2 k_{\text{NL}}^2}  \ , \nonumber \\
  & e_3^{\rm St} = 0 \ , \nonumber \\
& e_4^{\rm St} = -\frac{f^2 k_1 \mu _1 \left(k_1^3 \mu _1 \left(2 f \mu _1^2-1\right)+4 f  k_1^2 \mu _1^2 k_2 \mu _2+k_1 \mu _1
   \left(k_2^2 \left(4 f \mu _2^2-1\right)+k_3^2\right)+2  \left(k_3^2-k_2^2\right) k_2 \mu _2\right)}{4 k_1^2
   k_{\text{NL}}^2}   \ ,   \nonumber  \\
& e_5^{\rm St} =  \frac{f^2 k_1 \mu _1 \left(4 f  k_1^2 \mu _1^2 k_2 \mu _2+k_1 \mu _1 \left(k_2^2 \left(4 f \mu
   _2^2-1\right)+k_3^2\right)+k_1^3 \mu _1+2  \left(k_3^2-k_2^2\right) k_2 \mu _2\right)}{4 k_1^2
   k_{\text{NL}}^2}   \ , \nonumber   \\
& e_6^{\rm St} =    2   \ , \quad  e_7^{\rm St} =    -\frac{k_2^2+k_3^2}{k_{\text{NL}}^2}     \ , \quad e_8^{\rm St} =    -\frac{k_1^4+\left(k_2^2-k_3^2\right){}^2}{2 k_1^2 k_{\text{NL}}^2}   \ , \nonumber   \\
& e_9^{\rm St} =    -\frac{k_1^2}{k_{\text{NL}}^2}   \ , \quad e_{10}^{\rm St} =    -\frac{f \left(k_1 \mu _1+2 k_2 \mu _2\right) \left(\left(k_1^2-k_2^2+k_3^2\right) k_1 \mu _1+2 k_1^2 k_2 \mu _2\right)}{4 k_1^2 k_{\text{NL}}^2}    \ , \nonumber \\
& e_{11}^{\rm St} =     \frac{f k_1 \mu _1 \left( \left(k_1^2+k_2^2-k_3^2\right)k_1  \mu _1+2  \left(k_2^2-k_3^2\right) k_2 \mu _2\right)}{2 k_1^2 k_{\text{NL}}^2}   \ , \quad e_{12}^{\rm St} =    -\frac{2 f k_2 \mu _2 \left(k_1 \mu _1+k_2 \mu _2\right)}{k_{\text{NL}}^2}  \nonumber  \ ,   \\
& e_{13}^{\rm St} =   \frac{f}{4 k_1^2 k_2^2 k_3^2 k_{\text{NL}}^2}  \Big(  \left(k_1^2-k_2^2+k_3^2\right){}^2 k_1^2 \mu _1^2 k_2^2+2   \left(k_1^2-k_2^2+k_3^2\right){}^2 k_1 \mu _1 k_2^3 \mu
   _2    \nonumber  \\
   & \hspace{1in} +\left(\left(k_2^2+k_3^2\right) k_1^4-2 \left(k_2^2-k_3^2\right){}^2 k_1^2+\left(k_2^2-k_3^2\right){}^2 \left(k_2^2+k_3^2\right)\right) k_2^2 \mu
   _2^2  \Big)  \ .       \nonumber 
\end{align}}
\endgroup
\hspace{-.06in}All of the above $e_i^{\rm St}$ are symmetric when swapping $\kvec_2$ and $\kvec_3$, as expected from \eqn{b321epdef2}.  To see it, one must swap $k_2 \leftrightarrow k_3$ and  $\mu_2 \leftrightarrow \mu_3$, and then replace $\mu_3 = - k_3^{-1}( k_1 \mu_1 + k_2 \mu_2)$.  

Since, for the stochastic terms, we match terms of order $k^0$ and $k^2$, there are non-zero contributions coming from expanding factors of $P_{11} ( | \kvec - \qvec|) $ for small $k/q$ inside of the loops in \eqn{bispexpressionsrssbias}.  Thus, UV matching includes terms proportional to $P_{11}'(q)$ and $P_{11}''(q)$.   {We give the full expressions for all terms below, apart from $c_7^{\rm St}$ and $c_8^{\rm St}$, which are too long to display here; all full values are given in the accompanying Mathematica notebook.}   For the UV matching, we find
\begingroup
\allowdisplaybreaks
{\begin{align}
& c_1^{\rm St} =  -\frac{ \omega ^2 \bar{n} \left(-b_1+b_2+b_5\right){}^2 }{\pi ^2}  \ ,  \\
&   c_2^{\rm St} =  \frac{   \bar{n} k_{\text{NL}}^2  \left(b_1-b_2-b_5\right)  \left[  \left(14 b_1- 16 b_2\right) \gamma^2 - 7 (b_1 - b_2 - b_5) (  2 \gamma_1^2+  \gamma_2^2 )  \right] }{42 \pi ^2}  \nonumber  \ ,  \\
& c_3^{\rm St} =   -\frac{ \gamma ^2 \bar{n} k_{\text{NL}}^2  \left(b_1-b_2-b_5\right)  \left(-7 b_1+7 f+9\right) }{21 \pi ^2}  \ , \nonumber   \\
& c_4^{\rm St} =  \frac{\gamma ^2 \bar{n}  k_{\text{NL}}^2 \left(b_1 (35 f+54)-35 \left(b_2+b_5\right) f-2 \left(19 b_2+8 b_3+23 b_5+8 b_6+22 b_8\right)\right) }{105 \pi
   ^2 f}   \ ,  \nonumber  \\
& c_5^{\rm St} =    0   \ , \nonumber \\
& c_6^{\rm St} =     -\frac{ \omega ^2 \bar{n} \left(b_1-b_2-b_5\right) \left(13 b_1+34 b_2-47 b_3+42 b_5-110 b_6-82 b_8-63 b_{10}\right) }{21 \pi ^2}  \ , \nonumber  \\
& c_7^{\rm St} =   -\frac{\gamma ^2 \bar{n} k_{\text{NL}}^2  }{1470 \pi ^2} \Big(301 b_1^2+\left(656 b_2-7 \left(183 b_3-124 b_5+282 b_6+638 b_8+105 b_{10}\right)\right) b_1\nonumber  \\
& \hspace{1in} - 982 b_2^2+14 b_5 \left(47 b_3-35 b_5+44
   b_6+236 b_8\right) \nonumber \\
   & \hspace{1in} +2 b_2 \left(653 b_3-784 b_5+1052 b_6+2228 b_8+420 b_{10}\right)\Big)  {+ \mathcal{O}(P_{11}'(q), P_{11}''(q))} \ ,   \nonumber \\
& c_8^{\rm St} =   \frac{ \gamma ^2 \bar{n} k_{\text{NL}}^2 }{4410 \pi ^2} \Big(-21 b_1^2+\left(1689 b_2-7 \left(122 b_3-51 b_5+168 b_6+192 b_8\right)\right) b_1 - 1188 b_2^2 \nonumber \\
& \hspace{.5in} +2 \left(\left(187 b_3-546 b_5+348 b_6+12
   b_8\right) b_2+7 b_5 \left(13 b_3+36 b_6-36 b_8\right)\right)\Big) {+ \mathcal{O}(P_{11}'(q), P_{11}''(q))}   \ ,   \nonumber \\
& c_9^{\rm St} =    -\frac{  \gamma ^2 \bar{n} k_{\text{NL}}^2 \left(7 b_1-6 b_2-14 b_5\right) \left(b_1-b_2-b_5\right)}{42 \pi ^2}   \ , \nonumber \\
& c_{10}^{\rm St} =    \frac{\gamma ^2 \bar{n}  k_{\text{NL}}^2}{735 \pi
   ^2}  \Big(7 b_1^2 (35 f+22)+b_1 \big(-7 b_2 (35 f+79)-7 b_5 (35 f+103)+399 b_3  \nonumber \\
   & \hspace{1in} +1134 b_6+546 b_8+735 b_{10}+224 f+256\big) \nonumber \\
   & \hspace{1in} +b_2
   (287 f+417)-7 \left(73 b_3-57 b_5+178 b_6+122 b_8+105 b_{10}\right) f  \nonumber \\
   & \hspace{1in} -673 b_3+553 b_5-1618 b_6-1142 b_8-945 b_{10}\Big)   \ ,    \nonumber\\
& c_{11}^{\rm St} =   \frac{\gamma ^2 \bar{n} k_{\text{NL}}^2 }{1470 \pi ^2} \Big(7 b_1^2 (6-35 f)  \nonumber\\
& \hspace{.5in} +b_1 \left(7 b_2 (35 f+123)+49 b_5 (5 f+13)-623 b_3-1358 b_6-1162 b_8-735 b_{10}-469 f-634\right)  \nonumber\\
&  \hspace{.5in} +511
   b_3 f-154 b_5 f-b_2 \left(280 b_5+42 f+151\right)+1246 b_6 f  \nonumber\\
   & \hspace{.5in}  +854 b_8 f+105 b_{10} (7 f+9)-280 b_2^2+785 b_3-231 b_5+1730 b_6+1450 b_8\Big)  \ , \nonumber \\
& c_{12}^{\rm St} =  \frac{ \gamma ^2 \bar{n} k_{\text{NL}}^2 \left(7 b_1^2+\left(-109 b_2+32 b_3-93 b_5+32 b_6+88 b_8\right) b_1+70 \left(b_2+b_5\right){}^2\right)}{210
   \pi ^2}     \ ,  \nonumber\\
& c_{13}^{\rm St} =   \frac{  \gamma ^2 \bar{n} k_{\text{NL}}^2\left(-97 b_1+b_2+96 b_3+49 b_5+96 b_6+264 b_8\right)}{1470 \pi ^2}   \ ,   \nonumber 
\end{align}}
\endgroup
\hspace{-.06in}where $\gamma^2 = \int dq \, P_{11}(q)^2$, $\gamma_1^2 = \int dq \, q P_{11}(q) P_{11}'(q)$, $\gamma_2^2 = \int dq \, q^2 P_{11}(q) P_{11}''(q)$, and $\omega^2 = \int d q \, q^2 \, P_{11} ( q)^2$. 

UV matching for \eqn{b222hstoch} is  
\begin{align}
\begin{split}
c_1^{(222)} &  = \frac{4  (b_1 - b_2 - b_5)^3 \bar n^2}{ \pi^2} \int dq \, q^2 P_{11}(q)^3 \ , \\
c_2^{(222)} &  = - \frac{ 2 (b_1 - b_2 -b_5)^2 \bar n^2  \knl^2}{63 \pi^2}\left( 3 ( 7 b_1 - 8 b_2) \vartheta^3 - 7 ( b_1 - b_2 -b_5) ( 4 \vartheta_1^3 + \vartheta_2^3 + 2 \vartheta_3^3 ) \right)  \ ,\\
c_5^{(222)} &  =  \frac{ 4 (b_1 - b_2 -b_5)^2 (-9 + 7b_1 -7 f )f \bar n^2 \knl^2 \vartheta^3}{21 \pi^2} \ . 
\end{split}
\end{align}
{where $\vartheta^3 = \int dq \,  P_{11}(q)^3$, $\vartheta_1^3 = \int dq \, q P_{11}(q)^2 P_{11}'(q) $, $\vartheta_2^3 = \int dq \, q^2 P_{11}(q) P_{11}' ( q )^2$, and $\vartheta_3^3 = \int dq \, q^2 P_{11}(q)^2 P_{11}''(q)$. }

%%%%%%%%%%%%%5
\subsection{Parameter matching with \texttt{PyBird}} \label{pybirdapp}

{      {For reference, we give the conversion between the parameters used here and those used in \texttt{PyBird} (which we write in the \texttt{typewriter font}).  For the counterterm parameters, we have 
\begin{align}
\begin{split}
& \texttt{Bc1} = c^h_1 \ , \quad \texttt{Bc2} = c^\pi_1 \ , \quad \texttt{Bc3} = c^{\pi v}_1 \ , \quad \texttt{Bc4} = c^{\pi v}_3 \ , \quad \texttt{Bc5} = c^h_2 \ , \quad \texttt{Bc6} = c^h_3 \ ,  \quad \texttt{Bc7} = c^h_4  \\
 &  \texttt{Bc8} = c^h_5 \ , \quad \texttt{Bc9} = c^{\pi}_5 \ , \quad \texttt{Bc10} = c^{\pi v}_2 \ , \quad \texttt{Bc11} = c^{\pi v}_4 \ , \quad \texttt{Bc12} = c^{\pi v}_5 \ , \quad  \texttt{Bc13} = c^{\pi v}_6 \ , \\
 & \texttt{Bc14} = c^{\pi v}_7  \ , \quad  \texttt{ce2} = (2/3) f c^{\rm St}_3 \ , \quad \texttt{Be1} = c^{\rm St}_1 \ , \quad \texttt{Be2} = c^{\rm St}_2 \ , \quad \texttt{Be3} = c^{\rm St}_4 \ , \quad \texttt{Be4} = c^{\rm St}_5 \ , \\
   & \texttt{Be5} = 2c^{\rm St}_6 \ , \quad \texttt{Be6} = 2 c^{\rm St}_9 \ , \quad \texttt{Be7} = - c^{\rm St}_7 - c^{\rm St}_9 \ , \quad \texttt{Be8} = - c^{\rm St}_8 - 2 c^{\rm St}_9 \ , \quad \texttt{Be9} = 2 c^{\rm St}_{12} \\
   & \texttt{Be10} = c^{\rm St}_{11}   \ , \quad  \texttt{Be11} = c^{\rm St}_{10}  \ , \quad  \texttt{Be12} = c^{\rm St}_{13}  \ , \\
   &   \texttt{Bd1} = c^{(222)}_1  \ , \quad  \texttt{Bd2} = c^{(222)}_2 -c^{(222)}_5/6  \andd  \texttt{Bd3} = - c^{(222)}_5  \ . 
\end{split}
\end{align}
For the bias parameters, we have $\texttt{Bbi} = b_i$ for $i = 1 , \dots, 6, 8, \dots , 11$, and $\texttt{Bb7} = b_7+15 b_{13}/2$.  Other parameters in the \texttt{PyBird} code are derived from the ones above, and were used in the power-spectrum-only analysis.  These are given by 
\begin{align}
\begin{split}
& \texttt{b1} = \texttt{Bb1} \ , \quad  \texttt{b2} = \texttt{Bb2} \ , \quad  \texttt{b3} = \texttt{Bb3} + 15 \, \texttt{Bb8} \ , \quad  \texttt{b4} = \texttt{Bb5} \ , \quad \texttt{cct} = - \texttt{Bc1} \ ,  \\
& \texttt{cr1} = f \texttt{Bc2} - f^2 \texttt{Bc4}/2  \ , \quad  \texttt{cr2} = - f^2 \texttt{Bc3}/2  \ , \quad \texttt{ce0} = \texttt{Be1}  \ , \quad \texttt{ce1} = \texttt{Be2} + \texttt{ce2}/2 \ , \\
& \texttt{c2} = ( \texttt{b2} + \texttt{b4}) / \sqrt{2} \andd \texttt{c4} = ( \texttt{b2} - \texttt{b4})/\sqrt{2} \ .
\end{split}
\end{align}
}

%%%%%%%%%%%%%
%
%
%
%

%\bibliographystyle{utphys_UPDATE}
 \bibliographystyle{JHEP}
 \small
\bibliography{references_3}

 \end{document}